\documentclass[10pt,journal,compsoc]{IEEEtran}

\usepackage{hyperref}
\usepackage{multirow}
\usepackage{siunitx}
\usepackage{balance}
\usepackage{soul,color}
\usepackage{ulem}[normalem] 
\usepackage{amssymb}
\usepackage{tikz}

\ifCLASSOPTIONcompsoc
  \usepackage[nocompress]{cite}
\else
  \usepackage{cite}
\fi

\usepackage{hyperref}
\usepackage{multirow}
\usepackage{siunitx}
\usepackage{balance}
\usepackage{adjustbox}

\ifCLASSOPTIONcompsoc
  \usepackage[nocompress]{cite}
\else
  \usepackage{cite}
\fi
\ifCLASSINFOpdf
   \usepackage{graphicx}
  \graphicspath{{../pdf/}{../jpeg/}}
  \DeclareGraphicsExtensions{.pdf,.jpeg,.png}
\else
   \usepackage{graphicx}
\fi
\usepackage{amsmath}

\ifCLASSOPTIONcompsoc
  \usepackage[caption=false,font=footnotesize,labelfont=sf,textfont=sf]{subfig}
\else
  \usepackage[caption=false,font=footnotesize]{subfig}
\fi

\begin{document}

\title{An Evaluation Testbed for Locomotion in Virtual~Reality}

\author{Alberto Cannav\`o, \textit{Student Member,~IEEE}, Davide Calandra, F. Gabriele Prattic\`o, \textit{Student Member,~IEEE}, Valentina Gatteschi and Fabrizio Lamberti, \textit{Senior Member,~IEEE}
\IEEEcompsocitemizethanks{\IEEEcompsocthanksitem The authors are with the GRAINS -- GRAphics And INtelligent Systems group at the Dipartimento di Automatica e Informatica of Politecnico di Torino, 10129 Torino, Italy. e-mail: (see http://grains.polito.it/people.php).}
\thanks{Manuscript received XXX XX, XXXX; revised XXX XX, XXXX.}}

\markboth{IEEE TRANSACTIONS ON VISUALIZATION AND COMPUTER GRAPHICS,~Vol.~XX, No.~X, XXXX~XXXX}%
{Shell \MakeLowercase{\textit{et al.}}: Bare Demo of IEEEtran.cls for Computer Society Journals}

\IEEEtitleabstractindextext{%
\begin{abstract}
A common operation performed in Virtual Reality (VR) environments is locomotion. Although real walking can represent a natural and intuitive way to manage displacements in such environments, its use is generally limited by the size of the area tracked by the VR system (typically, the size of a room) or requires expensive technologies to cover particularly extended settings. A number of approaches have been proposed to enable effective explorations in VR, each characterized by different hardware requirements and costs, and capable to provide different levels of usability and performance. However, the lack of a well-defined methodology for assessing and comparing available approaches makes it difficult to identify, among the various alternatives, the best solutions for selected application domains. To deal with this issue, this paper introduces a novel evaluation testbed which, by building on the outcomes of many separate works reported in the literature, aims to support a comprehensive analysis of the considered design space. An experimental protocol for collecting objective and subjective measures is proposed, together with a scoring system able to rank locomotion approaches based on a weighted set of requirements. Testbed usage is illustrated in a use case requesting to select the technique to adopt in a given application scenario.   
\end{abstract}

\begin{IEEEkeywords}
Virtual Reality, virtual environments, locomotion, performance, user experience, requirements, evaluation, testbed.
\end{IEEEkeywords}}

\maketitle

\IEEEdisplaynontitleabstractindextext

\IEEEpeerreviewmaketitle

\definecolor{l1color}{RGB}{192,0,0}
\definecolor{l2color}{RGB}{55,86,35}
\definecolor{l3color}{RGB}{37,42,247}
\definecolor{l4color}{RGB}{100,100,100}
\definecolor{l5color}{RGB}{112,48,160}

\definecolor{l1_1color}{RGB}{246,180,100}
\definecolor{l1_2color}{RGB}{255,155,37}
\definecolor{l1_3color}{RGB}{255,0,0}
\definecolor{l1_4color}{RGB}{192,0,0}

\definecolor{l2_1color}{RGB}{152,216,157}
\definecolor{l2_2color}{RGB}{93,237,124}
\definecolor{l2_3color}{RGB}{103,243,7}
\definecolor{l2_4color}{RGB}{84,130,53}
\definecolor{l2_5color}{RGB}{55,86,35}

\definecolor{l3_1color}{RGB}{101,205,237}
\definecolor{l3_2color}{RGB}{42,156,226}
\definecolor{l3_3color}{RGB}{37,42,247}

\definecolor{l4_1color}{RGB}{217,217,217}
\definecolor{l4_2color}{RGB}{166,166,166}
\definecolor{l4_3color}{RGB}{100,100,100}

\definecolor{l5_1color}{RGB}{220,197,237}
\definecolor{l5_2color}{RGB}{170,114,212}
\definecolor{l5_3color}{RGB}{112,48,160}


\IEEEraisesectionheading{\section{Introduction}\label{sec:introduction}}

\IEEEPARstart{T}{he} continuous grow of VR (Virtual Reality) technology and its applications is posing developers a number of challenges, concerning how to provide users with virtual experiences capable to mimic as much as possible the real ones. When virtual environments are large enough, one of the tasks to be supported is locomotion, which allows users to freely move in the 3D world and explore it \cite{bowman20043d}. When scenarios are not purely exploratory, locomotion is often performed together with other tasks that may require the users to interact with objects populating the virtual world, e.g., to grab and manipulate them. In these cases, locomotion can be regarded as a \textit{secondary} task, and aspects such as intuitiveness and ease of use become essential for the outcome of the \textit{primary} task
\cite{Nilsson:2018:NWV:3181320.3180658}. 

Previous works showed that the most natural locomotion experience can be achieved by providing the users with the possibility to physically walk in the real world while being immersed in VR \cite{ruddle2006efficient, waller2013sensory, suma2007comparison}. 
%
Unfortunately, the use of physical, or real, walking is often constrained by the amount of required space  and/or the limited area that can be tracked by consumer-level VR products based on \textit{outside-in tracking} technology.
%
Recently, an alternative technology known as \textit{inside-out tracking} has been introduced to tackle the latter limitation, as well as to lower the prices and increase the portability of VR systems. However, devices based on this technology are usually characterized by a reduced tracking accuracy and by a limited region in front of the headset in which the hand controllers can be tracked. Indeed, systems specifically designed to support high accuracy, wide-area tracking also exist, but they are generally very expensive and require large amounts of obstacle-free space \cite{suma2007comparison}.
 
Because of the above scenario, there is a growing body of literature that recognizes the importance of designing \textit{stationary} locomotion techniques able to make the users explore virtual worlds regardless of the amount of space available in the real world \cite{garg2017ares}. Many alternatives have been proposed, each characterized by different hardware requirements, costs and levels of effectiveness in deceiving the human sensory system, i.e., in making it perceive the virtual movements as realistic. %
%
Notwithstanding, finding those which can perform better in a given application domain (or set of domains) is not an easy task. This situation is confirmed by the large number of studies that presented custom evaluation approaches (both in terms of experiments to perform, as well as of data to collect and analyze), often motivated by the need to compare the performance of a newly-proposed technique with related ones \cite{nilsson2013perceived, bowman1997travel, whitton2005comparing, schuemie2005effect}, etc. As a result, the literature does not offer researchers and practitioners a standard methodology for investigating the problem, but rather provides many fragmented partial solutions, often limited to a particular and non-generalizable use case. 

For these reasons, the purpose of the present work is to propose a comprehensive testbed \cite{testbed} designed to compare VR locomotion techniques. In particular, the contribution of this paper can be split in three parts. A \textit{methodology} supporting the experimental investigation of various techniques from many different perspectives is first proposed. The methodology, grounded on available literature, supports the collection of both objective and subjective measures (metrics) concerning users' performance and experience in the execution of a number of locomotion-related tasks arranged in a set of representative scenarios. This methodology is accompanied by a \textit{scoring system}, which is designed to combine collected experimental data  with a weighted set of requirements and provide a ranking of selected locomotion techniques based on intended usage conditions. Finally, the paper includes a \textit{use case}, i.e., an example on how the testbed could be exploited to compare several techniques based on the characteristics of a possible VR application they would be used into. The testbed is released as open source at \href{https://github.com/VRatPolito/LET-VR}{\textit{https://github.com/VRatPolito/LET-VR}} with the aim to let other researchers extend it by adding new scenarios/tasks, requirements/metrics and experimental data.

\section{Related Work}
\label{sec:related_work}
Locomotion in immersive environments is a well-studied research area, as confirmed by the large number of works in the literature which present surveys and systematic reviews on available techniques \cite{Nilsson:2018:NWV:3181320.3180658, boletsis2017new, AlZayer2018Virtual, anthes2016state, cardoso2019survey}. This section firstly introduces locomotion techniques that have been proposed so far. Afterwards, it discusses works that presented methodologies and studies  aimed to evaluate these techniques, by considering both the case in which locomotion represents the primary task to perform in the virtual environment, as well as the case in which it is combined with other tasks. 

\subsection{Locomotion Techniques}
As said, real walking represents the most natural  technique, since it allows the user to control the position of the avatar in the virtual environment by physically moving in the tracked area.
An alternative that is commonly adopted also in desktop or console 3D applications and videogames relies on gamepads and joysticks. 

Since these devices may cause disorientation in immersive virtual environments \cite{lathrop2002perceived}, locomotion techniques specifically tailored to VR started to be experimented by the research community (and to be implemented by the industry, in some cases) \cite{Nilsson:2018:NWV:3181320.3180658}. These techniques are generally designed to leverage users' spatial orientation abilities and minimize the distance between movements they perform with their body and movements produced in the virtual environment. The taxonomy defined in \cite{templeman1999virtual} roughly splits locomotion techniques into ``magical'' and ``mundane''. Magical techniques make the users move in a way that is not feasible in the real word, e.g., through so-called \textit{teleportation} \cite{bozgeyikli2016point}, using the \textit{world-in-miniature} metaphor \cite{stoakley1995virtual}, or via \textit{hand-based manipulations} of the virtual world \cite{stoev2001two}. In contrast, mundane techniques rely on real-world metaphors, and can be either vehicle- or body-centric. The former techniques involve the use of virtual vehicles \cite{fiore2013towards}\cite{wang2012comparing, beckhaus2007chairio}, 
whereas the latter ones are meant to recreate physical walking, running, swimming, etc., by making the user perform the same or similar movements \cite{fels2005swimming}. In particular, body-centric techniques rely on approaches based on \textit{repositioning}, \textit{proxy gestures}, and \textit{redirected walking} \cite{Nilsson:2018:NWV:3181320.3180658}.

Repositioning systems are solutions in which the users' forward movement due to the physical execution of the walking gesture is counterbalanced by a device that constrains them in a fixed position. Repositioning could rely on active components, like motorized platforms and floor tiles \cite{souman2011cyberwalk, iwata2005circulafloor}, human-size hamster balls \cite{medina2008virtusphere}, etc., or exploit passive elements like low-friction
omnidirectional treadmills or slippery shoes-based interfaces able to reduce the friction generated by the users' steps \cite{avila2014virtual, swapp2010implementation, iwata1996virtual}. Compared to active systems, passive systems represent a less expensive and simpler solution. The Cyberith's Virtualizer \cite{cakmak2014cyberith} and KatVR's Kat Walk are examples of commercial systems belonging to the latter category.
Solutions based on proxy gestures allow the users to navigate virtual environments through movements performed either with the upper or the lower part of their body. 
An example of proxy gesture-based solutions are the \textit{leaning} interfaces. They rely on side and forward leaning of the torso for controlling the locomotion, do not require additional hardware, and proved to be characterized by a high usability \cite{guy2015lazynav, kitson2017comparing, kruijff2016your, de2008using, harris2014human}.
Another example is \textit{walking-in-place} (WIP), which has been shown to be slightly more expensive that the above interfaces (since it requires some additional hardware) \cite{feasel2008llcm}, but also proved to be capable of providing some of the proprioceptive stimuli of real walking. A quite comparable user experience, which is generally perceived also as less fatiguing and comes at no extra cost, is obtained by exploiting upper-body gestures in so-called \textit{arm swinging} (AS). In this case, locomotion is obtained through the rhythmic swinging of the arms, similarly to what occurs in a real walk or in cross country skiing \cite{nilsson2013perceived, pai2017armswing}. Lastly, redirected walking includes techniques that influence the user's path through the physical environment by manipulating stimuli offered through/by the virtual environment itself, e.g., changing the user's virtual point of view by applying continuous, hopefully imperceptible, transformations to the mapping between real and virtual movement \cite{suma2012impossible}.

\subsection{Assessment of Locomotion as Primary Task}
Although it is only thanks to rather recent advancements in technology that VR has become largely accessible to various user categories, works that analyze the performance of different locomotion techniques are not new. A first example is represented by the framework proposed more than 20 years ago in \cite{bowman1997travel} to perform a comparison between alternative techniques. 
The main contributions were the proposal of a taxonomy to categorize available methods and the identification of the most relevant quality factors to evaluate their usability (speed, accuracy, spatial awareness, ease of learning, ease of use, information gathering, and presence) together with relevant scenarios for testing them. Because of the technology available at that time, the framework was limited by the fact that, besides head tracking, interaction could only rely on a spatial input device (like a 3D mouse), and the focus was basically on direction and speed control in absolute and relative displacements. 

More recently, a user study was performed \cite{nilsson2013perceived} with the aim to evaluate the perceived naturalness offered by four different locomotion methods, i.e., keyboard, AS, WIP, and \textit{hip movement} (HM), a technique that lets a user move in a virtual environment by swinging its hip left and right. Experiments also investigated the user's sense of presence and the amount of Unintended Positional Drift (UPD), that is, the physical movement in the forward direction occurring while performing the walking gesture. Another aspect whose evaluation was proposed in this work is the similarity in terms of physical strain between real walking and the gestures that the considered locomotion techniques use as a proxy for it. Similarly, in \cite{wilson2016vr}, AS and a diverse implementation of the WIP method based on a consumer electronics bracelet placed on the user's ankles were compared with real walking in terms of spatial orientation and ability to estimate travelled distances. 

Although body-centric techniques were included in the evaluation, the above studies still did not consider the other major approach to locomotion in VR mentioned in the previous section, i.e., repositioning systems. Works like \cite{calandra2018eg} and \cite{calandra2019icce} aimed to tackle this lack, by comparing a passive repositioning system (namely, a slippery shoes-based interface), AS and WIP in terms of usability and motion sickness. Evaluation was performed by asking users to carry out a complex task in a realistic immersive VR scenario simulating an emergency procedure, and by collecting objective and subjective measurements. Unfortunately, since experiments included a single task requesting the users to carry out different actions, authors were not able to isolate the actual contribution of the specific locomotion technique to the success of the task and of individual operations. 

A comparable result was achieved by the authors of \cite{nilsson2013tapping}. In this case, several variations of a single locomotion approach were analyzed. Specifically, the goal was to compare in terms of naturalness, sense of presence and UPD three WIP-based methods, leveraging respectively marching, wiping and tapping gestures. The task used in the evaluation requested the users to walk along a predefined path for a certain amount of time. The simplicity of the task did not allow the authors to find any difference in usability and performance, making them conclude that other tasks had to be designed, e.g., to investigate object avoidance capabilities, analyze the promptness in starting and stopping movements, measure movements accuracy, etc.

An example of accuracy task was presented, e.g., in \cite{whitton2005comparing}, where the users were requested to be as fast as possible in getting close to different targets positioned on walls, without hitting them. The locomotion techniques used in the experiments encompassed real walking, WIP and a joystick-based method, and combined three possible visual conditions: head-mounted display in a computer-generated environment, unrestricted, as well as field of view (FOV)-restricted natural vision in a corresponding real environment. The combination of the above elements led to the generation of five experimental configurations (some setups were deemed as not relevant). To study the impact of a given technique on performance, the authors used the final distance to target as well as a time-to-collision metric, defined as the target-user distance divided by the user's velocity. 

Given the variety of environments and tasks used in these studies, several works focused on the design of systems capable to support the creation of customized VR scenarios where locomotion techniques can be tested into. For instance, in \cite{sarupuri2018lute}, the authors identified a number of key attributes (like environment size, path length, path complexity, presence of obstacles, etc.) that should be controlled in a study on locomotion in VR. Then, they presented a tool that supports the manipulation of these parameters to generate different virtual environments which can stress the techniques of interest in terms of fatigue, motion sickness and usability over short-, medium- and long-distance travels. Although the tool is characterized by a high flexibility in the creation process, it does not come with ways to analyze the performance of the selected techniques, and has not been used yet in comparative studies. 

Finally, there are also works that tried to define methods for ranking techniques based on their ability to address the fundamental requirements of locomotion in VR. For example, in \cite{albert2018user}, a ranking tool considering three dimensions, namely, motion sickness, presence, and fatigue was presented. The proposed approach, referred to as a User-Centric Classification (UCC), can be exploited to plot in a 3D space the performance of a given technique, with each axis representing one of the above requirements. Thanks to the adopted visualization, it is quite easy to compare the techniques of interest along the selected dimensions; unfortunately, a considerable number of aspects which, based on previous works, shall be taken into account, are not considered in the devised evaluation scale.

\subsection{Assessment of Locomotion as Secondary Task}
In the works reviewed so far, the focus was mainly on locomotion itself. However, as said, in many applications like, for instance, videogames, locomotion could be just a secondary task, which might have to be performed together with other (primary) tasks such as searching for an object in the virtual environment, grabbing it, etc. 

Thus, in \cite{pai2017armswing}, a user study was conducted to evaluate the effectiveness of AS compared to WIP for a grabbing task combined with locomotion. Participants were asked to travel different paths while carrying a virtual object, which had to be released into a basket at the end of the path. Techniques were evaluated in terms of immersion, motion sickness, and physical demand. The mental workload dimension, which was not considered in previously cited works, was additionally investigated by using the NASA-TLX questionnaire \cite{nasatlx}. The work in \cite{ferracani2016locomotion} presented a framework to assess naturalness and effectiveness of four techniques including WIP, AS, \textit{Tap} (a metaphorical gesture that allows the users to move by tapping with the index finger in the direction they want to walk) and \textit{Push} (a gesture consisting in closing and opening the hand while dragging it, similar to moving a lever) by considering interaction during locomotion. In particular, during the tests, users were requested to navigate along predefined paths avoiding obstacles and interacting with virtual objects by relocating them. 

A further example of user studies investigating the impact of locomotion on interaction with objects is presented in \cite{wilson2018object}. Authors focused on the effect that different values of translational gain, i.e., the mapping between the physical and the virtual movement, can have on tasks that require picking and placing of virtual objects. Results showed that, although increasing translational gain could represent a seamless way for speeding up locomotion since it does not require any additional hardware, it may negatively affect accuracy, motion sickness, and mental workload for values greater than a given threshold. More recently, another study \cite{mayor2019comparative} investigated the relationships between interaction and locomotion in virtual environments in terms of presence, cybersickness, and usability. A classification of interactions in VR was proposed, based on the following categories: \textit{simple} (interactions performed by manipulating only the three positional degrees of freedom, DOFs) vs \textit{complex} (involving six DOF manipulations), and \textit{sequential} (only one hand involved) vs \textit{parallel} (both hands used simultaneously).

Another frequent task is object search. For instance, in \cite{lapointe2009comparative}, an experiment was designed to compare three bimanual locomotion techniques for desktop virtual walkthroughs. The considered techniques relied on a joystick with different numbers of DOFs which had to be manipulated for controlling an avatar's position in the virtual environment with the dominant hand, combined with a mouse to control the gaze with the other hand. The experiment consisted in a primed search task, where users already knew the position of the target and had to reach it in the shortest time possible. At the end of the experiment, the users were requested to judge ease of use, fatigue, accuracy and speed of the three techniques. In \cite{loup2018effects}, the authors defined a similar task for comparing a mundane technique (AS) with a magical technique (teleporting), whose performance was evaluated in terms of efficacy, effectiveness, motion sickness, user experience and cognitive load. 

Works above mostly focused on performance of the studied locomotion techniques and on perceived user experience, but there are also works that analyzed the impact of techniques on psychological and/or cognitive aspects. For instance, in \cite{schuemie2005effect}, objective and subjective metrics were defined to analyze the influence of a given technique in virtual experiences designed for treating phobias (specifically, the fear of heights), whereas in \cite{suma2010evaluation}, two user studies were performed to observe the effects of four techniques on the users' ability to gather and remember information from/on the virtual world. 

\begin{table*}[!ht]
\caption{Review of the most relevant related works according to the following dimensions (columns): work presents an evaluation testbed (possibly referred to as framework), i.e., a set of scenarios/tasks including also the evaluation metrics (1); collects objective (2) or subjective measures (3); considers other kinds of interactions during locomotion (4); tackles aspects concerning accuracy, input sensitivity, responsiveness, and/or level of control (5), operation speed (6), error-proneness (7), usability (8), presence and immersion (9), motion sickness (10), physical effort, V/R physical strain similarity, comfort, self-motion compellingness, and/or acclimatisation (11), mental effort/workload, and/or cognitive demand (12), ease of use, intuitiveness, and/or naturalness (13), appropriateness and/or effectiveness (14), learnability (15), enjoyability, satisfaction and/or frustration (16).} 
\label{tab:related_work}
\centering
\begin{adjustbox}{max width=0.8\textwidth}
\begin{tabular}{l|l|l|l|l|l|l|l|l|l|l|l|l|l|l|l|l|l}
\cline{2-17}
                                                              & (1) & (2) & (3) & (4)& (5) & (6) & (7) & (8) & (9) & (10) & (11) & (12) & (13) & (14) & (15) & (16)  \\ \hline
\multicolumn{1}{|l|}{\cite{bowman1997travel} Bowman 1997}         & \multicolumn{1}{c|}{\checkmark} & \multicolumn{1}{c|}{\checkmark} &   &   & \multicolumn{1}{c|}{\checkmark} & \multicolumn{1}{c|}{\checkmark} &   &   & \multicolumn{1}{c|}{\checkmark} &    &    &    &    & \multicolumn{1}{c|}{\checkmark}  &    &       \\ \hline
\multicolumn{1}{|l|}{\cite{nilsson2013perceived} Nilsson 2013}             & \multicolumn{1}{c|}{\checkmark} & \multicolumn{1}{c|}{\checkmark} & \multicolumn{1}{c|}{\checkmark} &   &   &   &   &   & \multicolumn{1}{c|}{\checkmark} &    & \multicolumn{1}{c|}{\checkmark}  &   & \multicolumn{1}{c|}{\checkmark} &    &    &       \\ \hline
\multicolumn{1}{|l|}{\cite{wilson2016vr} Wilson 2016}                     &   & \multicolumn{1}{c|}{\checkmark} &   &   & \multicolumn{1}{c|}{\checkmark} &   & \multicolumn{1}{c|}{\checkmark} &   & \multicolumn{1}{c|}{\checkmark} &    &    &    &    &    &    &       \\ \hline
\multicolumn{1}{|l|}{\cite{calandra2018eg} Calandra 2018}                   &   & \multicolumn{1}{c|}{\checkmark} & \multicolumn{1}{c|}{\checkmark} & \multicolumn{1}{c|}{\checkmark} & \multicolumn{1}{c|}{\checkmark} & \multicolumn{1}{c|}{\checkmark} & \multicolumn{1}{c|}{\checkmark} & \multicolumn{1}{c|}{\checkmark} & \multicolumn{1}{c|}{\checkmark} & \multicolumn{1}{c|}{\checkmark}  &    &    &    &    &    &        \\ \hline
\multicolumn{1}{|l|}{\cite{calandra2019icce} Calandra 2019}                 &   & \multicolumn{1}{c|}{\checkmark} & \multicolumn{1}{c|}{\checkmark} & \multicolumn{1}{c|}{\checkmark} & \multicolumn{1}{c|}{\checkmark} & \multicolumn{1}{c|}{\checkmark} & \multicolumn{1}{c|}{\checkmark} & \multicolumn{1}{c|}{\checkmark} & \multicolumn{1}{c|}{\checkmark} & \multicolumn{1}{c|}{\checkmark}  &    &    &    &    &    &        \\ \hline
\multicolumn{1}{|l|}{\cite{nilsson2013tapping} Nilsson 2013}               &   & \multicolumn{1}{c|}{\checkmark} & \multicolumn{1}{c|}{\checkmark} &   &   &   &   &   & \multicolumn{1}{c|}{\checkmark} &    & \multicolumn{1}{c|}{\checkmark} &   &  \multicolumn{1}{c|}{\checkmark}   &    &    &        \\ \hline
\multicolumn{1}{|l|}{\cite{whitton2005comparing} Whitton 2005}             & \multicolumn{1}{c|}{\checkmark} & \multicolumn{1}{c|}{\checkmark} & \multicolumn{1}{c|}{\checkmark} &   &   &   &   & \multicolumn{1}{c|}{\checkmark} &  &   &    &     &    & \multicolumn{1}{c|}{\checkmark} &   &         \\ \hline
\multicolumn{1}{|l|}{\cite{sarupuri2018lute} Sarapuri 2018}                 &   &   &   &   &   &   &   &   &   & \multicolumn{1}{c|}{\checkmark}  & \multicolumn{1}{c|}{\checkmark}  &    &    &    &    &        \\ \hline
\multicolumn{1}{|l|}{\cite{albert2018user} Albert 2018}                   & \multicolumn{1}{c|}{\checkmark} &   & \multicolumn{1}{c|}{\checkmark} &   &   &   &   &   & \multicolumn{1}{c|}{\checkmark} & \multicolumn{1}{c|}{\checkmark} & \multicolumn{1}{c|}{\checkmark} &    &    &    &    &        \\ \hline
\multicolumn{1}{|l|}{\cite{pai2017armswing} Pai 2017}                  &  & \multicolumn{1}{c|}{\checkmark} & \multicolumn{1}{c|}{\checkmark} & \multicolumn{1}{c|}{\checkmark} &   &   &   &   & \multicolumn{1}{c|}{\checkmark} & \multicolumn{1}{c|}{\checkmark} & \multicolumn{1}{c|}{\checkmark} & \multicolumn{1}{c|}{\checkmark} &    &    &    &        \\ \hline
\multicolumn{1}{|l|}{\cite{ferracani2016locomotion} Ferracani 2016}          & \multicolumn{1}{c|}{\checkmark} & \multicolumn{1}{c|}{\checkmark} & \multicolumn{1}{c|}{\checkmark} & \multicolumn{1}{c|}{\checkmark} & \multicolumn{1}{c|}{\checkmark} & \multicolumn{1}{c|}{\checkmark} & \multicolumn{1}{c|}{\checkmark} &   &   &    &    &    & \multicolumn{1}{c|}{\checkmark} &    &    &        \\ \hline
\multicolumn{1}{|l|}{\cite{wilson2018object} Wilson 2018}                 & \multicolumn{1}{c|}{\checkmark} & \multicolumn{1}{c|}{\checkmark} & \multicolumn{1}{c|}{\checkmark} & \multicolumn{1}{c|}{\checkmark} & \multicolumn{1}{c|}{\checkmark} & \multicolumn{1}{c|}{\checkmark} &   &   &   & \multicolumn{1}{c|}{\checkmark} & \multicolumn{1}{c|}{\checkmark} & \multicolumn{1}{c|}{\checkmark} &    &    &    &        \\ \hline
\multicolumn{1}{|l|}{\cite{mayor2019comparative} Mayor 2019}             & \multicolumn{1}{c|}{\checkmark} & \multicolumn{1}{c|}{\checkmark} & \multicolumn{1}{c|}{\checkmark} &   &   & \multicolumn{1}{c|}{\checkmark} &   & \multicolumn{1}{c|}{\checkmark} & \multicolumn{1}{c|}{\checkmark} & \multicolumn{1}{c|}{\checkmark} &    &    &    &    &    &        \\ \hline
\multicolumn{1}{|l|}{\cite{lapointe2009comparative} Lapointe 2009}          & \multicolumn{1}{c|}{\checkmark} & \multicolumn{1}{c|}{\checkmark} & \multicolumn{1}{c|}{\checkmark} &   & \multicolumn{1}{c|}{\checkmark} & \multicolumn{1}{c|}{\checkmark} &   &   &   &    & \multicolumn{1}{c|}{\checkmark} &   & \multicolumn{1}{c|}{\checkmark} &    &   & \multicolumn{1}{c|}{\checkmark}      \\ \hline
\multicolumn{1}{|l|}{\cite{loup2018effects} Loup 2018}                  & \multicolumn{1}{c|}{\checkmark} & \multicolumn{1}{c|}{\checkmark} & \multicolumn{1}{c|}{\checkmark} &   & \multicolumn{1}{c|}{\checkmark} &   & \multicolumn{1}{c|}{\checkmark} & \multicolumn{1}{c|}{\checkmark} &   & \multicolumn{1}{c|}{\checkmark}  &    & \multicolumn{1}{c|}{\checkmark} &    &    &    &        \\ \hline
\multicolumn{1}{|l|}{\cite{schuemie2005effect} Schuemie 2005}               & \multicolumn{1}{c|}{\checkmark} & \multicolumn{1}{c|}{\checkmark} & \multicolumn{1}{c|}{\checkmark} &   & \multicolumn{1}{c|}{\checkmark} &   & \multicolumn{1}{c|}{\checkmark} &   & \multicolumn{1}{c|}{\checkmark} & \multicolumn{1}{c|}{\checkmark} &    &    &    &    &    &       \\ \hline
\multicolumn{1}{|l|}{\cite{suma2010evaluation} Suma 2010}               & \multicolumn{1}{c|}{\checkmark} & \multicolumn{1}{c|}{\checkmark} & \multicolumn{1}{c|}{\checkmark} &   &   &   &   &   & \multicolumn{1}{c|}{\checkmark} & \multicolumn{1}{c|}{\checkmark} &    &    &    &    &    &      \\ \hline
\multicolumn{1}{|l|}{This work}             & \multicolumn{1}{c|}{\checkmark} & \multicolumn{1}{c|}{\checkmark} & \multicolumn{1}{c|}{\checkmark} & \multicolumn{1}{c|}{\checkmark} & \multicolumn{1}{c|}{\checkmark} & \multicolumn{1}{c|}{\checkmark} & \multicolumn{1}{c|}{\checkmark} & \multicolumn{1}{c|}{\checkmark} & \multicolumn{1}{c|}{\checkmark} & \multicolumn{1}{c|}{\checkmark}  & \multicolumn{1}{c|}{\checkmark}  & \multicolumn{1}{c|}{\checkmark}  & \multicolumn{1}{c|}{\checkmark}  & \multicolumn{1}{c|}{\checkmark}  & \multicolumn{1}{c|}{\checkmark}  & \multicolumn{1}{c|}{\checkmark}      \\ \hline
\end{tabular}
\end{adjustbox}
\end{table*}

\subsection{Considerations}
From the above analysis it is rather clear that a number of methods for evaluating and/or comparing locomotion methods were developed, each aimed to stress the features of the particular techniques considered. Experiments designed in the various works generally request the users to perform different tasks, each characterized by diverse levels of complexity and evaluated with a varying set of objective and subjective metrics. This heterogeneity is quite evident also from Table~\ref{tab:related_work}, which summarizes the most relevant works discussed in this section from the many considered perspectives (in columns). Only studies on psychological/cognitive aspects are neglected, as not in the scope of this work. 

By moving from the above observations and by leveraging findings in the literature, the present work proposes a comprehensive evaluation testbed that can be used to study locomotion techniques from different viewpoints and rank them based on users' preference and performance in the execution of a variety of tasks (commonly requested in typical VR applications and videogames). As it can be seen in Table~\ref{tab:related_work} (last row), the proposed testbed covers all the dimensions addressed in previous works.


\section{Evaluation Methodology}
\label{sec:evaluation_framework}
The design of the evaluation started with the analysis of experimental studies in the existing literature, which led to the identification of a set of important and recurrent tasks involving locomotion in VR. This set was then enriched with additional tasks, less considered in the literature, but very frequent in common applications. These tasks correspond to functionalities to be possibly supported by a given VR locomotion technique: hence, in the following they will be referred to as \textit{Functional Requirements} (FRs). Tasks were clustered in five scenarios, designed to group tasks/functionalities sharing similar features. 

The execution of the tasks in a given scenario is evaluated through a set of metrics, which can be either objective (when based on measurements automatically collected by the testbed application), or subjective (when based on user-provided answers). The metrics, either derived from dimensions explored by previous works or designed ad-hoc, refer to general characteristics that the technique should offer to the users of a given VR application: thus, they will be later referred to as \textit{Non-Functional Requirements} (NFRs). 

In the following, FRs and NFRs will be illustrated in detail, by referring to works they have been derived from, when appropriate. Instructions for performing the experiments will be also provided.

\subsection{Scenarios and Tasks}
\label{sec:scenarios_and_tasks}
\begin{figure*}%
\captionsetup[subfloat]{farskip=1pt}
\centering
\begin{tabular}{|l|l|l|l|ll}
\hline
\textbf{Scenario} & \multicolumn{5}{l|}{\textbf{Task}} \\ \hline
\multirow[t]{2}{*}{\begin{tabular}[c]{@{}l@{}}\tikz \fill [l1color] (0,0) rectangle (.25,.25); \textit{S1. Straight}\\ \textit{movements} \vspace{3.5cm} \end{tabular}}
& \multicolumn{1}{c|}{\subfloat{\includegraphics*[height=0.95in]{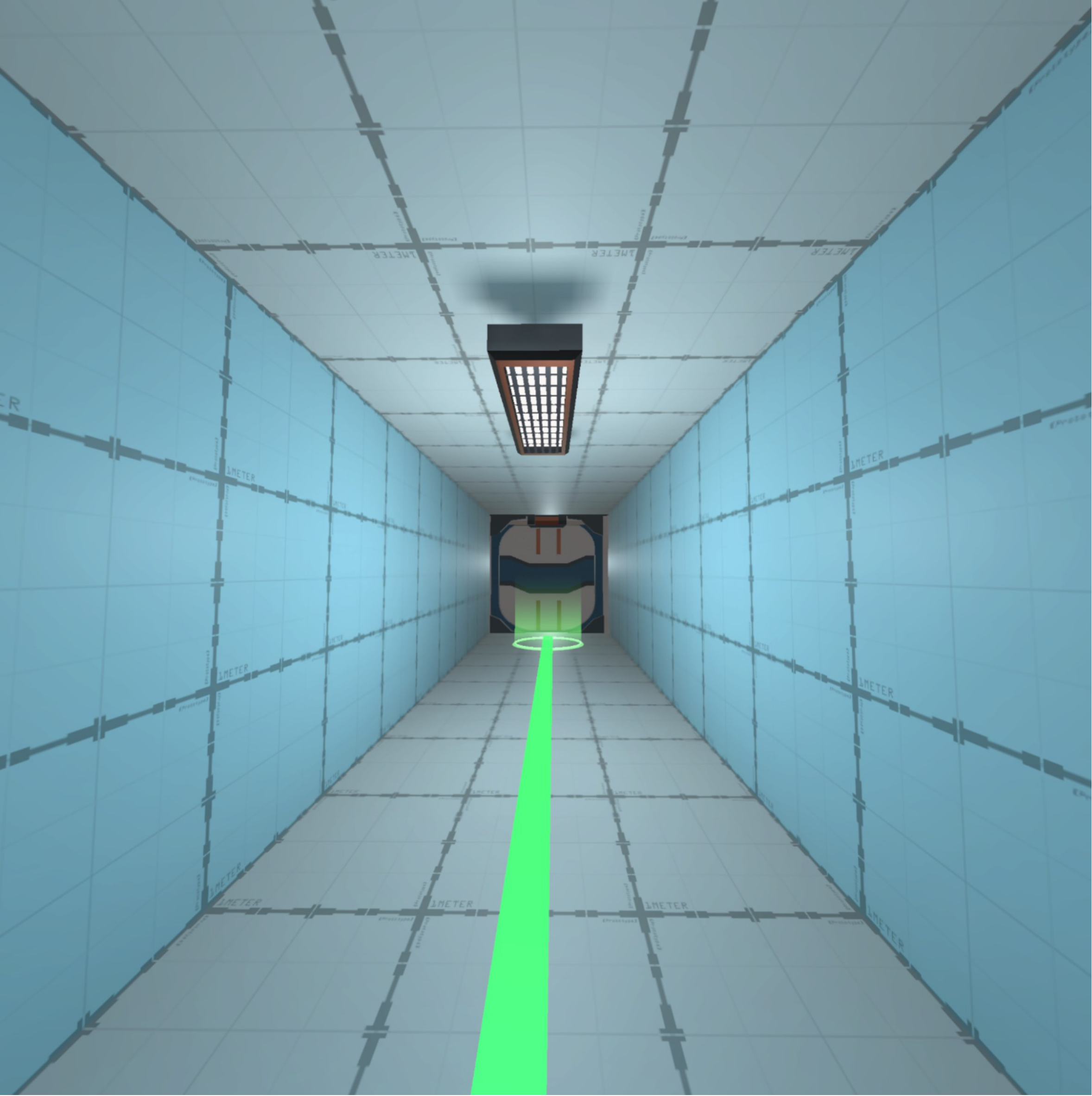}\label{fig:Fig1_a}}}
& \multicolumn{1}{c|}{\subfloat{\includegraphics*[height=0.95in]{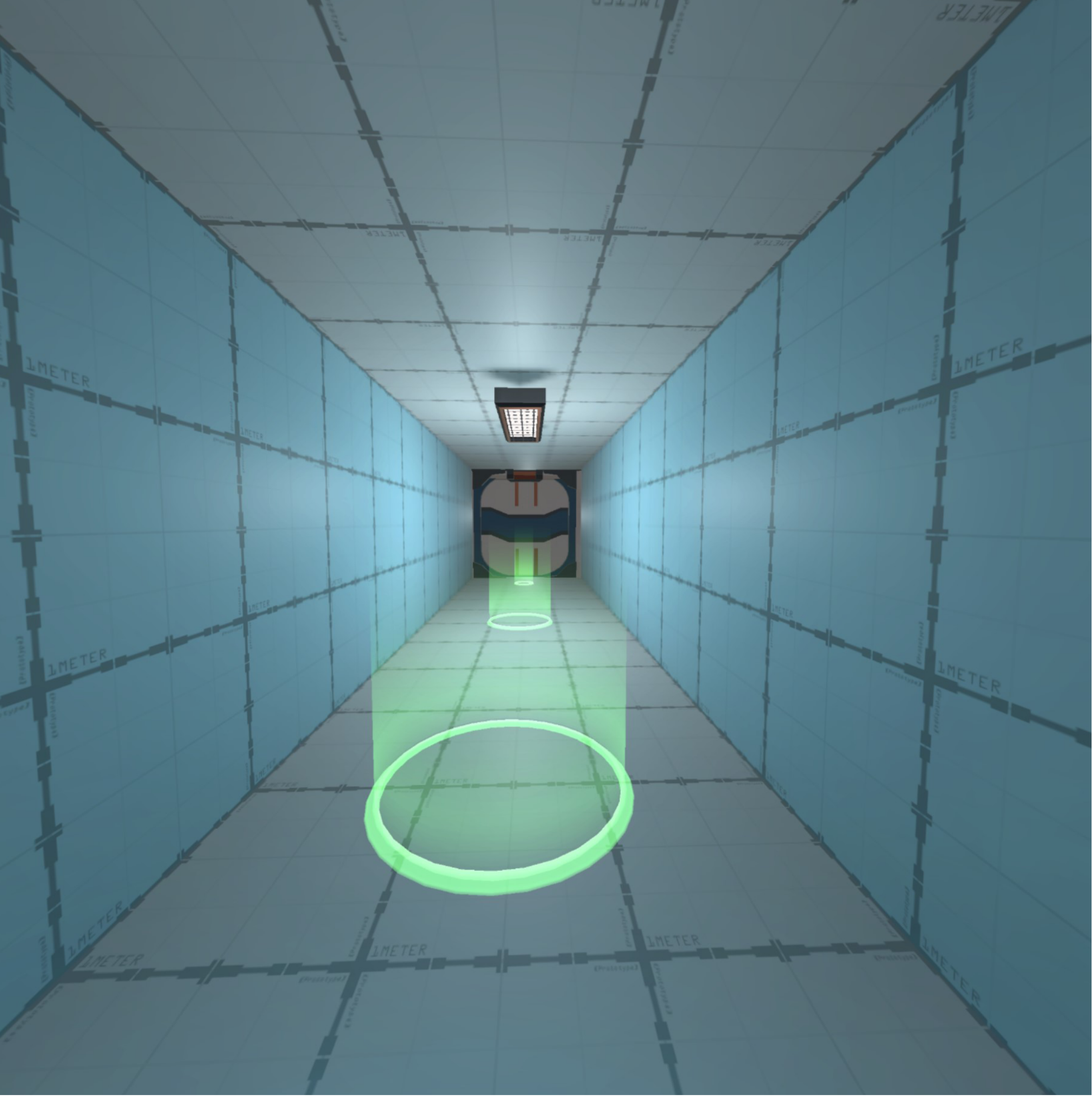}\label{fig:Fig1_b}}}
& \subfloat{\includegraphics*[height=0.95in]{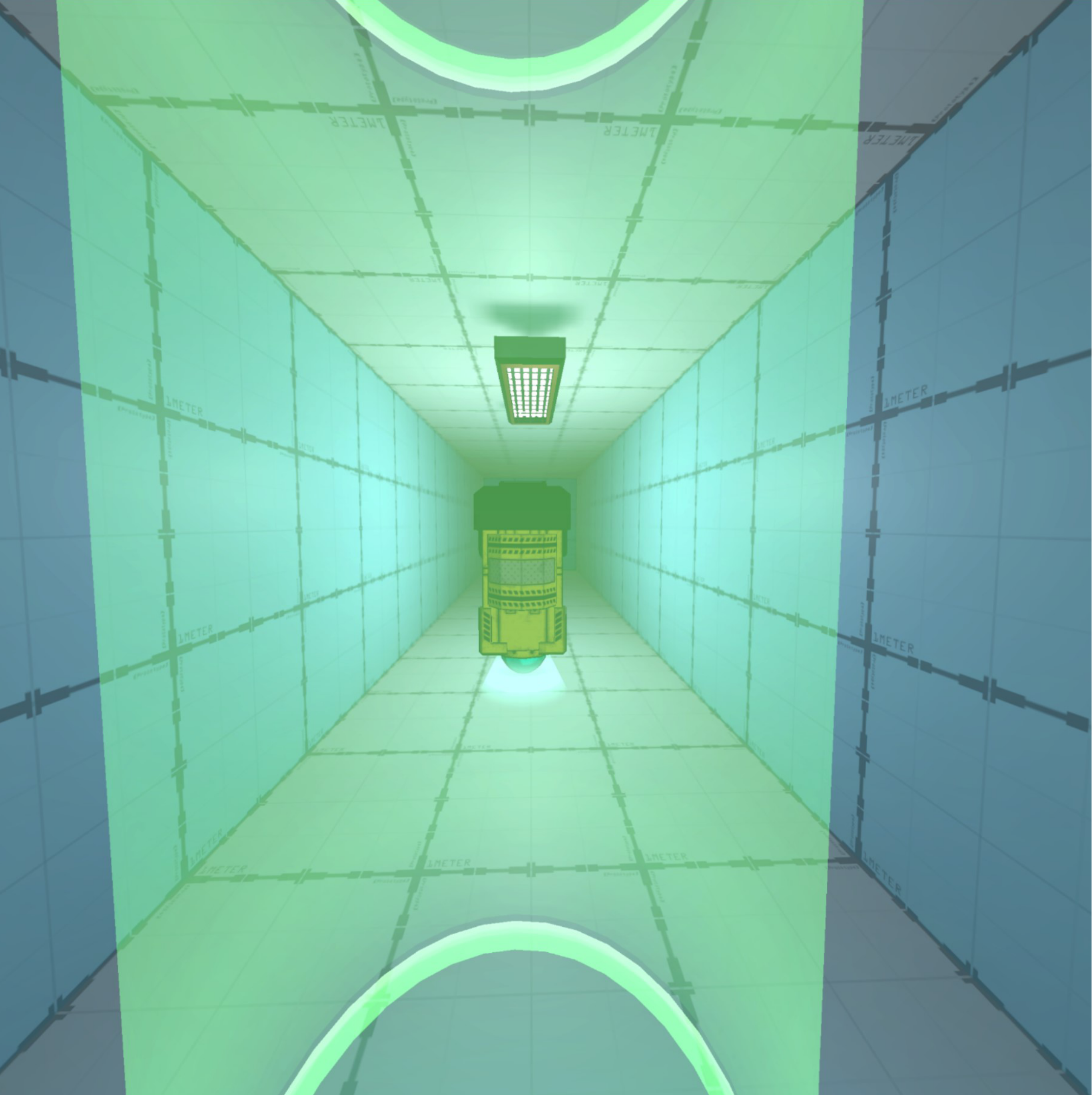}\label{fig:Fig1_c}}
& \multicolumn{1}{c|}{\subfloat{\includegraphics*[height=0.95in]{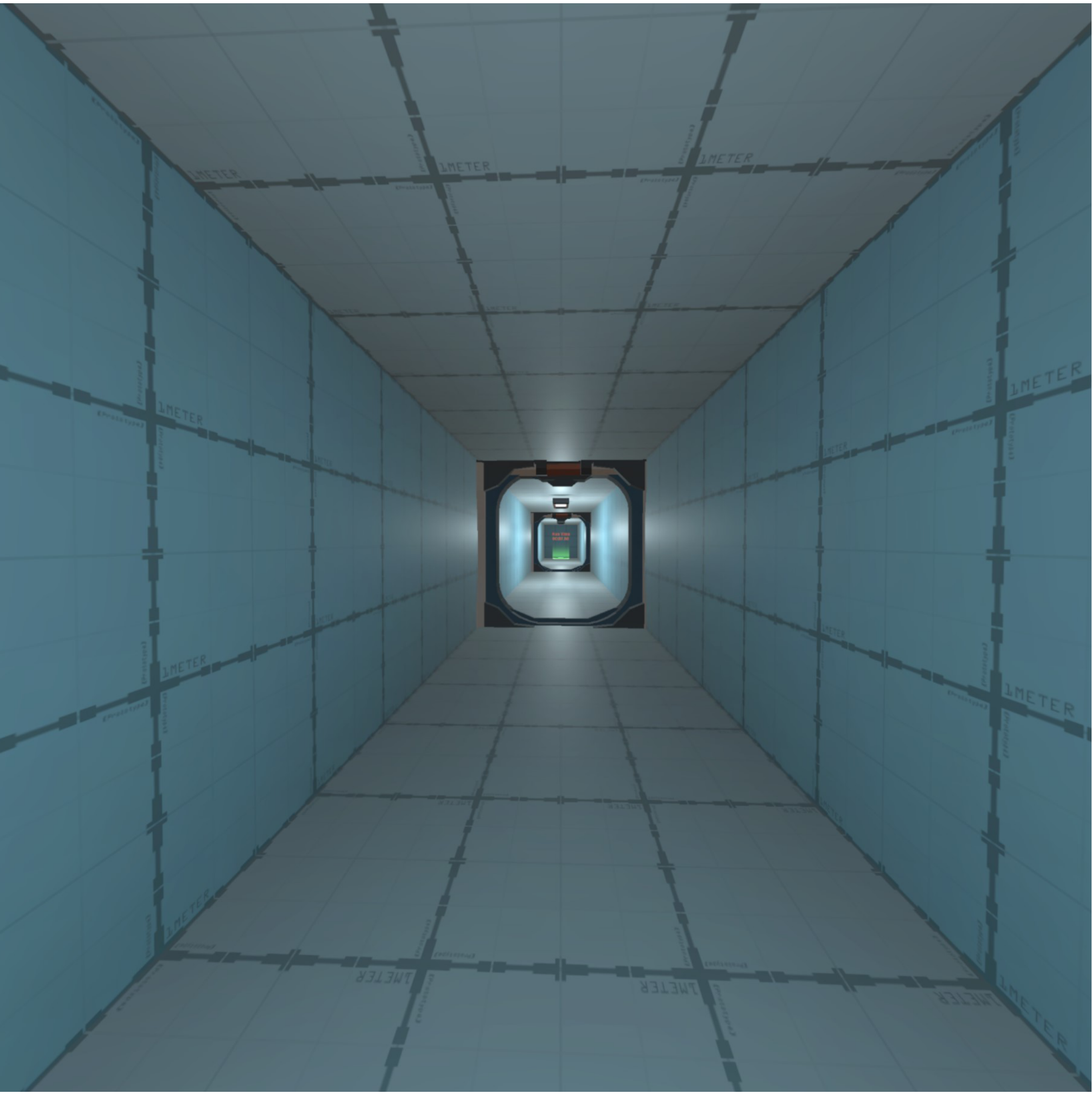}\label{fig:Fig1_d}}}
& \multicolumn{1}{c|}{}{}\\
& \multicolumn{1}{c|}{\begin{tabular}[c]{@{}c@{}} \tikz \fill [l1_1color] (0,0) rectangle (.25,.25); \textit{T1. Straight line} \\ \textit{walking (a)}\end{tabular}}
& \multicolumn{1}{c|}{\begin{tabular}[c]{@{}c@{}} \tikz \fill [l1_2color] (0,0) rectangle (.25,.25); \textit{T2. Over/Under-}\\ \textit{shooting (b)}\end{tabular}}
& \multicolumn{1}{c|}{\begin{tabular}[c]{@{}c@{}} \tikz \fill [l1_3color] (0,0) rectangle (.25,.25); \textit{T3. Chasing}\\ \textit{(c)}\end{tabular}}
& \multicolumn{1}{c|}{\begin{tabular}[c]{@{}c@{}} \tikz \fill [l1_4color] (0,0) rectangle (.25,.25); \textit{T4. Sprinting}\\ \textit{(d)}\end{tabular}}
& \multicolumn{1}{c|}{}\\ \hline
\multirow[t]{2}{*}{\begin{tabular}[c]{@{}l@{}}\tikz \fill [l2color] (0,0) rectangle (.25,.25); \textit{S2. Direction}\\ \textit{control} \vspace{3.5cm} \end{tabular}}
& \multicolumn{1}{c|}{\subfloat{\includegraphics*[height=0.95in]{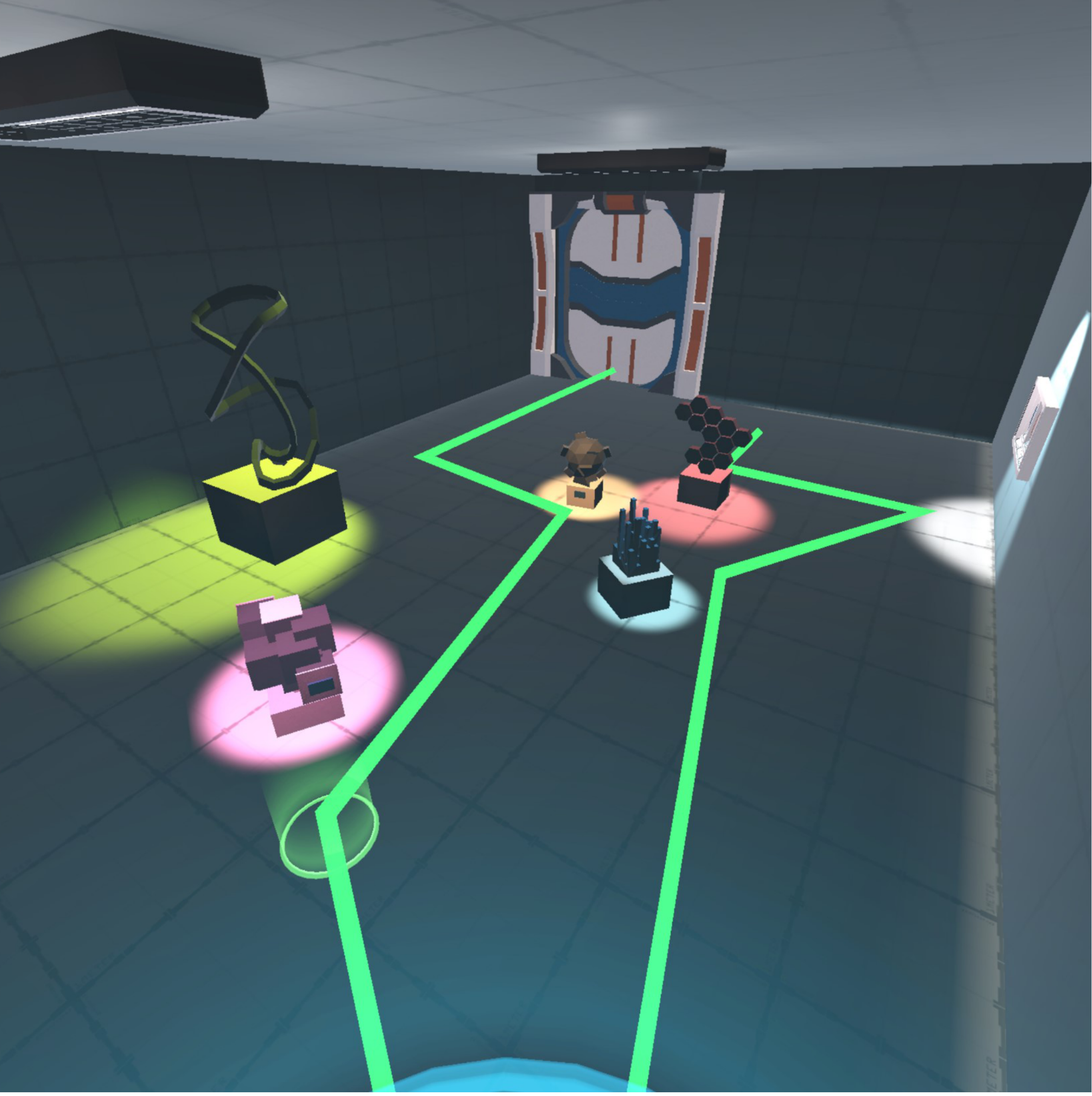}\label{fig:Fig1_e}}}
& \multicolumn{1}{c|}{\subfloat{\includegraphics*[height=0.95in]{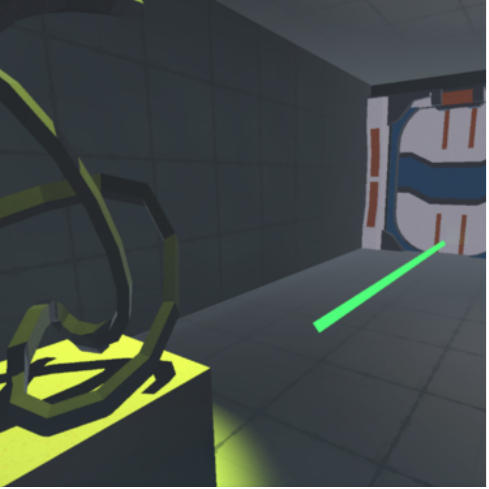}\label{fig:Fig1_f}}}
& \subfloat{\includegraphics*[height=0.95in]{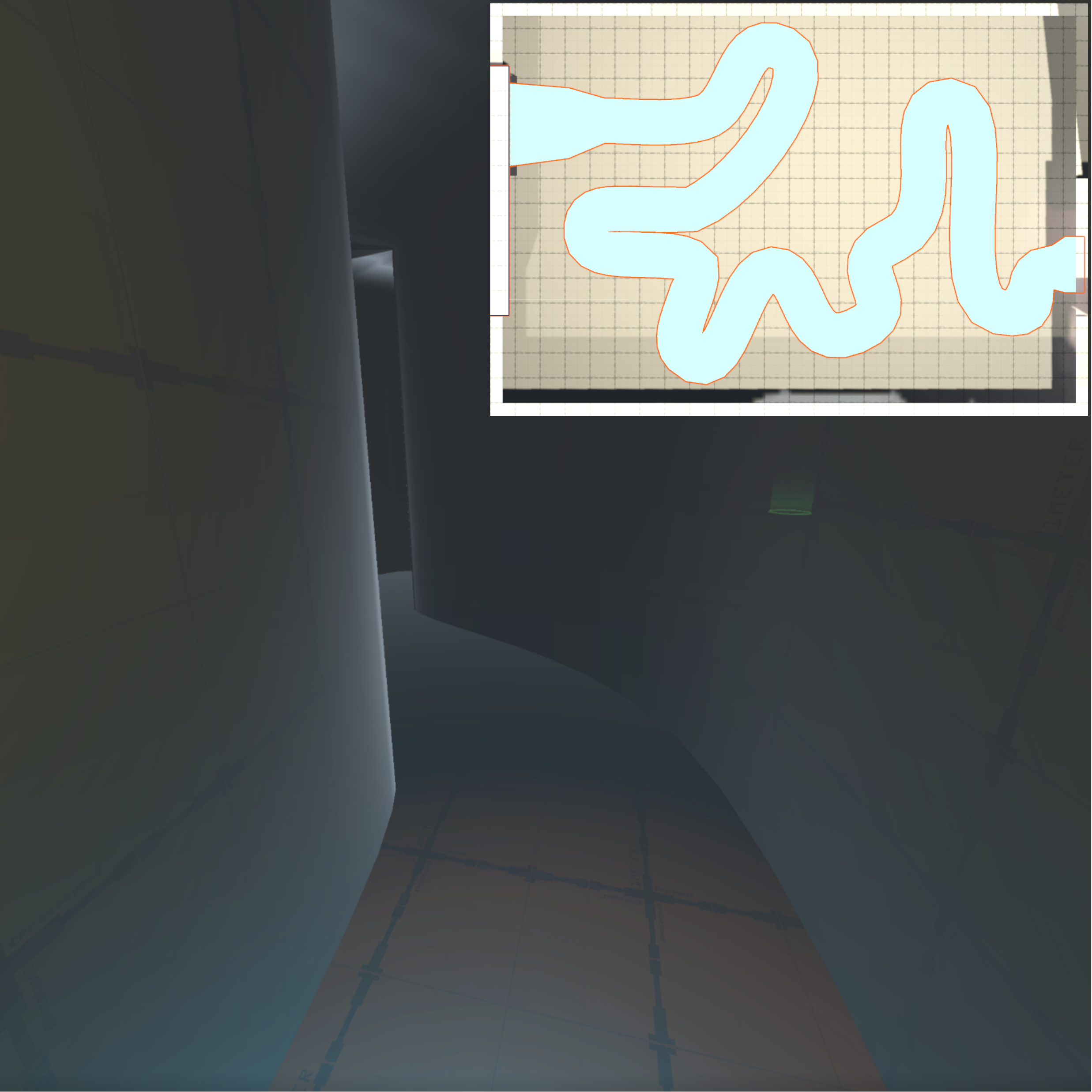}\label{fig:Fig1_g}}
& \multicolumn{1}{c|}{\subfloat{\includegraphics*[height=0.95in]{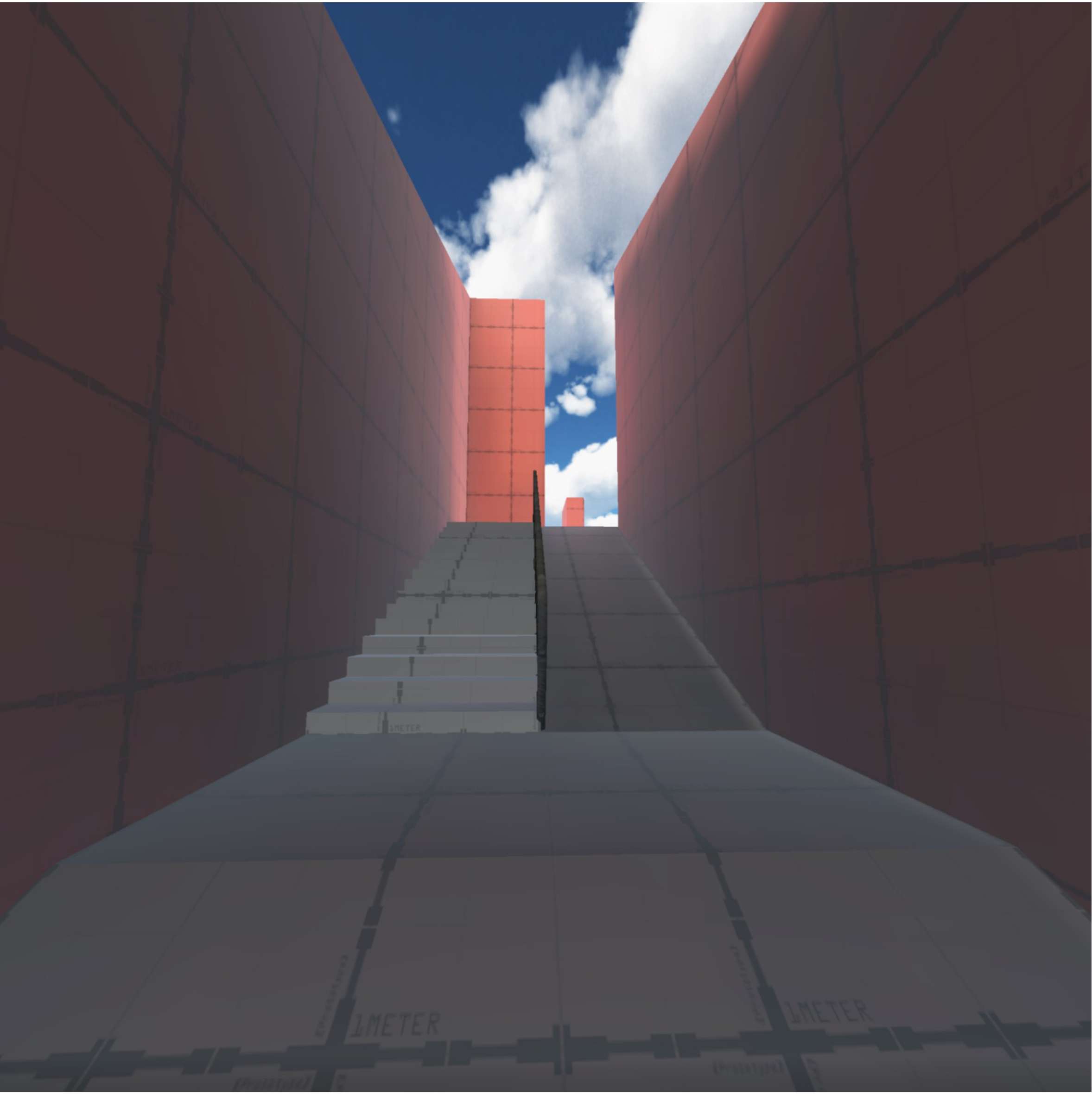}\label{fig:Fig1_h}}}
& \multicolumn{1}{c|}{\subfloat{\includegraphics*[height=0.95in]{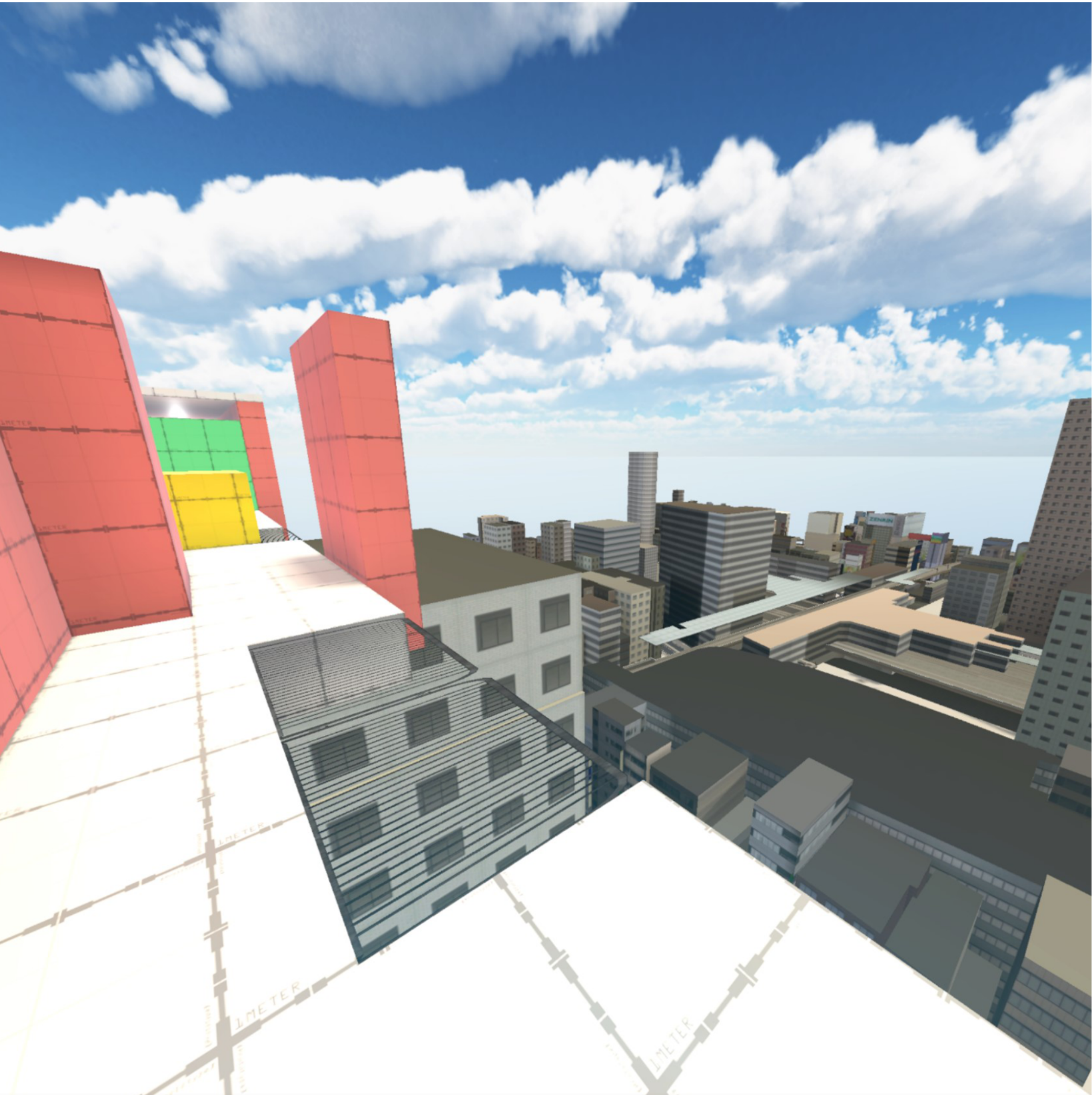}\label{fig:Fig1_i}}}\\
& \multicolumn{1}{c|}{\begin{tabular}[c]{@{}c@{}} \tikz \fill [l2_1color] (0,0) rectangle (.25,.25); \textit{T1. Multi-straight}\\ \textit{line walking (e)}\end{tabular}}
& \multicolumn{1}{c|}{\begin{tabular}[c]{@{}c@{}} \tikz \fill [l2_2color] (0,0) rectangle (.25,.25); \textit{T2. Backward}\\ \textit{walking (f)}\end{tabular}}
& \multicolumn{1}{c|}{\begin{tabular}[c]{@{}c@{}}\tikz \fill [l2_3color] (0,0) rectangle (.25,.25); \textit{T3. Curved}\\ \textit{walking (g)}\end{tabular}}
& \multicolumn{1}{c|}{\begin{tabular}[c]{@{}c@{}}\tikz \fill [l2_4color] (0,0) rectangle (.25,.25); \textit{T4. Stairs} \& \\ \textit{ramps (h)}\end{tabular}}
& \multicolumn{1}{c|}{\begin{tabular}[c]{@{}c@{}}\tikz \fill [l2_5color] (0,0) rectangle (.25,.25); \textit{T5. Fear}\\\textit{(i)}\end{tabular}}\\\hline
\multirow[t]{2}{*}{\begin{tabular}[c]{@{}l@{}} \tikz \fill [l3color] (0,0) rectangle (.25,.25); \textit{S3. Decoupled}\\\textit{movements} \vspace{3.5cm} \end{tabular}}
& \multicolumn{1}{c|}{\subfloat{\includegraphics*[height=0.95in]{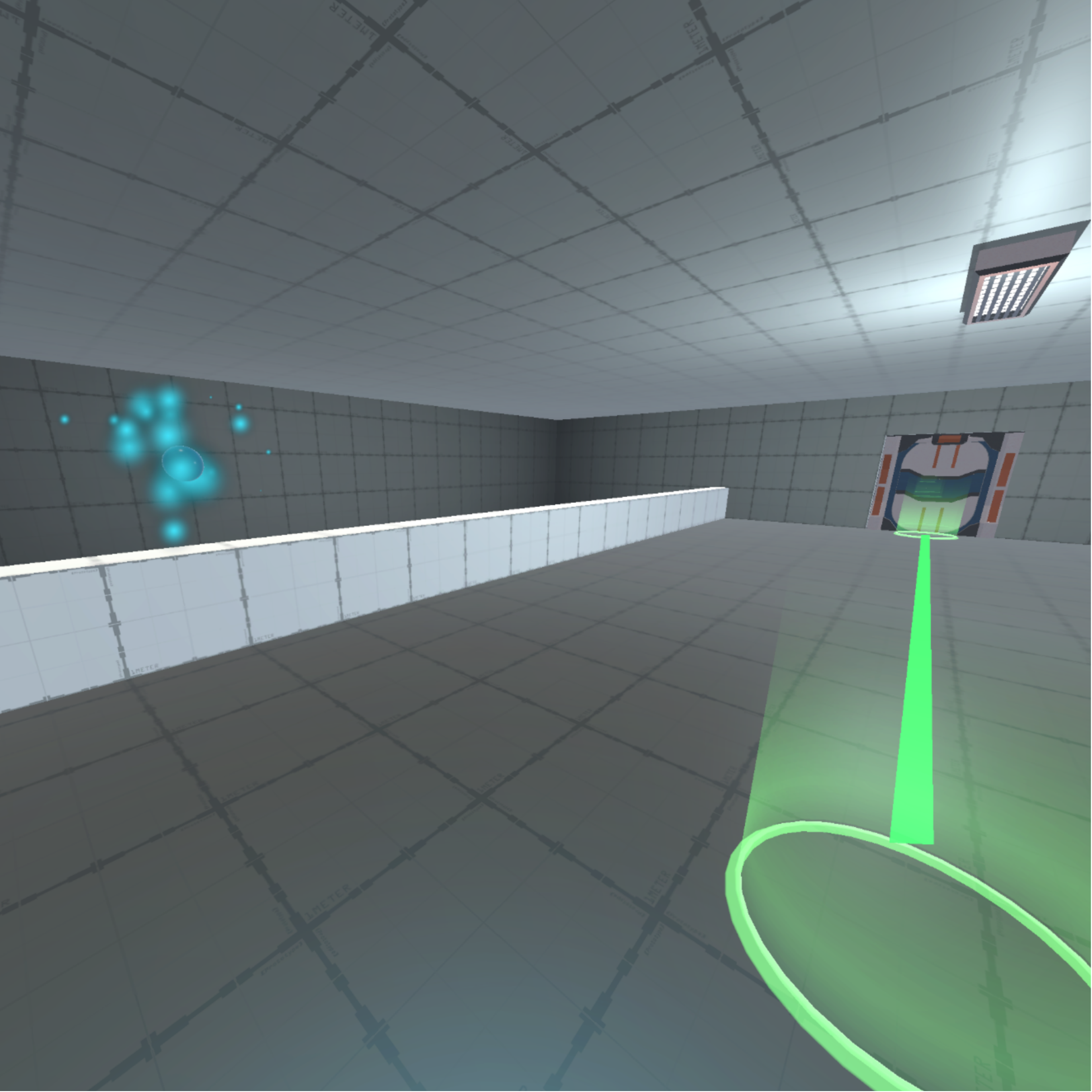}\label{fig:Fig1_j}}}
& \multicolumn{1}{c|}{\subfloat{\includegraphics*[height=0.95in]{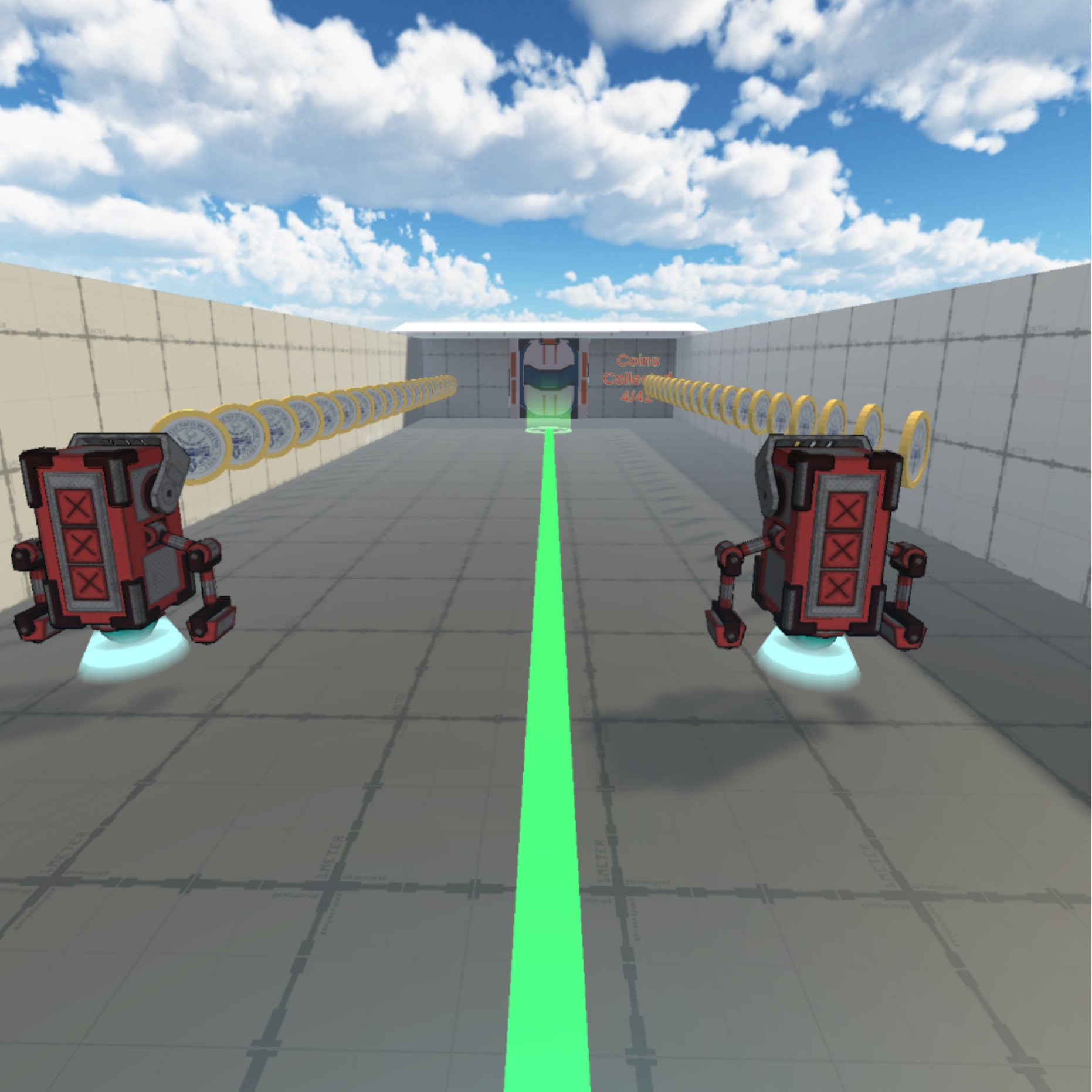}\label{fig:Fig1_k}}}
& \subfloat{\includegraphics*[height=0.95in]{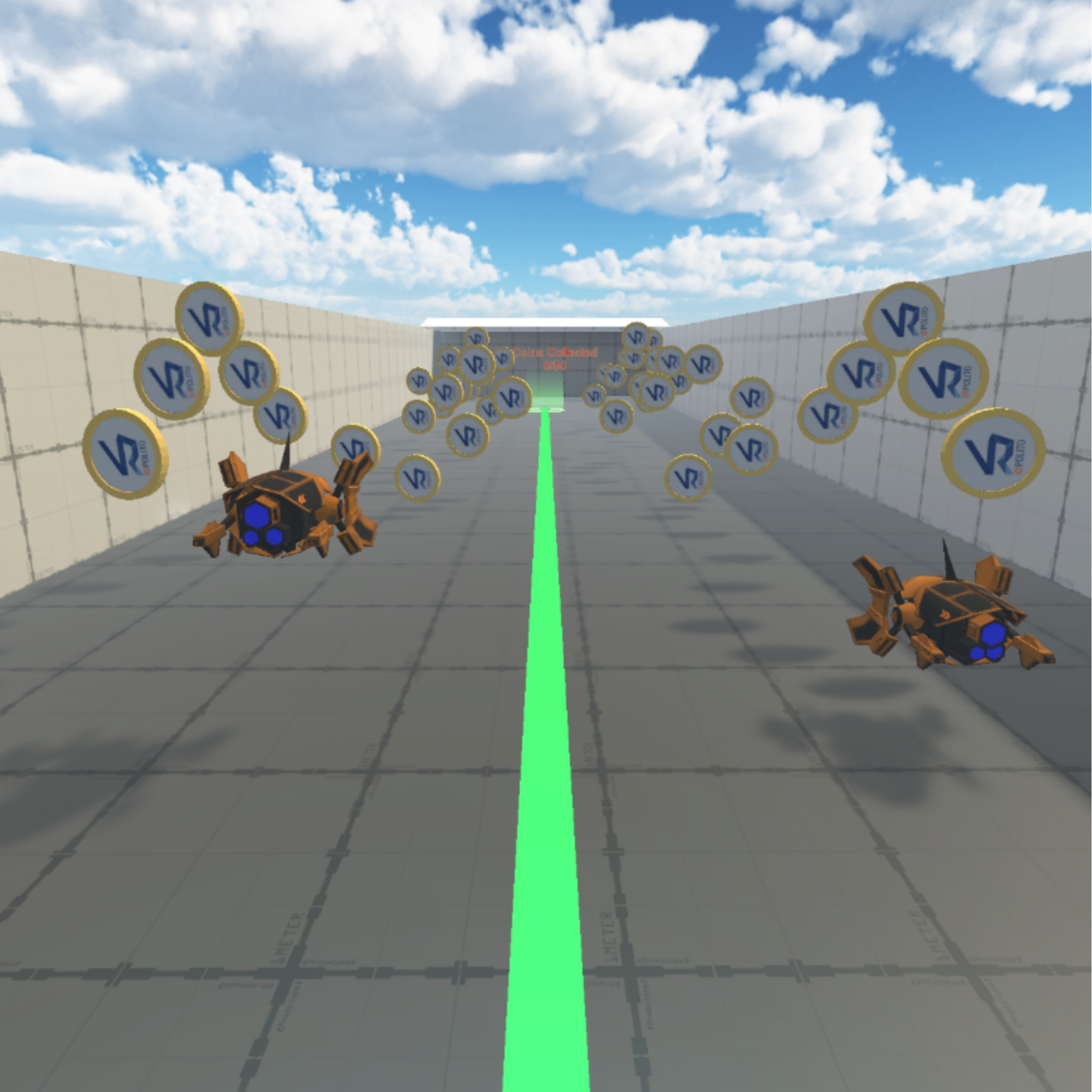}\label{fig:Fig1_l}}
&
& \\
& \multicolumn{1}{c|}{\begin{tabular}[c]{@{}c@{}} \tikz \fill [l3_1color] (0,0) rectangle (.25,.25); \textit{T1. Decoupled} \\ \textit{gaze (j)}\end{tabular}}
& \multicolumn{1}{c|}{\begin{tabular}[c]{@{}c@{}} \tikz \fill [l3_2color] (0,0) rectangle (.25,.25); \textit{T2. Stretched-}\\ \textit{out hands (k)}\end{tabular}}
& \multicolumn{1}{c|}{\begin{tabular}[c]{@{}c@{}} \tikz \fill [l3_3color] (0,0) rectangle (.25,.25); \textit{T3. Decoupled}\\ \textit{hands (l)}\end{tabular}}
& \multicolumn{1}{c}{}
& \multicolumn{1}{c}{}     \\ \cline{1-4}
\multirow[t]{2}{*}{\begin{tabular}[c]{@{}l@{}}\tikz \fill [l4color] (0,0) rectangle (.25,.25); \textit{S4. Agility} \\ \vspace{3.5cm} \end{tabular}}
& \multicolumn{1}{c|}{\subfloat{\includegraphics*[height=0.95in]{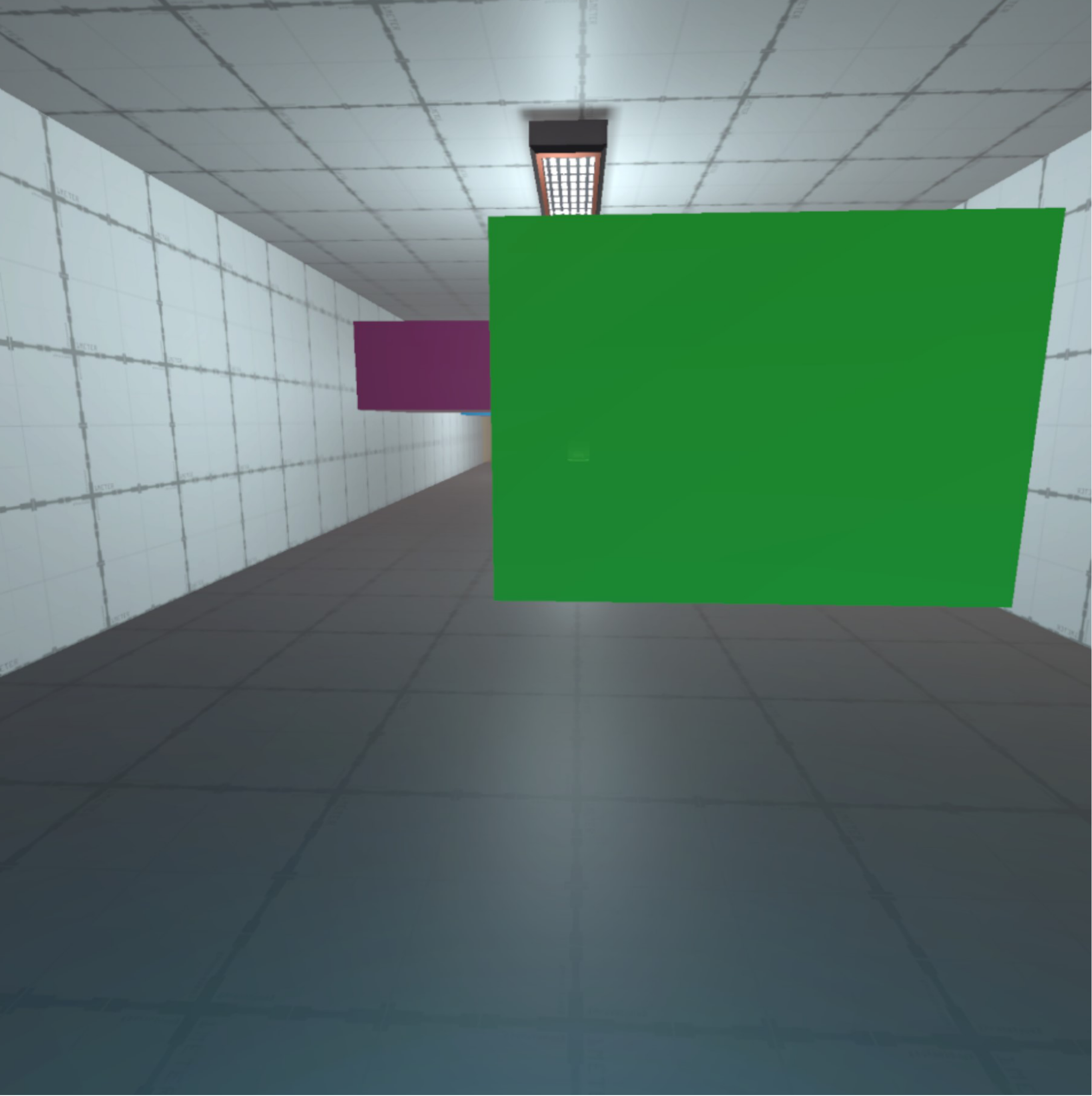}\label{fig:Fig1_m}}}
& \multicolumn{1}{c|}{\subfloat{\includegraphics*[height=0.95in]{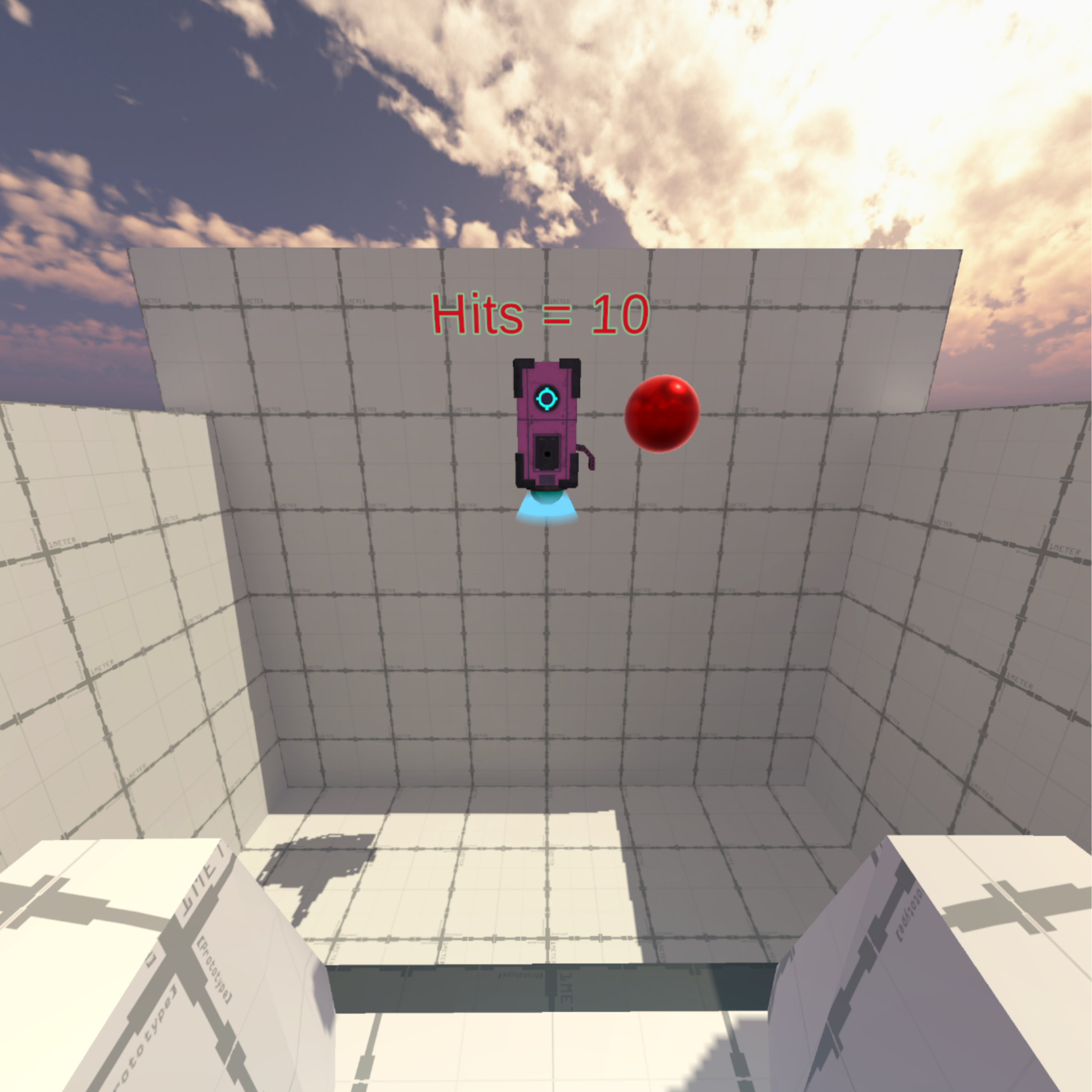}\label{fig:Fig1_n}}}
& \subfloat{\includegraphics*[height=0.95in]{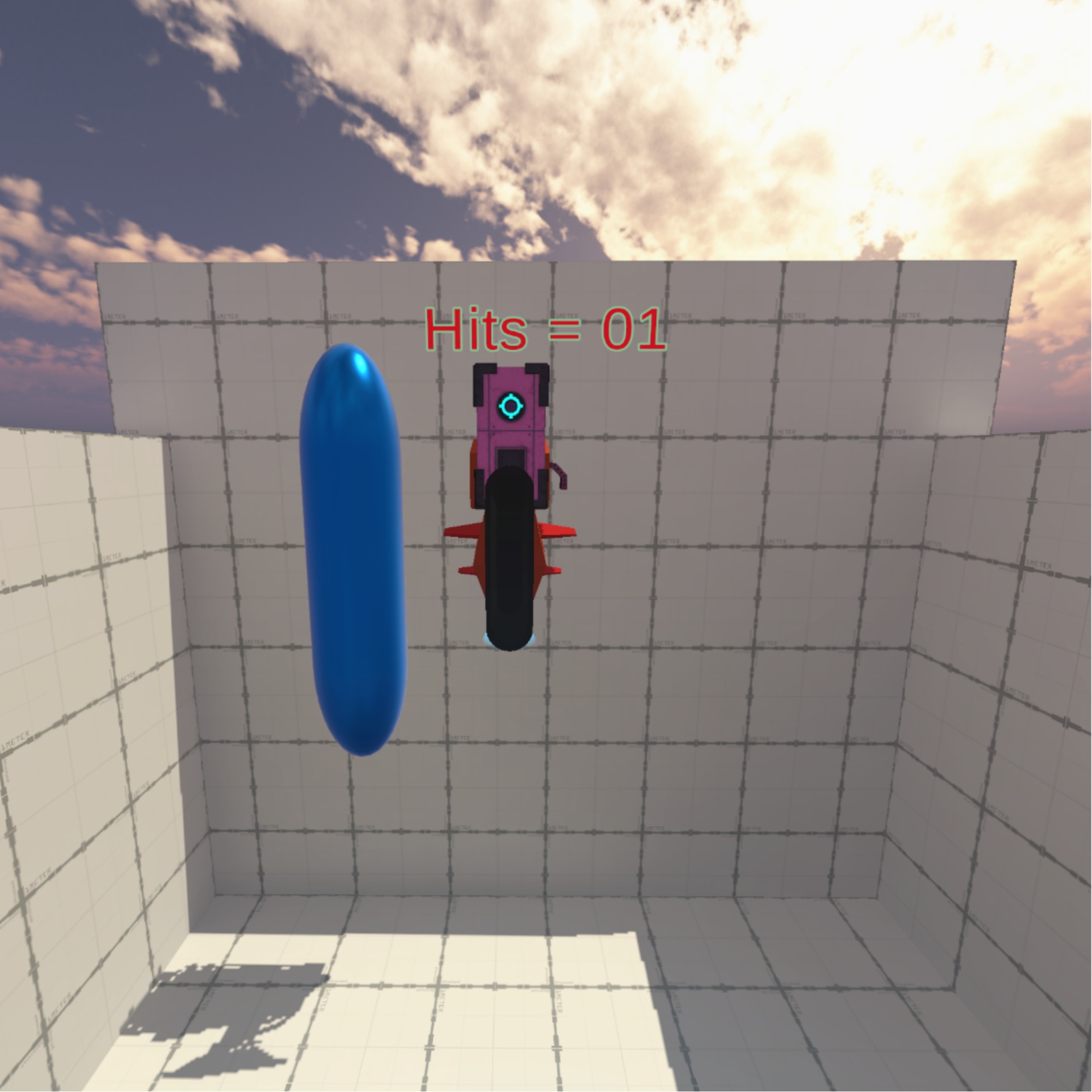}\label{fig:Fig1_o}}
&
& \\
& \multicolumn{1}{c|}{\begin{tabular}[c]{@{}c@{}} \tikz \fill [l4_1color] (0,0) rectangle (.25,.25); \textit{T1. Dynamic} \\ \textit{agility (m)}\end{tabular}}
& \multicolumn{1}{c|}{\begin{tabular}[c]{@{}c@{}} \tikz \fill [l4_2color] (0,0) rectangle (.25,.25); \textit{T2. Stationary} \\ \textit{agility (n)}\end{tabular}}
& \multicolumn{1}{c|}{\begin{tabular}[c]{@{}c@{}} \tikz \fill [l4_3color] (0,0) rectangle (.25,.25); \textit{T3. Evasion} \\ \textit{(o)}\end{tabular}}
& \multicolumn{1}{c}{}
& \\ \cline{1-4} \cline{6-6}
\multirow[t]{2}{*}{\begin{tabular}[c]{@{}l@{}} \tikz \fill [l5color] (0,0) rectangle (.25,.25); \textit{S5. Interaction}\\ \textit{with objects} \vspace{3.5cm} \end{tabular}}
& \multicolumn{1}{c|}{\subfloat{\includegraphics*[height=0.95in]{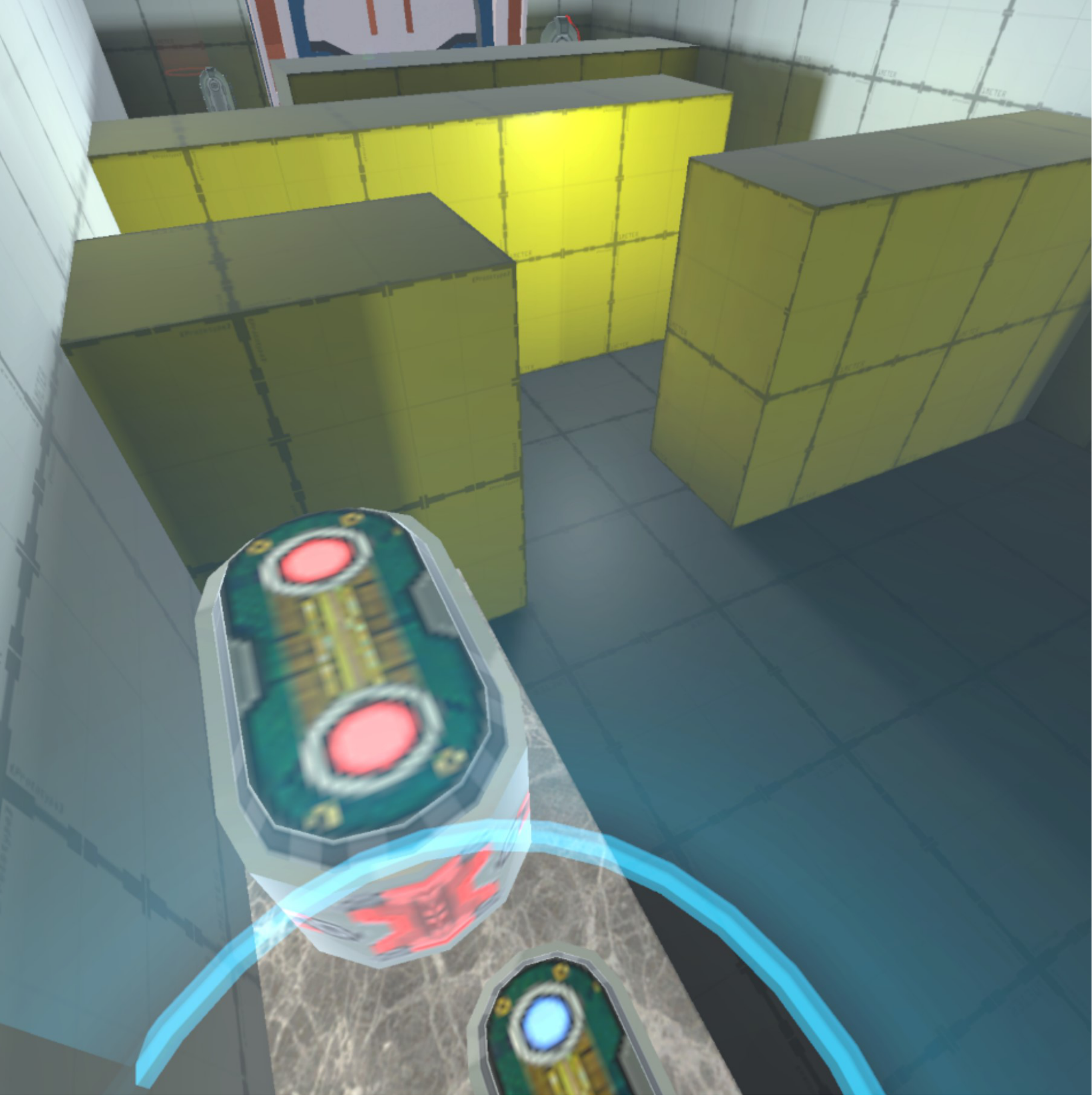}\label{fig:Fig1_p}}}
& \multicolumn{1}{c|}{\subfloat{\includegraphics*[height=0.95in]{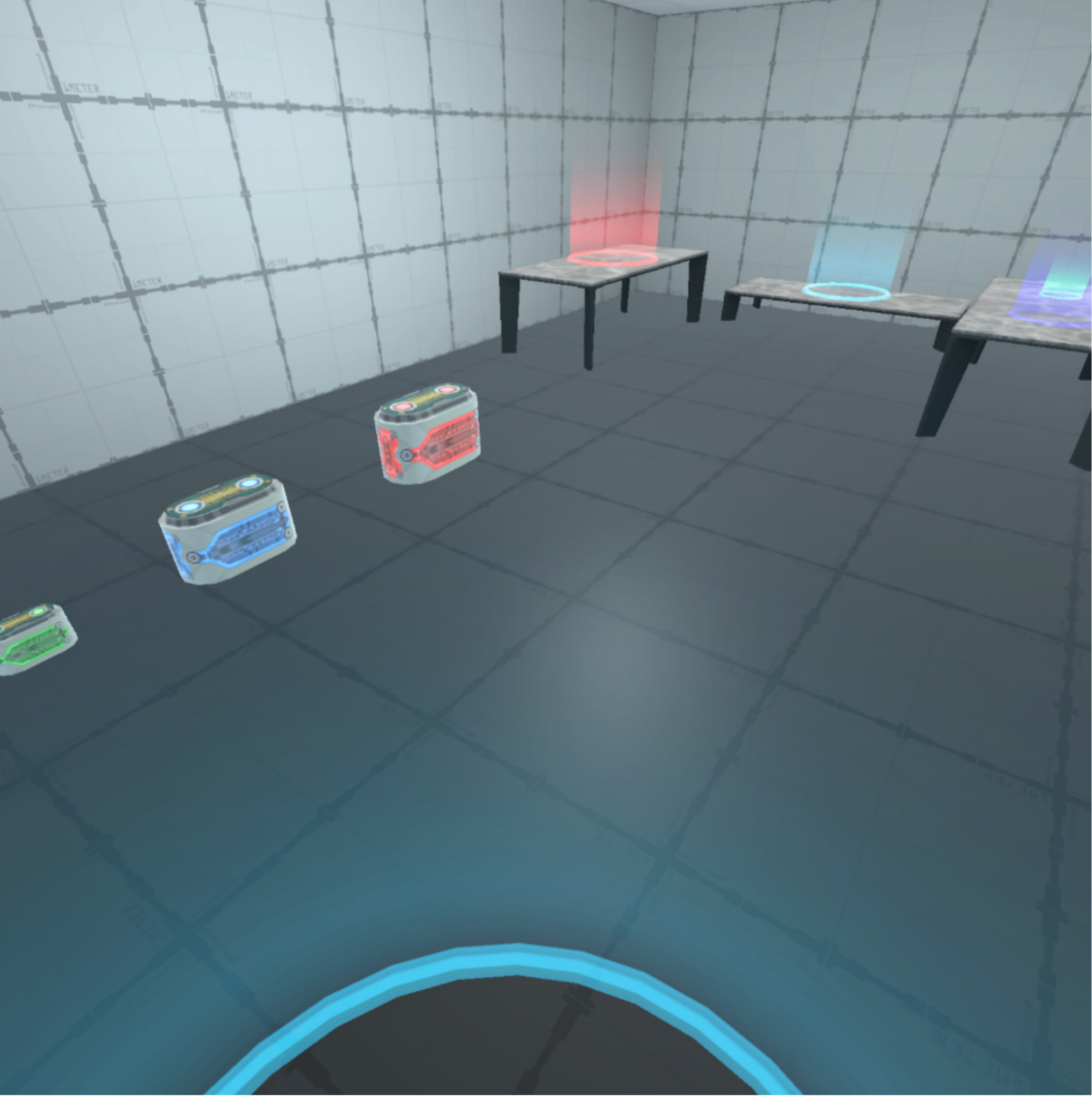}\label{fig:Fig1_q}}}
& \subfloat{\includegraphics*[height=0.95in]{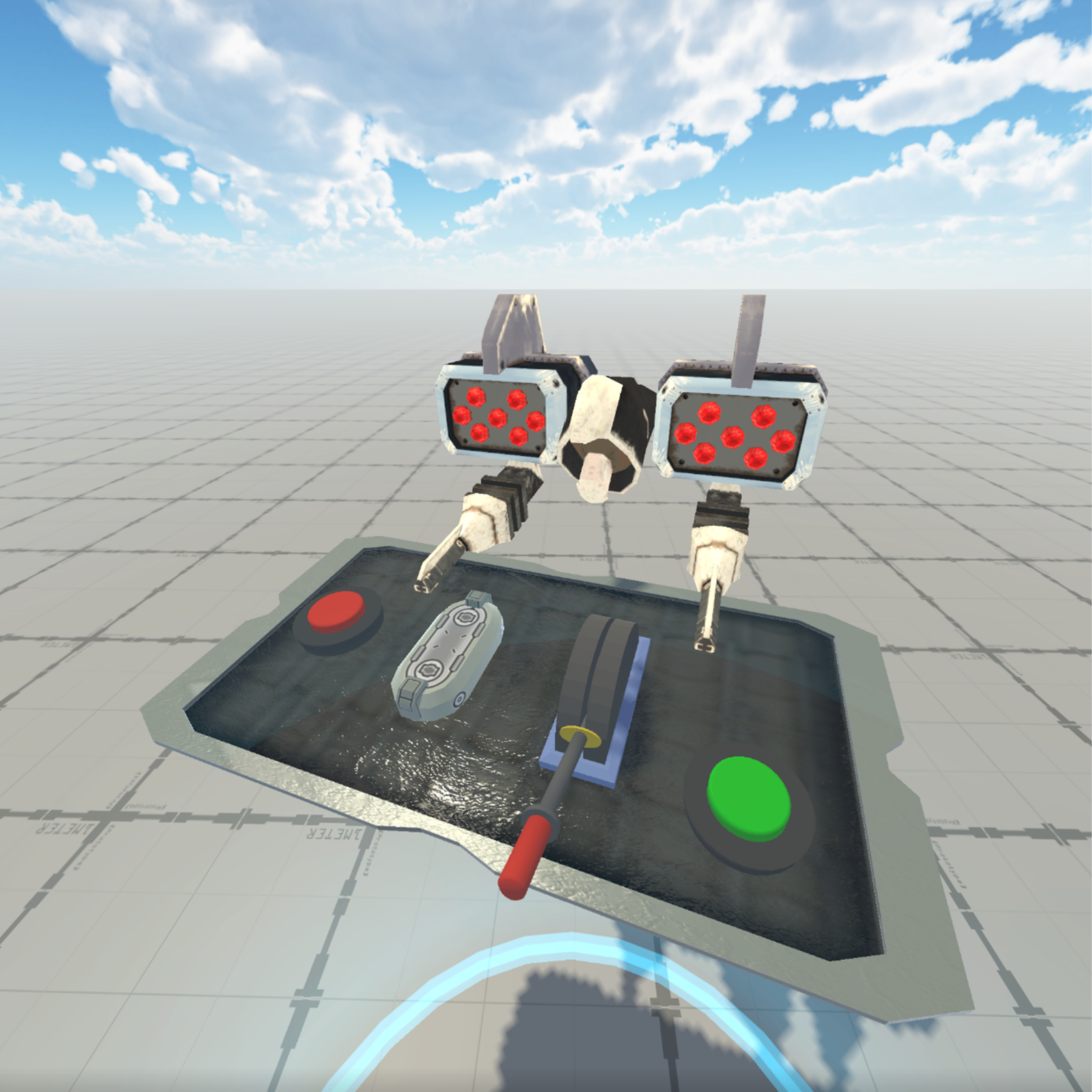}\label{fig:Fig1_r}}
&
& \multicolumn{1}{|c|}{\subfloat{\includegraphics*[height=0.95in]{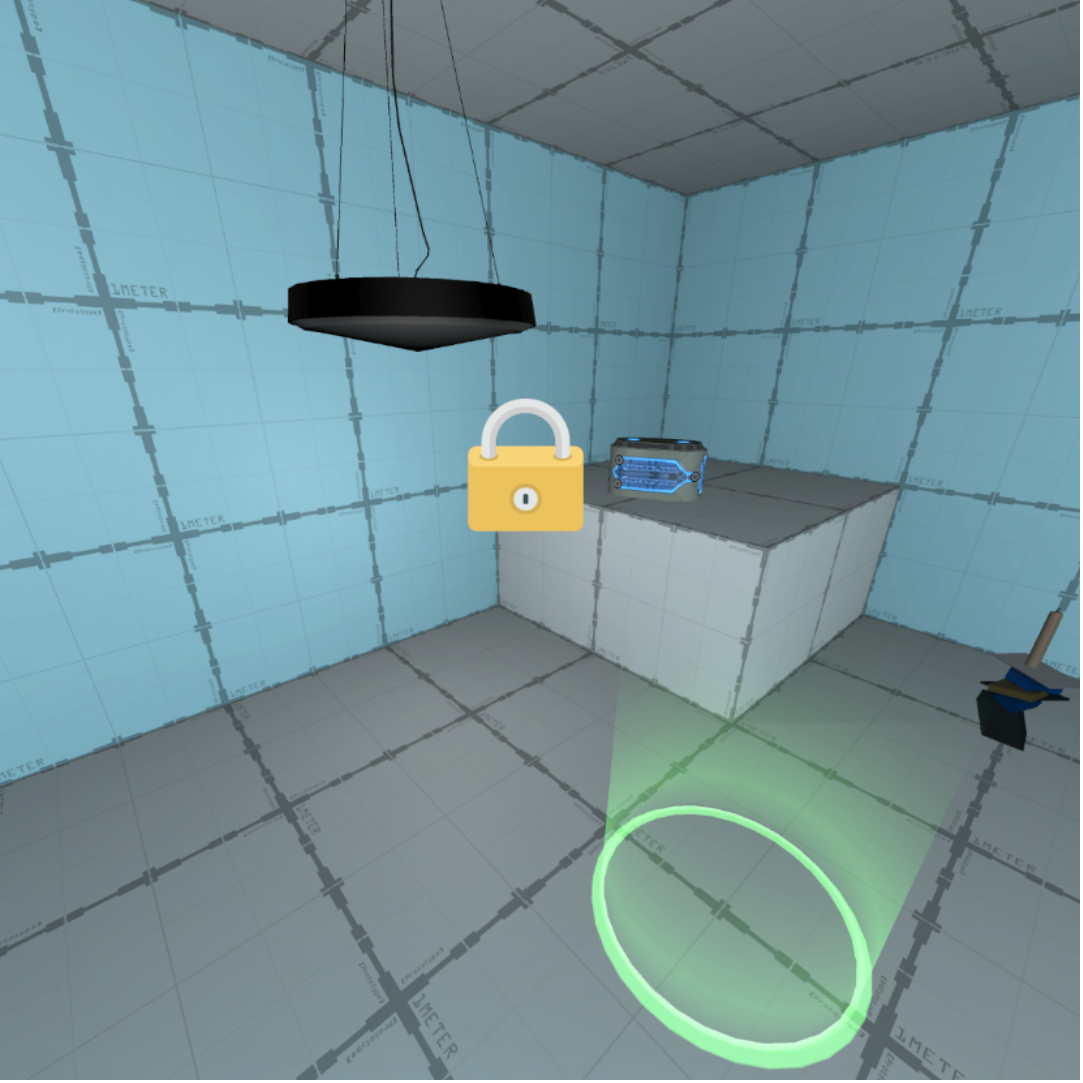}\label{fig:Fig1_s1}}}\\
& \multicolumn{1}{c|}{\begin{tabular}[c]{@{}c@{}} \tikz \fill [l5_1color] (0,0) rectangle (.25,.25); \textit{T1. Grabbing} \\ \textit{(p)}  \end{tabular}}
& \multicolumn{1}{c|}{\begin{tabular}[c]{@{}c@{}} \tikz \fill [l5_2color] (0,0) rectangle (.25,.25); \textit{T2. Manipulation} \\ \textit{(q)} \end{tabular}}
& \multicolumn{1}{c|}{\begin{tabular}[c]{@{}c@{}} \tikz \fill [l5_3color] (0,0) rectangle (.25,.25); \textit{T3. Interaction} \\ \textit{in motion (r)}\end{tabular}}
& \multicolumn{1}{c}{}
& \multicolumn{1}{|c|}{\begin{tabular}[c]{@{}c@{}} \textit{Training}\\\textit{(s)}\end{tabular}} \\ \cline{1-4} \cline{6-6}
\end{tabular}
\caption{ Testbed application: a)--r) devised scenarios and tasks (functional requirements of a locomotion technique) considered in the evaluation, and s) training scenario (depicted for sake of completeness). Green circles/cylinders represent destinations to be reached in order to complete the task or parts of it, whereas green lines indicate to the user the path to follow (when needed).}
\label{tab:TabTasks}
\end{figure*}

The devised scenarios separately address five major aspects that an effective locomotion technique should support: \textit{straight movements}, \textit{direction control}, \textit{decoupled movements}, \textit{agility} and \textit{interaction with objects}.
Several screenshots showing operations (requirements) to be performed (supported) in the above scenarios are provided in Fig.~\ref{tab:TabTasks}. Some videos showing tasks execution with different locomotion techniques are available at \href{http://tiny.cc/8uxlsz}{\textit{http://tiny.cc/8uxlsz}}.

\subsubsection{S1. Straight movements}
\label{sec:straight_movements}
The first scenario is designed to assess a given technique under the most simple locomotion conditions, i.e., with movements that do not require directional changes. Hence, it focuses on different tasks one may have to perform while moving on a straight path, from taking a simple stroll, to stopping at exact points, reaching a given target and sprinting. These tasks are quite common in VR simulations where the user impersonates a character and an appropriate level of physicality is required (e.g., \cite{calandra2018eg} and \cite{calandra2019icce}).

In the first task named \textit{T1. Straight line walking} and illustrated in Fig.~\ref{fig:Fig1_a}, the user has to reach a destination in front of it by following the green path on the floor as much as possible (like in \cite{nabiyouni2015comparing}). Afterwards, in task \textit{T2. Over/Under-shooting}, it has to stop within three green circular areas which are characterized by a decreasing radius; during the virtual experience, the new area appears only when the previous one has been reached (Fig.~\ref{fig:Fig1_b}). This task is a simplified version of the task proposed in \cite{whitton2005comparing}. Once the last destination has been reached, the user is asked to follow a moving object, namely a robot, while always remaining within a green area behind it (Fig.~\ref{fig:Fig1_c}). Since the robot periodically changes its speed, this is referred to as task \textit{T3. Chasing}. Finally, the user is requested to cover a long path (shown in Fig.~\ref{fig:Fig1_d}) as quickly as possible (\textit{T4. Sprinting} task). Both following targets at a fixed distance and running are very common operations in VR videogames \cite{UnbreakableVRRunner}.

\subsubsection{S2. Direction control}
\label{sec:direction_control}
The second scenario aims to evaluate locomotion techniques in the execution of tasks that involve different ways of performing direction changes, which may be requested by the actual goals of the task being executed, or by constraints set by the simulated environment. The devised tasks require the user to change direction multiple times while proceeding on straight lines, moving backward, following a curved path, walking uphill under different conditions and in presence of a dangerous obstacle (a chasm). The scenario includes aspects that may characterize, for instance, simulations of walkable natural environments, but also of urban settings, in which it could be necessary to navigate multi-story buildings featuring obstacles and stairs.

In the first task (\textit{T1. Multi-straight line walking}), the user has to follow a path made up of a sequence of straight lines (a polyline). This task was inspired by \cite{nabiyouni2015comparing}, where a user study on locomotion techniques was carried out by simulating a visit to a museum. The user is requested to move between artworks and stand for a while in front of them (until lights switch off) before moving to the next one (as done in \cite{paris2017acquisition}, though in a different environment). The virtual environment and the complete path to be followed are depicted in Fig.~\ref{fig:Fig1_e}. As said, the emphasis here is on direction control, rather than on speed: hence, a new section of the path appears only when the operation on the current artwork has been completed. In the second task (\textit{T2. Backward walking}), the user is requested to leave the museum room by looking at the last artwork while walking backward to reach the door behind it (Fig.~\ref{fig:Fig1_f}). This is one of the tasks proposed in \cite{ferracani2016locomotion} (the other tasks in that work are taken into account in different scenarios of the proposed testbed). Afterwards, the user has to traverse a tunnel (Fig.~\ref{fig:Fig1_g}, with path shown) that requires it to make continuous direction adjustments while moving (\textit{T3. Curved walking}), like in \cite{nilsson2013tapping}. The fourth task (\textit{T4. Stairs \& ramps}) focuses on direction control from a different perspective, as it first requests the user to climb up a stair and a ramp, then analyzes its preferences by making it choose between a stair or a ramp for a final, long climb (as shown in Fig.~\ref{fig:Fig1_h}). This task has been included to evaluate the disturbance introduced by movements applied to the user's point of view: stairs generate a sussultatory pattern, ramps a more uniform motion, and their effects on the user's experience (e.g., on motion sickness) may vary depending on the locomotion technique being used. Lastly, the degree of confidence in controlling direction with the given technique is investigated in a so-called \textit{T5. Fear} task, in which the user needs to cross a hazardous area on a high building roof, where it risks to fall down in case of wrong movements (Fig.~\ref{fig:Fig1_i}). User's interactions with stairs and hazardous areas were originally investigated in \cite{schuemie2005effect}.

\subsubsection{S3. Decoupled movements}
\label{sec:decoupled_movements}
This scenario aims to investigate to what extent a given locomotion technique lets the user decouple the control of walking (and its direction) from gaze and hand movements; this ability is particularly important in scenarios which are not just exploratory, but request the user to perform also gaze or hand interactions. In these cases, allowing the user to move freely while avoiding interference from or onto the interaction technique could prevent a possible deterioration of the user experience.

In the first task, shown in Fig.~\ref{fig:Fig1_j}, the user needs to walk forward on a straight path while keeping the gaze fixed on an object (a blue sphere) that moves besides it (\textit{T1. Decoupled gaze}). The ability to look elsewhere while walking is used in many VR experiences \cite{Fallout4VR}, e.g., to gather information on the surrounding environment while moving, keep the gaze fixed on a specific spot while waiting for a given event, avoid obstacles coming from aside, etc. 
In the second task (\textit{T2. Stretched-out hands}), there are two robots to the left and right sides of the user. The user can control the robots' relative position by moving the hands close/far from its body. The goal here is to make the robots collect a number of floating coins while walking on a straight path (Fig.~\ref{fig:Fig1_k}). The coins are lined up so that the user has to keep its arms completely stretched out to collect them. This task is designed to stress locomotion techniques that rely on hand controllers, which could possibly restrict some arm movements. The last task (\textit{T3. Decoupled hands}) is similar to the previous one, but in this case coins are placed at different heights and distances from the straight path (Fig.~\ref{fig:Fig1_l}). Thus, the user needs to move its hands in the proper way to control the attached robots (which in this case can be moved also vertically) and collect the coins while following the path. This task is meant to further investigate the performance of locomotion techniques that could interfere with arm movements. 
The ability to effectively maintain a high level of control over the intended motion (preventing undesired/accidental movements) while the hands are busy in other interactions is key in many VR applications and videogames, e.g., for shooting \cite{EchoCombat} or collecting objects \cite{CatchandRelease}.

\subsubsection{S4. Agility}
\label{sec:agility} 
The fourth scenario focuses on evaluating the level of virtual and physical agility enabled by the locomotion technique being studied; in particular,  aspects pertaining agility while walking freely or being constrained in some of the movements are evaluated. These abilities turn out to be quite important in simulations of chaotic events, as well as in applications requiring an avoidance reaction to sudden threats.

In the first task (\textit{T1. Dynamic agility}), the user has to reach the opposite side of a room by avoiding collisions with blocks of different shapes that are moving towards it by controlling its direction and crouching when needed (Fig.~\ref{fig:Fig1_m}) \cite{BeatSaber}. 
In the second task (\textit{T2. Stationary agility}), the user is facing a robot that periodically shoots balls at it with different speeds and directions, following a pseudo-random pattern (Fig.~\ref{fig:Fig1_n}). The user is requested to avoid the balls by only moving the head and hip: in this way, it is possible to evaluate the encumbrance caused by the considered technique to movements that are performed on the spot, i.e., which do not actually involve the use of locomotion, as requested in applications such as, e.g., \cite{SuperHotVR}. The last task, (\textit{T3. Evasion}), is a variation of the previous one, where the user has to dodge a series of larger, human-sized capsule-shaped bullets: in this case, it needs to use the locomotion technique to rapidly move left or right, as bullets can no more be avoided just with on-the-spot movements (Fig.~\ref{fig:Fig1_o}).

\subsubsection{S5. Interaction with objects}
\label{sec:interaction_with_objects}
The last scenario is designed to evaluate the impact of the locomotion technique on precise interaction with objects available in the virtual environment, which is a fundamental feature in applications simulating procedures that involve the use of virtual tools \cite{calandra2019icce}. In particular, the scenario investigates pick-transport-place interactions with objects under various conditions (one or more objects, presence or absence of obstacles, occluded sight), manipulation and positioning of objects requiring precise movements and continuous displacements, and impact of complex interactions performed during locomotion.

In the first task, named \textit{T1. Grabbing}, the user is asked to grab objects (batteries) and place them into several target positions (sockets, to open  doors) based on their colors, sizes and shapes, like in \cite{pai2017armswing, ferracani2016locomotion}. At the beginning of the task, there is only one object to move at a time. Afterwards, two objects need to be grabbed at the same time, one per hand (the size of one of the objects is intentionally larger than the other one, thus occluding the user's sight). In the second part of the task, complexity is further increased by asking the user to carry objects while traversing a maze and avoiding collisions with walls, like in \cite{suma2010evaluation} (Fig.~\ref{fig:Fig1_p}). In the second task (\textit{T2. Manipulation}), fine-grained control of locomotion and manipulation is requested to the user in order to reposition objects scattered in the environment at given locations and then assemble a tower by precisely stacking them (Fig.~\ref{fig:Fig1_q}). This kind of interaction is used, for example, in \cite{JobSimulator}. The last task, \textit{T3. Interaction in motion}, is the most demanding one, as it is designed to stress the interference of the locomotion technique on the way the user performs compulsory interactions with objects while moving. The user is placed in front of a moving element (a robot holding a panel) on which it will have to perform four interactions in a specific order (Fig.~\ref{fig:Fig1_r}). Once the user gets close to the robot, the latter starts to move at a (constant) speed close to the user's walking speed, and continuously adjusts its direction to keep moving away from the user. The user will need to modulate appropriately its speed in order to be able to perform the requested interactions on the robot: if it stays too far, then it will not be allowed to interact; if it gets too close or tries to overtake the robot, the latter will wriggle out of it with a rapid direction change. Regarding interactions, the user will have to press a button, insert a battery in a socket, push a lever, and press another button. The order was chosen so that similar interactions are not placed close to each other.

\subsubsection{Training}
\label{sec:training}
As it will be discussed in the following, the above scenarios were designed to be modular, so that one could possibly pick just some of them in a given study, depending on the specific aspects to be investigated. Hence, the users need to be trained on the functioning of the particular locomotion technique being investigated before taking the real test. To this purpose, a dedicated scenario was developed and populated with challenges useful to learn how to master the considered technique with the assistance of an external human guidance (see also Section \ref{sec:test_execution}). The training scenario is depicted in Fig.~\ref{fig:Fig1_s1}. Several videos showing the execution of the training are also provided at \href{http://tiny.cc/8uxlsz}{\textit{http://tiny.cc/8uxlsz}}.\\
It is worth noticing that, in order to generalize the testbed scenarios to the characteristics of the user and to the features of the selected locomotion technique, some of the tasks (e.g., \textit{S3.T3}, \textit{S5.T3}, etc.) were designed to adapt to the particular configuration being tested by leveraging a calibration file. The file includes information about the maximum height of the user's head and the maximum distance between its hands, thus allowing the testbed to account both for user’s stature and for possible constraints set by the locomotion technique (e.g., seated configuration, interface harness, etc.). Information can be collected either manually or automatically (using a procedure included in the training scenario); in the latter case, before leaving the training, the user is invited to reach the calibration spot and keep a particular calibration pose (straighten up as much as possible, with arms laterally outstretched) for a few seconds.

\subsection{Metrics}
\label{sec:metrics}
The proposed testbed has been designed to analyze NFRs for a VR locomotion technique (reported in Table \ref{tab:player_req}) on given tasks in objective or subjective terms. 

\subsubsection{Objective Metrics}
\label{sec:objective_metrics}
Some of the NFRs are analyzed by defining, for each task, several objective metrics. Some of these metrics are taken from previous works; where necessary, new metrics have been defined ad hoc for individual tasks, based on the particular aspects to be investigated. All these ``per-task'' metrics are reported in Table~\ref{tab:Tab1}, together with the references to works they have been derived from (when available).

Objective metrics are grouped in one of the following categories, corresponding to three major NFRs possibly relevant for a locomotion technique: the \textit{Operation speed} at which tasks are performed, the \textit{Accuracy} obtained in carrying them out, and the \textit{Error-proneness}, respectively marked as \textit{OS}, \textit{AC} and \textit{EP} in Table \ref{tab:Tab1}. 
The only exception is the \textit{StairsChoice} metric (marked with \textit{OT}, for \textit{Other}, in Table \ref{tab:Tab1}); this metric actually represents a subjective measure which is indirectly obtained through an objective measure, and will contribute in a different way to the evaluation. 

Some of the metrics are defined as \textit{elementary}, as they directly represent the corresponding requirement, e.g., for tasks in which the user has a single goal. As a matter of example, in \textit{S1.T1}, spatio-temporal path deviation \textit{STPathDev} is an immediate accuracy measure.
This measure was inspired by the metric used in \cite{whitton2005comparing}, which was slightly modified to better fit the purpose of the evaluation. The metric was originally intended for tasks not considering the possibility for the user to stop or to walk backwards; however, for some of the tasks in the testbed (e.g., \textit{S1.T1}) the act of staying still far from the target path is considered as an error. Hence, \textit{STPathDev} for user \textit{i} was defined as:

\begin{equation}
    STPathDev_{i} = \int_{0}^{ComplTime_{i}} PathDev_{i}(t) dt\
\label{eq:stpd}
\end{equation}
where $PathDev_{i}(t)$ is the deviation from the target path as a function of time. Basically, it corresponds to the area between $PathDev_{i}(t)$ and the time axis. 

Other metrics are named \textit{cumulative}, as the requirement for the task is defined by multiple independent measures like, e.g., in task \textit{S5.T1}, where Error-proneness is given by \textit{NumItemFalls}, \textit{NumBodyColl} and \textit{NumItemColl}. 

Finally, metrics are referred to as \textit{compound} when the user's goal is complex and the requirement for the given task needs to be expressed by combining multiple dependent measures. For example, if the user is asked to walk straight while maintaining the gaze on a specific target, like in task \textit{S3.T1},
then \textit{STPathDev} could be used to measure the Accuracy related to the first goal, whereas the percentage of time the user looked at the target could be used for second goal, but the combined metric should be directly proportional to both of them. 

\begin{table}[t!]
\caption{Non-functional requirements considered in the testbed: those evaluated via objective metrics are marked with \textsuperscript{+} if per-tasks, and with \textsuperscript{\#} if per-scenario; those assessed subjectively are marked with $^{\star}$ if assessed per-scenario, with $^{\circ}$ if assessed on the overall experience.}
\label{tab:player_req}
\centering
\begin{tabular}{|l|l|}
\hline
Accuracy\textsuperscript{+}, abbr. AC          &  Subjective Units of Discomfort (AQ)$^{\star}$   \\ \hline
Operation speed\textsuperscript{+}, abbr. OS                      &  Self-motion compellingness$^{\circ}$\\ \hline
Error-proneness\textsuperscript{+}, abbr. EP           & Acclimatisation$^{\circ}$ \\ \hline
Physical effort\textsuperscript{\#}           & Control$^{\circ}$ \\ \hline
Input sensitivity$^{\star}$               & Presence$^{\circ}$  \\ \hline
Input responsiveness$^{\star}$                & Learnability$^{\circ}$  \\ \hline
Ease of use$^{\star}$             &   Intuitiveness$^{\circ}$      \\ \hline
Perceived errors$^{\star}$          &  Comfort$^{\circ}$ \\ \hline
Appropriateness$^{\star}$ & Enjoyability$^{\circ}$  \\ \hline
Satisfaction$^{\star}$ & Overall system usability$^{\circ}$ \\ \hline
 Mental effort$^{\star}$             &   Mot. sickness: Nausea (SSQ)$^{\circ}$  \\ \hline
  Perceived physical effort$^{\star}$           &   Mot. sickness: Oculomotor (SSQ)$^{\circ}$\\ \hline
Naturalness$^{\star}$         &   Mot. sickness: Disorientation (SSQ)$^{\circ}$\\ \hline
  V/R Phys. str. similarity$^{\star}$   & Mot. sickness: Total  (SSQ)$^{\circ}$    \\ \hline
\end{tabular}
\end{table}

The compound metrics that have been defined are reported in the following. In particular, for the Accuracy requirement of tasks \textit{S2.T2}, \textit{S3.T1}, \textit{S3.T2}, and \textit{S3.T3}, a normalized  \textit{STPathDev} for user \textit{i} is first defined as:
\begin{equation}
    NrSTPathDev_{i} = \frac{STPathDev_{i}}{MaxDist\cdot ComplTime_{i}}
\label{eq:nrstpd}
\end{equation}
where \textit{MaxDist} is the maximum distance available for the user to the left or to the right of the path. Then, this measure is combined with the percentage of time the user was actually looking at the target behind it while performing the task (given by $LookAtRate_{i}$):
\begin{equation}
    AccuracyBkw_{i} = LookAtRate_{i}(1 - NrSTPathDev_{i})
\label{eq:bcka}
\end{equation}

Similarly, for \textit{S3.T1}, $NrSTPathDev_{i}$ is combined with the percentage of time the user was effectively uncoupling the gaze direction from the walk direction (given by $GazeUncRate_{i}$):
\begin{equation}
    AccuracyGazeUnc_{i} = GazeUncRate_{i} (1 - NrSTPathDev_{i})
\label{eq:gua}
\end{equation}

For \textit{S3.T2}, the normalized metric is combined with the percentage of time the user kept its hands stretched-out while walking ($StrcRate_{i}$):
\begin{equation}
    AccuracyStrc_{i} = StrcRate_{i}(1 - NrSTPathDev_{i})
\label{eq:stra}
\end{equation}

For \textit{S3.T3}, the \textit{NrSTPathDev} is multiplied by the percentage of coins collected by the user ($ScoreRate_{i}$):
\begin{equation}
    AccuracyHandsUnc_{i} = ScoreRate_{i}(1 - NrSTPathDev_{i})
\label{eq:hua}
\end{equation}
Finally, in order to describe in quantitative terms a further NFR representing the physical effort associated with the use of a given locomotion technique, the variation in user's heart rate before and after each scenario (to be measured using, e.g., an optical sensor) is accounted by a \textit{Physical effort} metric. 
Although heart rate variability could be a challenging and noisy measure, it already proved to be a rather accurate estimation of the energy consumption associated with a given physical exercise \cite{fischer2004energy, westerterp2009assessment, hills2014assessment, pai2017armswing}.

\begin{table*}[]
\caption{Metrics defined to evaluate in objective terms the user's experience with the given locomotion technique in the particular task, classified with respect to the corresponding NFR they contribute to, i.e., \textit{OS} for Operation speed, \textit{AC} for Accuracy, \textit{EP} for Error-proneness and \textit{OT} for Other (for cumulative metrics, numbers are used after the letter identifying the requirement to list contributing elements).}
\label{tab:Tab1}
\centering
\begin{adjustbox}{max width=\textwidth}
\begin{tabular}[t]{|p{0.15\textwidth}|p{0.2\textwidth}|p{0.77\textwidth}|}
\hline
\textbf{Scenario}  & \textbf{Task}    & \textbf{Per-task metric}\\ \hline

\multirow{3}{*}{\begin{tabular}[t]{@{}p{0.15\textwidth}@{}}\tikz \fill [l1color] (0,0) rectangle (.25,.25); \textit{S1. Straight movements}\end{tabular}} 
&  \tikz \fill [l1_1color] (0,0) rectangle (.25,.25); \textit{T1. Straight line walking}
& \begin{tabular}[t]{@{}p{0.77\textwidth}@{}} \textit{OS: ComplTime (s)}: task completion time. \\ 
\textit{AC: STPathDev (m$\cdot$s)}: spatio-temporal path deviation, calculated as the time integral of the distance from the line to follow as a function of time. 
\\
\textit{EP: NumWallColl}: num. of collisions with the walls of the corridor.\\
\end{tabular}
\\ \cline{2-3}

& \begin{tabular}[t]{@{}p{0.2\textwidth}@{}} \tikz \fill [l1_2color] (0,0) rectangle (.25,.25); \textit{T2. Over/Under-shooting} (separate values for the three  target destinations) \end{tabular}
& \begin{tabular}[t]{@{}p{0.77\textwidth}@{}}
\textit{OS: ComplTime (s)}: see above.
\\
\textit{AC: TargetDist (m)}: final distance between the user and the center of the  target in which the user stops \cite{whitton2005comparing}.\\ 
\textit{EP: NumExits}: num. of times the user exits the target destination.
\end{tabular}
\\ \cline{2-3}

& \tikz \fill [l1_3color] (0,0) rectangle (.25,.25); \textit{T3. Chasing}
& \begin{tabular}[t]{@{}p{0.77\textwidth}@{}}
\textit{AC1: InsideTargetRate (\%)}: percentage of time the user remained inside the reference area while following the moving robot.\\ 
\textit{AC2: AvgDist (m)}: average distance between the user and the center of the reference area while following the moving robot.
\\
\textit{EP: NumInterr}: num. of walking interruptions.
\end{tabular}
\\ \cline{2-3}
& \tikz \fill [l1_4color] (0,0) rectangle (.25,.25); \textit{T4. Sprinting}
& \begin{tabular}[t]{@{}l@{}} \textit{OS: ComplTime (s)}: see above.
\\
\textit{EP: NumWallColl}: see above.
\end{tabular}
\\ \hline

\multirow{5}{*}{\begin{tabular}[t]{@{}p{0.15\textwidth}@{}}\tikz \fill [l2color] (0,0) rectangle (.25,.25); \textit{S2. Direction control}\end{tabular}}
& \begin{tabular}[t]{@{}p{0.2\textwidth}@{}} \tikz \fill [l2_1color] (0,0) rectangle (.25,.25); \textit{T1. Multi-straight line walking} (values averaged on the six target destinations) \end{tabular} 
& \begin{tabular}[t]{@{}p{0.77\textwidth}@{}}
\textit{OS: ComplTime (s)}: see above.
\\
\textit{AC1: InitAngErr (deg)}: initial angular error, i.e., difference in angles between the target and the walking direction   after one meter \cite{paris2017acquisition}.
\\ 
\textit{AC2: EstPathLen (m)}: distance traveled to reach the target \cite{paris2017acquisition}.
\\
\textit{AC3: RecallTime (s)}: time spent at determining the new direction and start walking after the lights were turned off \cite{paris2017acquisition}.
\end{tabular} 
\\ \cline{2-3} 

& \tikz \fill [l2_2color] (0,0) rectangle (.25,.25); \textit{T2. Backward walking}
& \begin{tabular}[t]{@{}p{0.77\textwidth}@{}}
\textit{OS: ComplTime (s)}: see above.
\\
\textit{AC: AccuracyBkw (\%)}: path deviation correlated with the percentage of time the user looked at the target in front of it while walking backwards.
\\
\textit{EP: NumLookOut}: num. of times the gaze was turned away from target.
\end{tabular}
\\ \cline{2-3}

& \tikz \fill [l2_3color] (0,0) rectangle (.25,.25); \textit{T3. Curved walking}
& \begin{tabular}[t]{@{}p{0.77\textwidth}@{}}
\textit{OS: ComplTime (s)}: see above.
\\
\textit{EP: NumInterr}: see above. \end{tabular}\\ \cline{2-3}

& \begin{tabular}[t]{@{}l@{}} \tikz \fill [l2_4color] (0,0) rectangle (.25,.25); \textit{T4. Stairs \& ramps} \end{tabular}
& \begin{tabular}[t]{@{}p{0.77\textwidth}@{}}
\textit{OT: StairsChoice (0/1)}: user's choice between ramp (0) and stairs (1).
\end{tabular}
\\ \cline{2-3}

& \tikz \fill [l2_5color] (0,0) rectangle (.25,.25); \textit{T5. Fear}
& \begin{tabular}[t]{@{}p{0.77\textwidth}@{}}
\textit{OS: ComplTime (s)}: see above.
\\
\textit{AC: Avoidance (m$\cdot$s)}: inspired by \cite{schuemie2005effect}, but calculated like  \textit{STPathDev}, in this case using the deviation between the walking path and the edge of the large drop.
\\
\textit{EP: NumFalls}: num. of falls from the roof.
\end{tabular}
\\ \hline

\multirow{2}{*}{\begin{tabular}[t]{@{}p{0.15\textwidth}@{}}\tikz \fill [l3color] (0,0) rectangle (.25,.25); \textit{S3. Decoupled movements}\end{tabular}}
& \tikz \fill [l3_1color] (0,0) rectangle (.25,.25); \textit{T1. Decoupled gaze}
& \begin{tabular}[t]{@{}p{0.77\textwidth}@{}}
\textit{OS: ComplTime (s)}: see above.
\\
\textit{AC: AccuracyGazeUnc (\%)}: path deviation correlated with the percentage of time the user looked at target on its side while walking.
\\
\textit{EP: NumInterr}: see above.
\end{tabular}
\\ \cline{2-3} 

& \tikz \fill [l3_2color] (0,0) rectangle (.25,.25); \textit{T2. Stretched-out hands}
& \begin{tabular}[t]{@{}p{0.77\textwidth}@{}}
\textit{OS: ComplTime (s)}: see above.
\\
\textit{AC: AccuracyStrc (\%)}: path deviation correlated with the percentage of time the user kept the arms stretched while walking.
\\
\textit{EP: NumInterr}: see above.
\end{tabular}
\\ \cline{2-3} 
& \tikz \fill [l3_3color] (0,0) rectangle (.25,.25); \textit{T3. Decoupled hands}
& \begin{tabular}[t]{@{}p{0.77\textwidth}@{}}
\textit{OS: ComplTime (s)}: see above.
\\

\textit{AC: AccuracyHandsUnc (\%)}: path deviation correlated with the percentage of coins collected  by the user.
\\
\textit{EP: NumInterr}: see above.
\end{tabular}
\\ \hline

\multirow{2}{*}{\begin{tabular}[t]{@{}p{0.15\textwidth}@{}}\tikz \fill [l4color] (0,0) rectangle (.25,.25); \textit{S4. Agility}\end{tabular}}
& \tikz \fill [l4_1color] (0,0) rectangle (.25,.25); \textit{T1. Dynamic agility}
& \begin{tabular}[t]{@{}p{0.77\textwidth}@{}}
\textit{OS: ComplTime (s)}: see above.
\\
\textit{EP: NumObsColl}: num. of collisions with the moving blocks \cite{ferracani2016locomotion}. 
\end{tabular}
\\ \cline{2-3} 

& \tikz \fill [l4_2color] (0,0) rectangle (.25,.25); \textit{T2. Stationary agility}
& \begin{tabular}[t]{@{}p{0.77\textwidth}@{}}
\textit{EP: NumHits}: num. of times the user was hit by a bullet. \end{tabular}
\\ \cline{2-3} 
& \tikz \fill [l4_3color] (0,0) rectangle (.25,.25); \textit{T3. Evasion}
& \begin{tabular}[t]{@{}p{0.77\textwidth}@{}}
\textit{EP: NumHits}: see above. \end{tabular}
\\ \hline

\multirow{2}{*}{\begin{tabular}[t]{@{}p{0.15\textwidth}@{}}\tikz \fill [l5color] (0,0) rectangle (.25,.25); \textit{S5. Interaction with objects}\end{tabular}} 
& \begin{tabular}[t]{@{}p{0.2\textwidth}@{}} \tikz \fill [l5_1color] (0,0) rectangle (.25,.25); \textit{T1. Grabbing} (separate values for the three parts) \end{tabular}
& \begin{tabular}[t]{@{}p{0.77\textwidth}@{}}
\textit{OS: ComplTime (s)}: see above.
\\
\textit{EP1: NumItemFalls}: num. of times grabbed objects fell from user's hands.  
\\
\textit{EP2: NumBodyColl}: num. of collisions of the user with walls while traveling the maze
\\
\textit{EP3: NumItemColl}: num. of collisions of the items grabbed with walls while traveling the maze\cite{ferracani2016locomotion}.\end{tabular}
\\ \cline{2-3} 
& \tikz \fill [l5_2color] (0,0) rectangle (.25,.25); \textit{T2. Manipulation}
& \begin{tabular}[t]{@{}p{0.77\textwidth}@{}}
\textit{OS: ComplTime (s)}: see above.
\\
\textit{AC1: AvgSetupAcc (\%)}: average positional and rotational accuracy of items placed on the table during the setup phase.
\\
\textit{AC2: AvgTowerAcc (\%)}: average positional and rotational accuracy of items placed on the table during the assembly phase. 
\end{tabular}
\\ \cline{2-3} 
& \tikz \fill [l5_3color] (0,0) rectangle (.25,.25); \textit{T3. Interaction in Motion}
& \begin{tabular}[t]{@{}p{0.77\textwidth}@{}}
\textit{OS: ComplTime (s)}: see above.
\\
\textit{AC: CloseToTargetRage (\%)}: percentage of time the user remained close to the moving robot.
\\
\textit{EP: NumErrors}: num. of times the user performed an interaction in the wrong order or the grabbed object fell from its hands.
\end{tabular}
\\ \hline

\end{tabular}
\end{adjustbox}
\end{table*}

\subsubsection{Subjective Metrics}
\label{sec:subjective_metrics}
The remaining NFRs are analyzed in a subjective way. Subjective evaluation relies on a questionnaire split in three sections. The first section, referred to as \textit{pre-test}, is meant to be delivered before running the experiments. The second section is meant to be delivered after the completion of each scenario (\textit{after-scenario}), with slight differences from one scenario to another. Questions in this section define so-called ``per-scenario'' metrics. Lastly, the third section is intended as a \textit{post-test} questionnaire, and includes questions contributing to the definition of so-called ``overall'' metrics. Like for objective metrics, subjective metrics can be \textit{elementary} (when described by a single question), or \textit{cumulative} (when analyzed through multiple questions). Questions (sometimes in the form of statements the user has to express its agreement/disagreement with) are to be scored on a Likert Scale from 1 to 5; for motion sickness symptoms, the scale ranges from 0 (none) to 3 (severe). The complete questionnaire is available at \href{http://tiny.cc/dzxlsz}{\textit{http://tiny.cc/dzxlsz}}. In the following, the three sections are described in detail.

\textit{Pre-test section}: the section starts with questions aimed to evaluate previous experience with technologies related to the experiments. Afterwards, the section incorporates all the questions from the Simulator Sickness Questionnaire (SSQ) tool \cite{kennedy1993simulator}, which is used to rate the severity of motion sickness symptoms before the experiment (if any).

\textit{After-scenario section}: at the completion of each scenario, the user is invited to evaluate the Input sensitivity, Input responsiveness, Ease of use, Perceived (Perception of) errors,  Appropriateness and Satisfaction of/with the given technique based on statements adapted from the questionnaire proposed in \cite{kalawsky1999vruse}. The user is also asked to judge the level of Mental and Perceived physical effort associated with the use of the particular technique by rating dimensions defined by the ISO 9241-400 standard. Furthermore, the Naturalness of the walking gesture and the Similarity of the real physical strain with the virtual gesture it is serving as a proxy for are assessed based on questions defined in \cite{nilsson2013tapping}. For task \textit{S2.T5} (\textit{Fear}) the user was also asked to rate the level of disturbance on the \textit{Subjective Units of Discomfort} (SUD) scale (Acrophobia Questionnaire, AQ), as proposed in \cite{schuemie2005effect}. 

\textit{Post-test section}: this section includes questions that refer to the experience as a whole. In particular, the user is requested to provide its evaluation of the considered technique in terms of Self-motion compellingness (i.e., the perceived physical movement during the experience) and Acclimatisation (i.e., how quickly the user forgot that it was not really walking), as defined in \cite{nilsson2013tapping}. Moreover, the user needs to judge the level/sense of Control, Presence, Learnability, Intuitiveness, Comfort, Enjoyability and Overall system usability provided by the given technique according to \cite{kalawsky1999vruse}. Lastly, severity of  motion sickness symptoms is collected again using the SSQ to study in detail changes from the beginning to the end of the experience.

\subsection{Experimental Protocol}
\label{sec:experimental_protocol}

This section illustrates how to customize the testbed and how to perform the testing in order to foster reproducibility. 

\subsubsection{Preparation}
\label{sec:test_preparation}

As said, the five scenarios were designed to cluster tasks with similar characteristics, in order to maintain the consistency of the per-scenario metrics and to let the testbed users skip the testing of unnecessary aspects. That is, the study participants may be asked to take, e.g., only the first four scenarios, dropping the last one if interaction with objects is not of interest.  
Reordering the scenarios is possible, though not recommended, as they have been designed to challenge the participants with tasks of increasing complexity. 

Differently than scenarios, tasks cannot be skipped, as such changes would influence the evaluation. In fact, metrics have been devised to assess user's performance and experience on the scenario as a whole, with dimensions that in some cases are investigated through more than one task. Moreover, aspects like Motion sickness, Acclimatisation or Physical effort, to name a few, can only be evaluated once the user has spent a sufficient amount of time in the scenario. This is also the reason why it was decided to move the delivery of the questionnaire at the end of each scenario, rather than at the end of each individual task. 

Each participant is expected to test the selected scenarios with just one locomotion technique. This is due to the fact that the testbed has been developed around a between-subjects design, in order to enable the reuse of data from previous experiments when new techniques have to be included in a study (as long as the conditions are reasonably comparable to the previous ones, e.g., in terms of demographics information). Hence, questionnaire does not contain any question asking the participants to directly compare two (or more) techniques. 

In principle, a within-subjects approach could be used too, but strategies for dealing with learning effects would be hard to design. It is also worth considering that the duration of the complete test experience is about 80 minutes  (all scenarios and questionnaire sections included). For a within-subjects study, time requested would explode as the number of techniques tested grows, not to mention that participants could be mentally and physically challenged by requested repetitions. In conclusion, a within-subjects design could probably be feasible for a small number of techniques, possibly limiting the number of tested scenarios. 

\subsubsection{Execution}
\label{sec:test_execution}
Participants must be instructed not to eat or drink anything two hours prior to the experiment and not to consume illicit drugs or caffeine for 12 hours prior to the experiment \cite{ nonmangiarenonbere}. Only participants without illnesses or visible altered state of consciousness shall be accepted as testers.

Every participant has to be briefly informed about the purpose of the experience, the structure of the experimental protocol, as well as the operation of the VR hardware and the locomotion technique used for the activity. After that, the participant fills in the pre-test section of the questionnaire and the administrator shall decide whether to proceed with the experiments or discard the participant, considering the match with the target audience set for the study and the SSQ scores.

The participant is then provided with the VR equipment, and let in the training environment. As said, this scenario is meant to let the participant familiarize with the selected locomotion technique, and to deliver all the information that may be necessary to carry out the various tasks (if necessary, further details on the operation of the locomotion technique could be provided also prior to wearing the VR equipment, especially if safety considerations are involved). In particular, participants must be introduced to the concept of ``destination'', i.e., the visual indicators they will have to reach in order to start, perform or conclude many of the tasks, as well as to the so-called ``blocked'' state, that is a temporary situation (signaled by a padlock, as in Fig. \ref{fig:Fig1_s1}) in which the locomotion technique is disabled, usually to let the administrator explain the following task before proceeding; this state is activated automatically at the end of each task, but the administrator can also block the participant manually in order to give additional instructions. 

During the training, the participant shall be asked to perform a set of actions that will be later requested in the execution of the tasks (i.e., walking straight, changing direction, adjusting the speed, walking backwards, etc.), as well as to experiment with hand interactions (by picking up some objects and transporting them to other locations). Operations above need to be repeated until the participant is feeling confident in using the locomotion technique and in control of the various interactions. Finally, calibration data for the participant have to be set, either automatically or manually. Information to be used for the guidance are provided with the testbed in the ``administrator script''.

After the training, the participant is ready to start the testing procedure, which can be summarized as follows:
\begin{enumerate}
    \item participant's heart rate is collected just before entering the scenario, with the VR headset on;
    \item scenario is started, participant is let in the first task;
    \item locomotion is blocked, so that task can be explained by the administrator;
    \item locomotion is unblocked, and the participant performs the task;
    \item at the end of the task, the system blocks the locomotion again;
    \item points 3)--5) are repeated for all the tasks;
    \item participant's heart rate is collected again at the end of the scenario;
    \item headset is removed, and participant is asked to fill in the after-scenario section of the questionnaire, supervised by the administrator;
    \item points 1)--8) are repeated for the scenarios of interest;
    \item participant fills in the post-test section of the questionnaire, supervised by the administrator.
\end{enumerate}

In case of interruptions due to internal (e.g., extreme motion sickness) or external reasons (blackout, force majeure, etc.), the activity has to be suspended. It is at the discretion of the administrator to decide whether to let the participant resume the experiment or not, and how to treat the possibly incomplete data, e.g., discarding them, filling missing values with the mean or worst scores, etc.


\section{Scoring System}
\label{sec:rss_tool}

As a result of the application of the methodology illustrated in the previous section, raw measurements corresponding to objective and subjective metrics are collected in a so-called \textit{Raw Database} (RDB) for a number of participants with all the locomotion techniques being studied. In order to use these data for comparing the various techniques, a scoring system based on the weighted sum model (WSM) is proposed as part of the testbed. WSM is a multi-criteria decision analysis (MCDA) method which was designed to evaluate a set of alternatives (techniques, in this case) based on a number of decision criteria (requirements) which are assigned a weight indicating the relative importance. 

\subsection{Metrics Normalization}
\label{sec:metric_contrib}

The contributions of the various metrics to the overall score need to be expressed in the same unit. Hence, a normalization step is required. In the devised system, normalization relies on testing the statistical significance of differences between the mean of the metric for a given technique and for all the other techniques being studied. 

A significance threshold equal to $5\%$ is used (the system does not mandate, however, the use of a particular statistic test or threshold). If the test gives a $p$-value less than or equal to the threshold, one point is assigned to the ``best'' technique. The proposed normalization approach does not consider the magnitude of the differences. However, the advantage is that it does not require the definition of lower and/or upper bounds for the metrics. As a matter of example, an arbitrary bound on the \textit{ComplTime} of a given task would not allow the testbed to account for future locomotion techniques providing largely different speed performance than those currently available. Similar considerations apply to \textit{NumFalls} or \textit{NumInterr}, which could theoretically grow unconstrained for poorly controllable techniques.

The system requires the testbed user to define, for each metric, a so-called \textit{direction}. Direction can be either positive or negative, and is used to determine whether a technique should be considered as better than another one when the mean value for the metric is greater or smaller than that of the other technique, respectively.

As a matter of example, assume that four techniques T1, T2, T3, and T4 are tested and that, for a given metric $m$, significant differences are found between T1 and T2, T1 and T3, as well as T3 and T4. The best technique in pairwise comparisons will get $1$ point. The selection of the best technique depends on the direction assigned to the metric, which, as said, could be either positive (so that the technique with the higher mean will be selected as the best, and will get the points), or negative (vice versa). Thus, if direction of metric $m$ is positive and the means for T1, T2, T3 and T4 are, e.g., $1.0$, $2.0$, $2.5$, and $1.5$, respectively, the score $S_m$ (in points) assigned by the system for that metric will be equal to 0 for T1, 1 for T2, 2 for T3, and 0 for T4.

For a cumulative metric 
%
, the score is computed as:
\textbf{\begin{equation}
S_m = \frac{1}{\hat{N_{e}}
} \sum_{e=1}^{N_e} S_{e}
\label{eq:cms}
\end{equation}}
\hspace{-3pt}where $S_{e}$ is the number of points of the $e$-th element, ${N_{e}}$ is the number of elements and $\hat{N_{e}}$ is the number of elements which showed at least one statistically significant difference.

\subsection{Weights Selection}
\label{sec:weight_Selection}
The scoring system requests the testbed user to assign a weight (in the range [0,1]) to a set of predefined dimensions. Weights are then combined with statistically processed data to compute an overall score for the selected techniques, and rank them based on their suitability to the specific application domain they will be used into. 

Dimensions to be weighted include both FRs and NFRs identified in the previous section. As said, FRs correspond to tasks and scenarios illustrated in Fig. \ref{tab:TabTasks}, whereas NFRs are aligned with the objective and subjective metrics collected through the testbed application and are reported in Table \ref{tab:player_req}. In a basic WSM implementation, weights would be directly combined with requirements. In this case, a two-level combination is used, due to existence of two types of requirements. 

For what concerns FRs, weights can be assigned with both a coarse (per-scenario) or a fine (per-task) granularity. If weights are defined per-scenario, the weight of each scenario is applied to all its tasks; if weights are assigned per-task, per-scenario weights are determined automatically by averaging per-task ones. 
NFR weights, on the contrary, are assigned directly to the corresponding requirements; the only exception pertains Motion sickness, for which weights can be either assigned to major components identified in the SSQ (namely Nausea, Oculomotor symptoms and Disorientation), or to the requirement as a whole (Total). 

\subsection{Scores Computation}
\label{sec:score_compute}

For the computation of the overall score for a given technique, every NFR-related score is first multiplied by the corresponding weight, obtaining a so-called \textit{weighted NFR score}. 
Then, the per-task weighted NFR scores (related to the first three objective metrics, i.e., Accuracy, Operation speed and Error-proneness), are multiplied by the fine FR weights, whereas the per-scenario weighted NFR scores (related to the questions in the after-scenario questionnaire and to the last objective metric, the \textit{Physical effort}) are multiplied by the coarse FR weights.
The weighted NFR scores related to the overall metrics (based on the questions in the post-test questionnaire) will not be influenced by any of the FR weights, as they refer to the experience as a whole. The overall score is obtained by summing up the weighted scores of the various metrics.

The weighted score associated with the objective metrics of task $T_i$ in scenario $S_j$ is calculated as:
\begin{equation}
        \begin{aligned}[b]
        S_{S_{j}.T_{i}} = w_{S_{j}.T_{i}}  \cdot  \hspace{5.7cm} \\
         \cdot (w_{OS} \cdot S_{OS_{S_{j}.T_{i}}} + w_{AC}\cdot S_{AC_{S_{j}.T_{i}}} + w_{EP} \cdot S_{EP_{S_{j}.T_{i}}}) 
        \end{aligned}
\label{eq:ss}
\end{equation}
where $w_{S_{j}.T_{i}}$ is the weight assigned to the task, $w_{OS}$, $w_{AC}$ and $w_{EP}$ are the weights assigned to the Operation speed, Accuracy and Error-proneness objective metrics, and  $S_{OS_{S_{j}.T_{i}}}$, $S_{AC_{S_{j}.T_{i}}}$ and $S_{EP_{S_{j}.T_{i}}}$ are the scores computed for the three metrics on the given task. 

For task \textit{S2.T4}, in order to handle the peculiarity of the \textit{StairsChoice} metric, its contribution is calculated as:
\textbf{\begin{equation}
\hat{S}_{S_{2}.T_{4}} = w_{S_{2}.T_{4}} (w_{ST}\cdot S_{ST} + w_{RA}\cdot S_{RA})
\label{eq:scs}
\end{equation}}
\hspace{-3pt}where $w_{S_{2}.T_{4}}$ is the weight of task \textit{S2.T4}, $w_{ST}$ and $w_{RA}$ are special mutually exclusive weights that can just be 0-0, 0-1 or 1-0, whereas $S_{ST}$ and $S_{RA}$ are the scores computed from the metric, but with opposite directions (stairs or ramps).

Similarly, the score of task \textit{S2.T5} has to be modified to consider the contribution of the \textit{SUD}, which is the only subjective metric related to a single task; thus, the additional score for the task is calculated as:
\textbf{\begin{equation}
\hat{S}_{S_{2}.T_{5}} = w_{S_{2}.T_{5}} (w_{SUD} \cdot S_{SUD})
\label{eq:fs}
\end{equation}}
\hspace{-3pt}which is basically the contribution of the weighted score for the additional metric combined with the task weight $w_{S_{2}.T_{5}}$.

The subjective, per-scenario part of the score for scenario $S_j$ will be calculated as:
\textbf{\begin{equation}
S_{s_{j}} = w_{S_{j}}\sum_{m=1}^{N_s} w_{s_{m}}\cdot  S_{s_{m_{j}}}
\label{eq:spss}
\end{equation}}
\hspace{-3pt}where $w_{S_{j}}$ is the weight of the scenario, $N_s$ is the number of subjective metrics defined for the scenario, $w_{s_{m}}$ and $S_{s_{m_{j}}}$ are respectively the weight of a subjective metric \textit{m} and its score for the given scenario.

The overall score for scenario $S_j$ will be computed by combining the (per-scenario) subjective component in (\ref{eq:spss}) with the (per-task) objective component in (\ref{eq:ss}) as:
\textbf{\begin{equation}
S_{S_{j}} = S_{s_{j}} + \sum_{i=1}^{N_t} S_{S_{j}.T_{i}}  + w_{PE} \cdot S_{PE_{j}}
\label{eq:oss}
\end{equation}}
\hspace{-3pt}where ${N_t}$ is the number of tasks in the scenario; the formulation also considers $w_{PE}$ and $S_{PE_{j}}$, which are the weight and the score for the \textit{Physical Effort} objective metric. 

For the overall metrics, the score is calculated as:
\textbf{\begin{equation}
S_{o} = \sum_{i=1}^{N_o}w_{o_{m}} \cdot S_{o_{m}}
\label{eq:os}
\end{equation}}
\hspace{-4pt}where $N_o$ is the number of overall metrics, $w_{o_{i}}$ is the weight of the $m$-th overall metric and $S_{o_{m}}$ is its score.

Based on this formulation, the overall score for the given locomotion technique is computed as:
\textbf{\begin{equation}
S = S_{o} + \sum_{j=1}^{N_S} S_{S_{j}} + \hat{S}_{S_{2}.T_{4}} + \hat{S}_{S_{2}.T_{5}} 
\label{eq:ots}
\end{equation}}
\hspace{-3.5pt}where $N_S$ is the number of scenarios included in the test. 

\subsection{Testbed Usage}
\label{sec:framework_usage}

The results obtained through the application of the scoring system to data in the RDB are recorded in the so-called \textit{Weighted Database} (WDB). The WDB basically contains the values computed for each metric in a given set of experiments, their mean values and significances, as well as the overall scores obtained by applying weights.

The WDB is characterized by a \textit{fixed} part, which is strictly related to the testing activity performed and encompasses:
\begin{itemize}
    \item the set of techniques included in the experiments;
    \item the set of scenarios included in the experiments;
    \item possibly applied demographic constraints.
\end{itemize}

Then, it contains a \textit{variable} part that needs to be configured by the testbed user in order to compute the overall scores. This part includes:
\begin{itemize}
    \item the weights assigned to the FR/NFR metrics;
    \item the direction assigned to each metric;
    \item the subset of techniques to be actually compared.
\end{itemize}

The variable part can be adjusted by the testbed user depending on the specific study to be performed, without the need to carry out any further experimental  activity. 

For the purpose of identifying the locomotion technique that represents the best match for a particular scenario given a set of possible alternatives, the user can either consider an existing WDB (if the fixed configuration is appropriate for the study at hand), or generate a new one (by performing the experiments and applying the statistical analysis). Then, it should adjust the weights for the various requirements depending on the specific application scenario. For instance, should the goal be to select the best technique for a VR application in which the users need to run and complete the assigned tasks in the shortest time possible, then FRs like, e.g., \textit{Sprinting}, and NFRs like, e.g., \textit{Operation speed}, would probably have to be assigned a high weight. The overall scores computed by the system would then provide the ranking of techniques for the specific application. 

If some of the techniques in the WDB are not of interest or are not compatible with the given scenario (e.g., the application requires more buttons than those that are available on the controllers when a particular technique is used), then they should be discarded in the RDB, and a new WDB could be generated by repeating the statistical analysis and recomputing weighted scores.  

Besides overall scores, the user could also analyze scores computed for individual metrics, with the aim to identify aspects of a given technique that are particularly effective and could be exploited, e.g., in the design of a new technique.


\section{Use Case}
\label{sec:use_case}
After having discussed the characteristics of the testbed, a use case is presented to provide potential users with an example of the workflow which is expected to be followed to compare a set of locomotion techniques. Various techniques coming from both industrial and academic research were preliminarily analyzed, by focusing on those capable to address the limited space constraints. Four techniques were ultimately selected for the comparison, chosen among those for which implementation details or ready-to-use packages were available (especially when more variants existed).  

In the following, selected techniques and their implementations will be first described. Then, the user study performed to collect experimental data will be introduced, and the use of the relative scoring system will be explained; in particular, it will be shown how to generate the WDB and use it to compare the selected techniques under the particular usage conditions set by a specific VR application.

\subsection {Considered Techniques}
\label{sec:techniques}

As said, four techniques were selected for the comparison: AS, WIP, Cyberith's Virtualizer (CV), and joystick (JS). The VR system used in the experiments was the HTC Vive.

\subsubsection{Arm Swinging}
\begin{figure}
\centering
\subfloat[]{\includegraphics*[width=0.5\columnwidth]{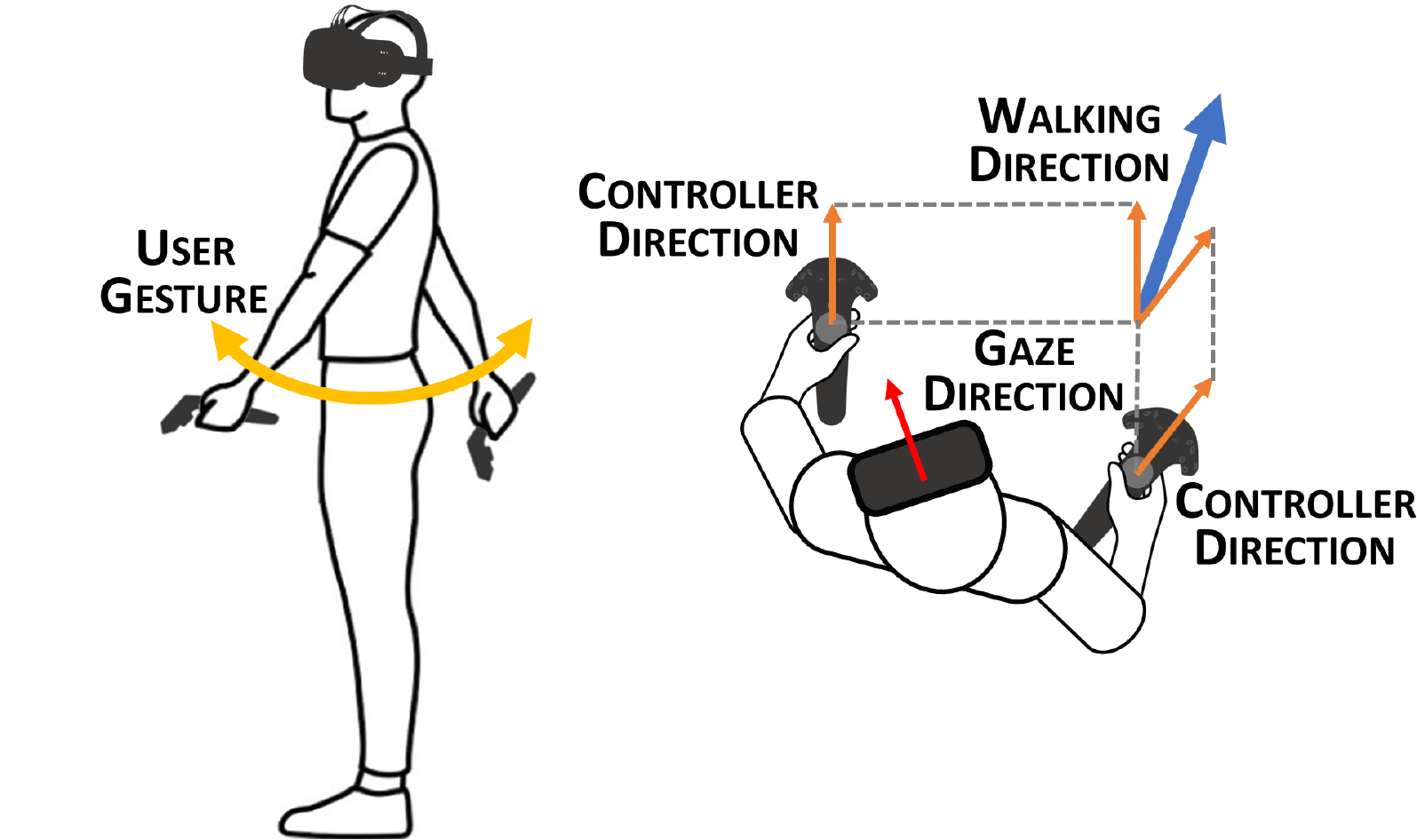}%
\label{fig:Fig2_a}}
\hfil
\subfloat[]{\includegraphics*[width=0.5\columnwidth]{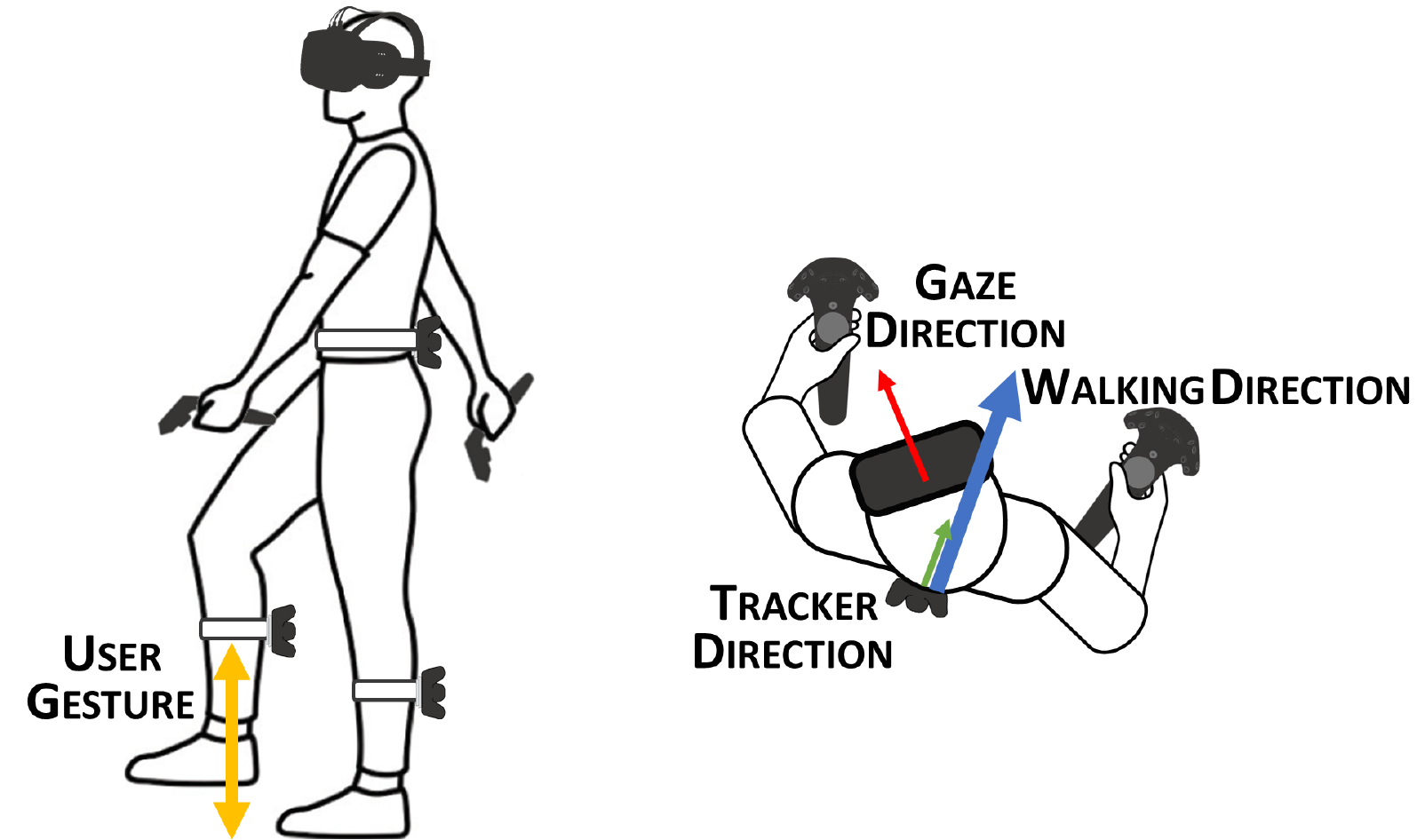}%
\label{fig:Fig2_b}}
\hfil
\subfloat[]{\includegraphics*[width=0.5\columnwidth]{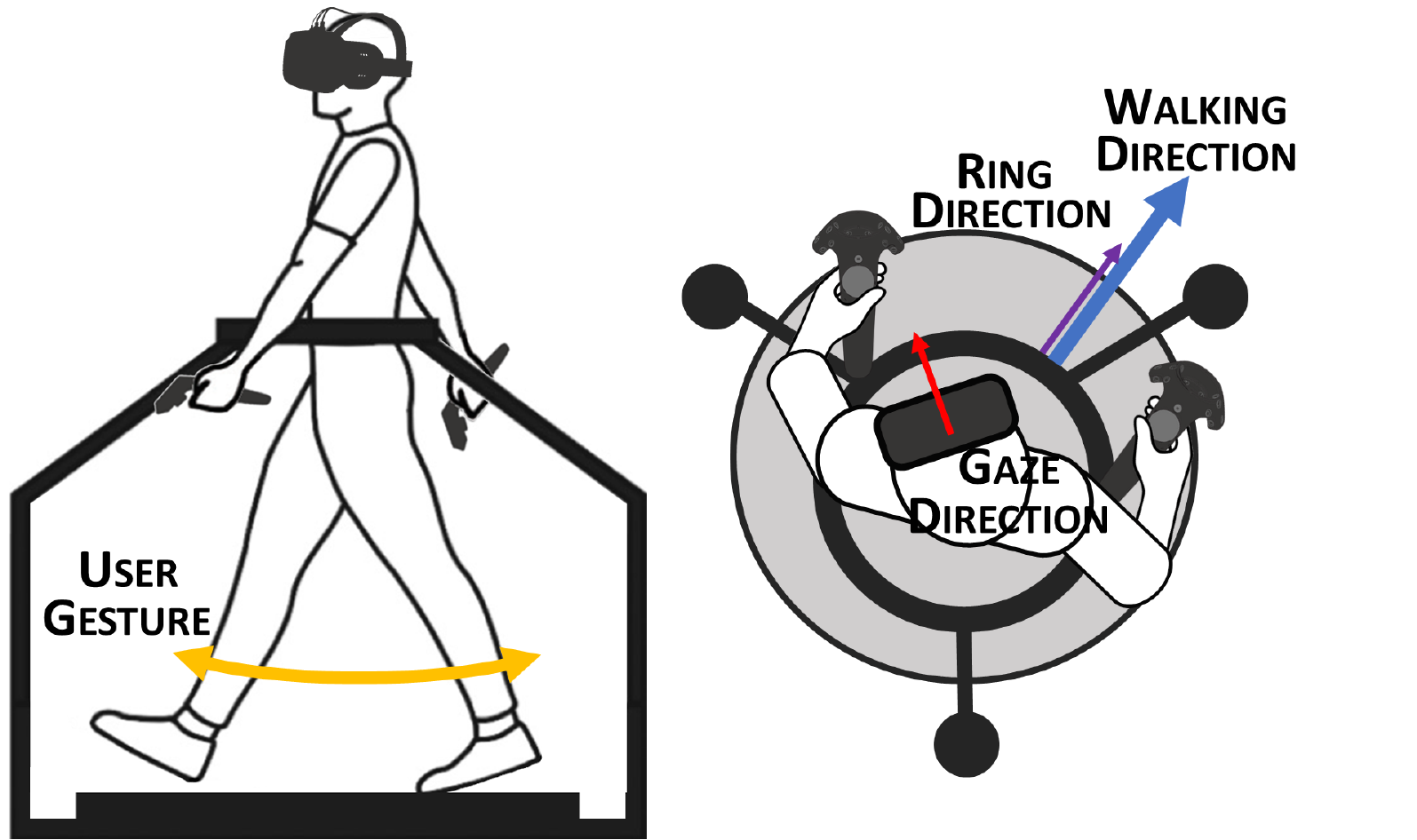}%
\label{fig:Fig2_c}}
\hfil
\subfloat[]{\includegraphics*[width=0.5\columnwidth]{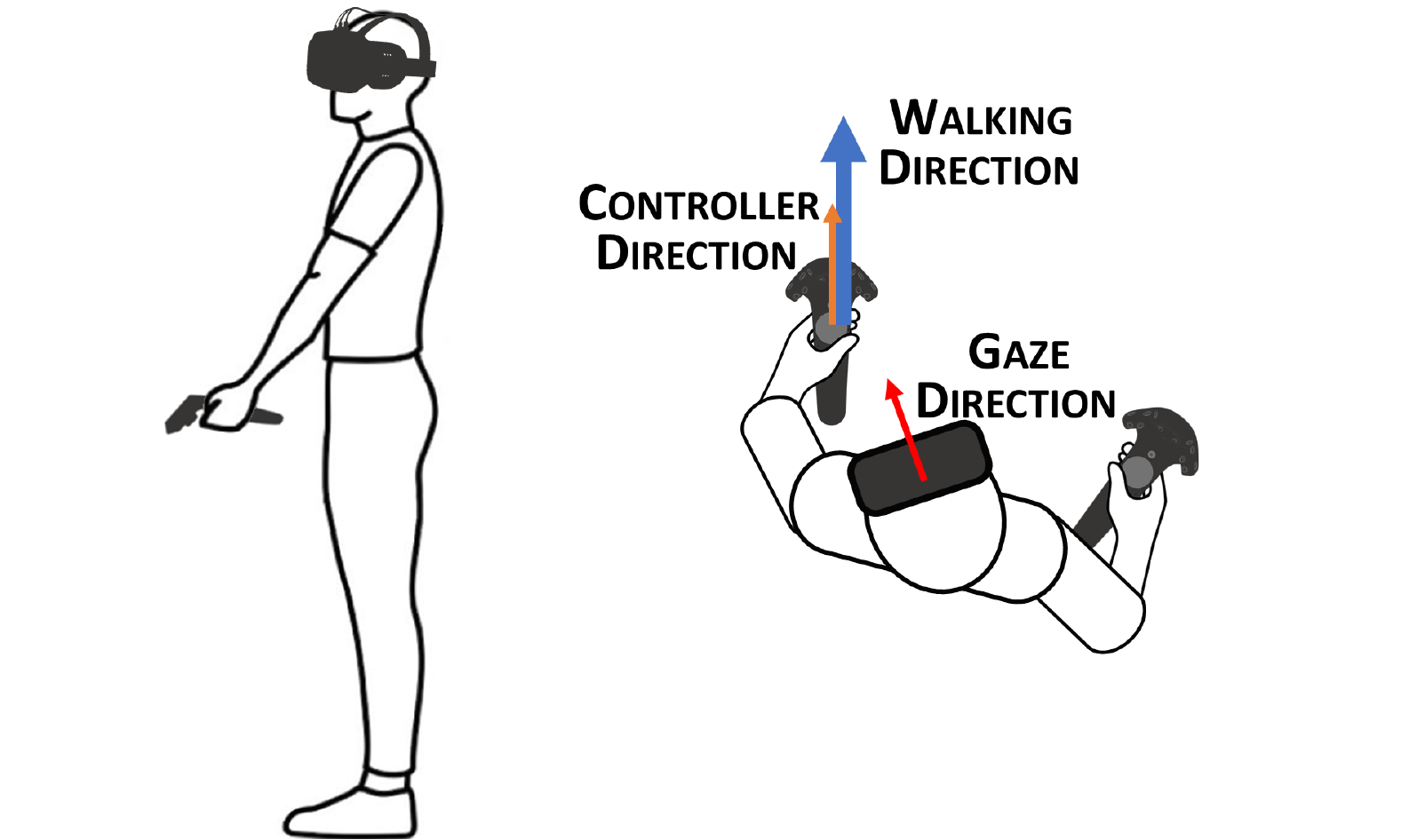}%
\label{fig:Fig2_d}}
\caption{Locomotion techniques considered in the use case: a) arm swinging (AS), b) walking-in-place (WIP), c) Cyberith's Virtualizer (CV), and d) joystick (JS).}
\label{fig:Fig2}
\end{figure}

According to the experiments performed in \cite{loup2018effects}, AS seems to leverage the most natural gesture among the approaches not involving the use of feet. To generate movement, the user must hold a button on the hand controllers (the grip, in the considered implementation), then swing the arms back and forth for walking/running. For this technique (as well as for the others), the trigger buttons on the hand controllers are used to interact with objects. In this work, the publicly available AS method in \cite{armswinger} is used, in which the direction of movement is determined by averaging the orientation of the two controllers (Fig.~\ref{fig:Fig2_a}), whereas the stride length is directly mapped on the arm swinging. This choice allows the user to decouple the walk direction from the gaze orientation.

\subsubsection{Walking-in-Place}
The second technique included in the study is WIP, which is another common locomotion method proposed in various versions and implementations. To generate movement, the user needs to perform a particular gesture with its legs while remaining in place. The direction of movement can be obtained from the head (by tying it to the gaze direction) or from other devices (when the gesture is recognized, e.g., using wearable sensors). In the use case, the LLCM-WIP variant proposed in \cite{feasel2008llcm} was chosen; according to this approach, the movement is generated through a direct mapping between the space covered by two sensors attached to the user's leg and the speed of the user in the virtual environment, whereas the direction is obtained through a third sensor placed on the user's back. In the exploited implementation, three Vive Trackers were used, two attached on the user's calves through custom 3D-printed supports, one tied to the back through a belt (Fig.~\ref{fig:Fig2_b}). With this solution it was possible to ``align'' this technique to the other ones (by acting on the configuration parameters), and to decouple the head/gaze orientation from the direction of movement. The gesture used is the marching one, which is considered as a kind of standard for this technique \cite{nilsson2013tapping}. A filter on horizontal leg movements was added, in order to mitigate unwanted motion when the user turns around.

\subsubsection{Cyberith's Virtualizer}
The third technique that was considered in the comparison is the CV, whose functioning is illustrated in Fig.~\ref{fig:Fig2_c}. This is one of the few commercial solutions for locomotion in VR that falls under the category of passive repositioning systems (thanks to the use of a low-friction walking surface and of a rotating containment ring, which prevents the user from displacing in the physical space). In this case, the walking direction depends on the orientation of the ring. This slippery shoes-based interface, which was originally presented in \cite{cakmak2014cyberith}, has been used already in previous research works and compared with other techniques, though only in broad terms \cite{Nilsson:2018:NWV:3181320.3180658, calandra2018eg}.

\subsubsection{Joystick}
For the JS, the implementation developed in \cite{boletsis2019vr} was used. In this implementation, movement is activated by pressing the pad button of any of the hand controllers and modulating the speed by moving up and down the thumb over the touch pad: the upper bound of the pad generates the maximum speed, which decreases linearly by moving down the finger towards the lower bound (zero speed). The direction is given by the controller whose touch pad is pressed to activate the motion. In order to transition from walking to running, the user needs to press the grip button (on the same controller) while performing the previous actions; this way, speed will be set to a fixed value, higher than the one reachable through the previously described modulation.

\subsubsection{Configuration}

During the experiments, the user is requested to stay within a predefined working area (since the testbed supports the study of a specific locomotion technique at a time, room-scale movements are disabled, although they could be studied as a separate technique). The working area is signaled in the virtual environment through a semi-transparent cyan cylinder with a 65cm radius around the initial user's position, which at the user's feet level is displayed as a brighter circle.

When, due to unintentional movements, the user enters a warning region that ranges from 90\% to 100\% of the working area radius, the controllers generate a haptic feedback in order to help it to remain at the center of the working area. When the user exits the working area, movement in the virtual environment is disabled, so that the only action allowed to the user is stepping back into it. The reason for this choice was to allow the user to crouch and perform on-the-spot head and hip movements while avoiding motion sickness and disorientation that would be introduced, e.g., by simply limiting the headset movements to three DOFs. In order to standardize the experience across the four techniques, maximum speed was set to 7m/s \cite{speed_max_alt}. For JS, the maximum value was used for running (when the grip button is pressed), whereas for walking the speed can be modulated between 0 and 3.5 m/s (slightly higher than common values for real walking, in the 1.5--2.6m/s range \cite{speed_walk}). For the other techniques, the user can modulate the speed between zero and the maximum value. 

Regarding interaction with objects, the implementation adopted in \cite{calandra2018eg} and \cite{calandra2019icce} was exploited, which proved to be characterized by a high usability regardless of the particular locomotion technique used in the experiments. In particular, the user needs to touch the object with the tip of the hand controller and, once the contact is confirmed by a visual (outline) and tactile (vibration) feedback, it can push and hold the grab button (the trigger in this case, visually signaled by a blinking outline) to bind the object to its hand and move it. The object can be dropped by releasing the trigger button, or passed to the other hand by grabbing it with the corresponding controller.

\subsection{Collecting Experimental Data}
\label{sec:experiment}
In order to produce the RDB, an experimental activity was carried out by involving 48 volunteers from the student and academic population at the authors' university (37 males and 11 females, aged between 19 and 37). Each participant was assigned to one of the four locomotion techniques.

According to information collected through the pre-test section of the questionnaire, participants had a very limited experience with VR and related concepts, since $12.5\%$ of them were regular users of the technology, whereas $31.25\%$ asserted to play 3D videogames often or very often. Only $8.33\%$ of the participants had used sometimes the assigned locomotion technique before the experiment, $20.83\%$ had a little knowledge of it, and $70.83\%$ had never used it. Regarding motion sickness, none of the participants reported particularly high symptoms before the experiments. Like in \cite{pre_ssq}, a statistical analysis was performed to identify possible differences among the groups with respect to the pre-test conditions (precisely, Kruskal-Wallis and Dunn's post-hoc tests were used, due to the observed data distribution); no statistically significant difference was found.

Participants who had never experienced a VR system were given time to familiarize with this technology (headset, controllers, etc.). They could then initiate the training. Afterwards, they were asked to complete the experimental protocol. All the scenarios were tested, in the default order. All the participants were able to complete the experiment.

\subsection{Computing Normalized Scores}
\label{sec:pdb_generation}
Data in the RDB were processed using the devised scoring system. Possible outliers (identified by using the Z-score) were removed first. Afterwards, normality distribution of the data was checked by applying the Shapiro-Wilk test on each metric. 

Since participants experimented all the scenarios using only one interface, it could be assumed that collected data are independent. Hence, in order to compute the $p$-values required by the scoring system, the following approach was pursued. For data presenting non-normal distributions, statistical significance of the differences was studied using the Kruskal-Wallis test, followed by the Dunn's post-hoc test for pairwise comparisons. For normally distributed data, statistical significance was tested using ANOVA, followed by one-to-one comparisons with the Tukey's HSD range test.

To implement the procedure illustrated in Section \ref{sec:rss_tool} and generate the WDB for the four techniques, an Excel spreadsheet was developed: pasting the experimental data and choosing the weights, the proper statistical tests can be applied to compute the scores (for all the metrics and the selected techniques). The spreadsheet including both raw data and weighted scores is available at \href{http://tiny.cc/hzxlsz}{\textit{http://tiny.cc/hzxlsz}}.

\begin{figure*}%
\centering
\subfloat[Accuracy]{\includegraphics*[width=0.6\columnwidth]{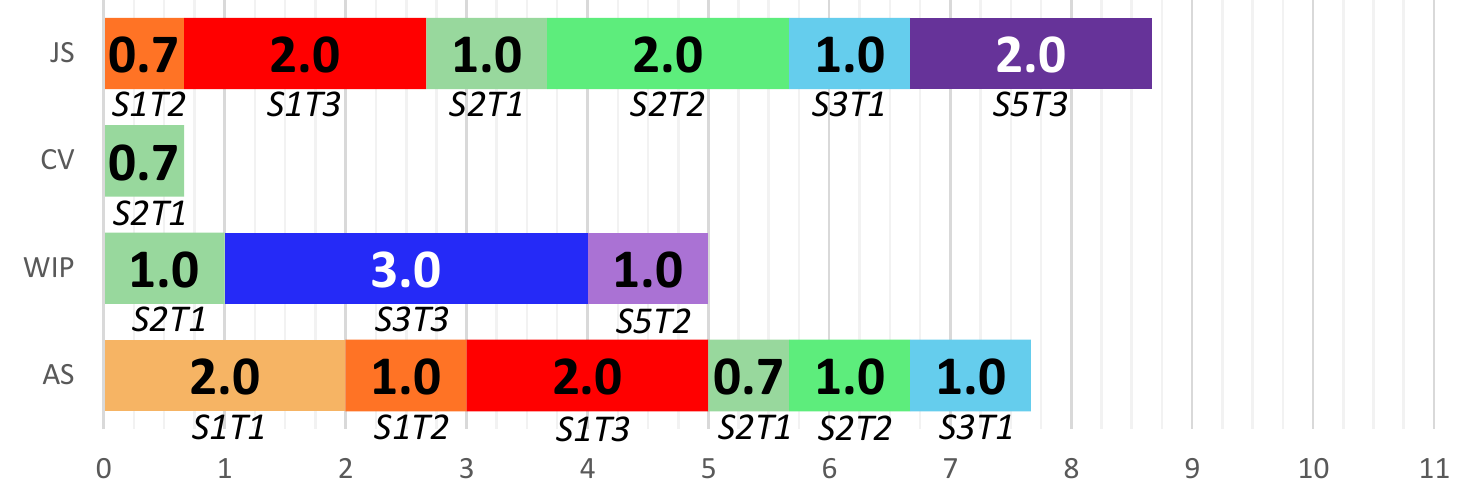}%
\label{fig:plot_2}}
\hfil
\subfloat[Operation Speed]{\includegraphics*[width=0.6\columnwidth]{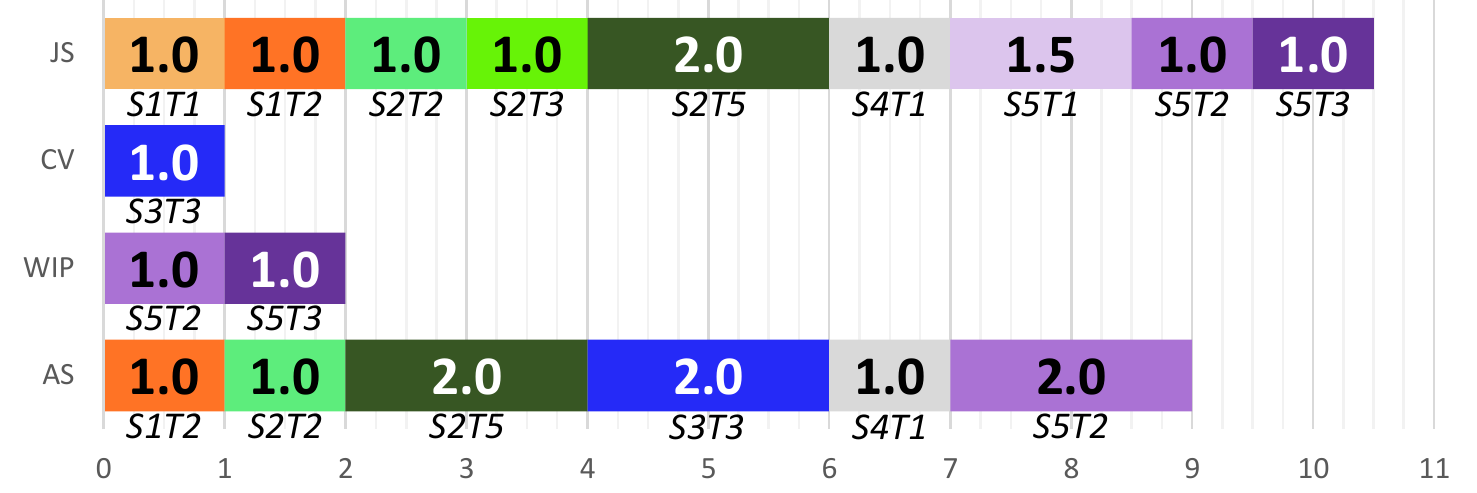}%
\label{fig:plot_1}}
\hfil
\subfloat[Error-proneness]{\includegraphics*[width=0.6\columnwidth]{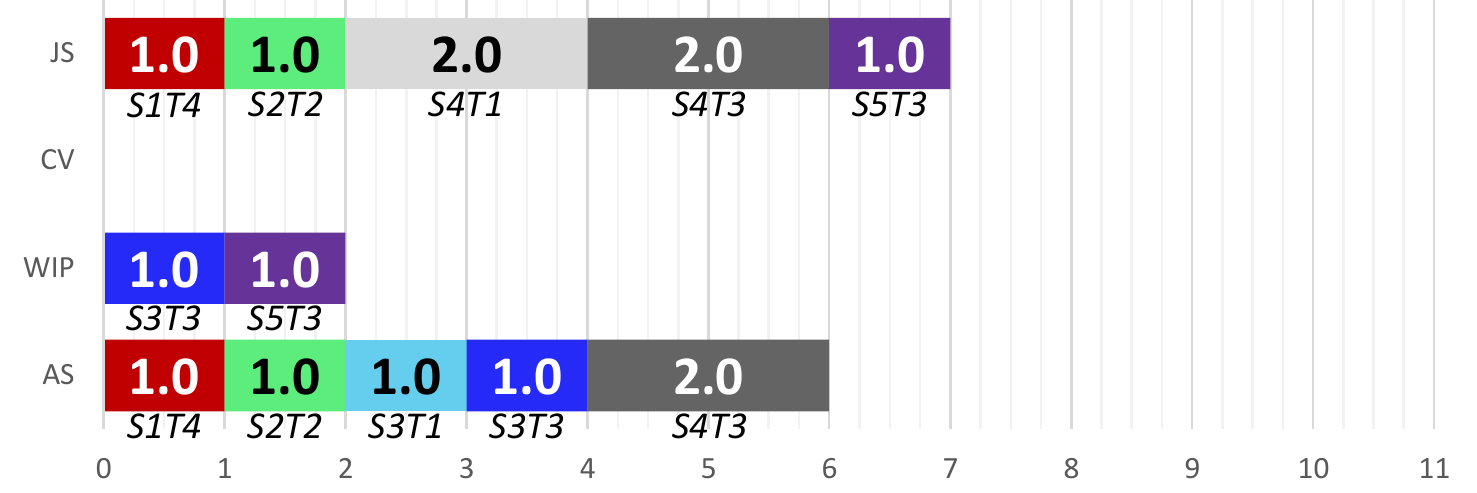}%
\label{fig:plot_3}}
\hfil
\subfloat[Physical effort]{\includegraphics*[width=0.6\columnwidth]{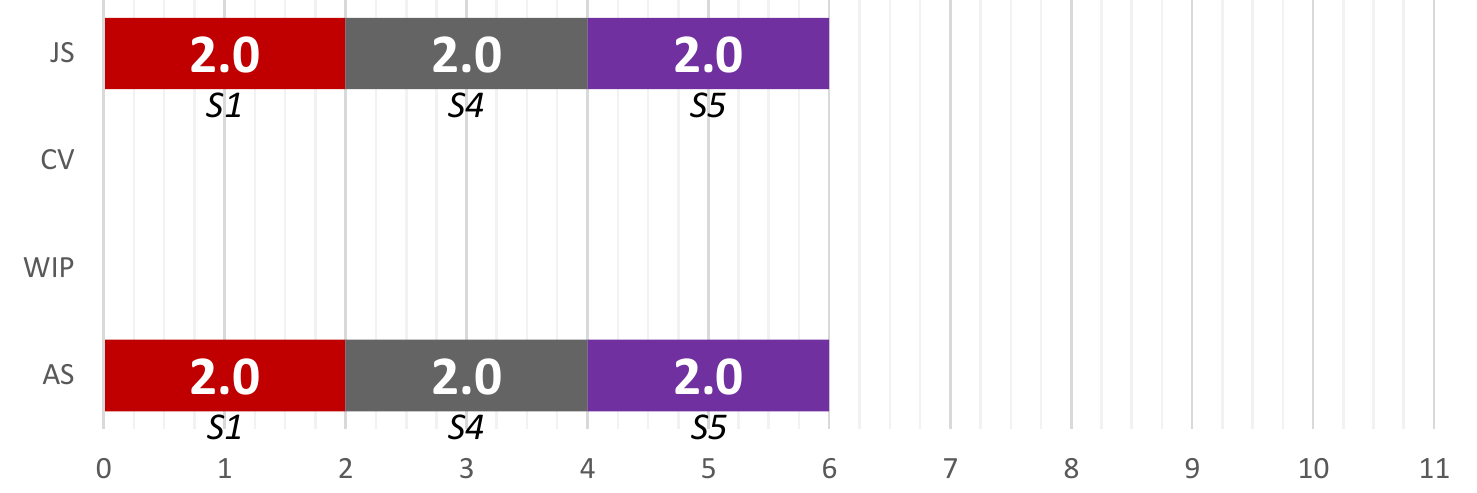}%
\label{fig:plot_14}}
\hfil
\subfloat[Input sensitivity]{\includegraphics*[width=0.6\columnwidth]{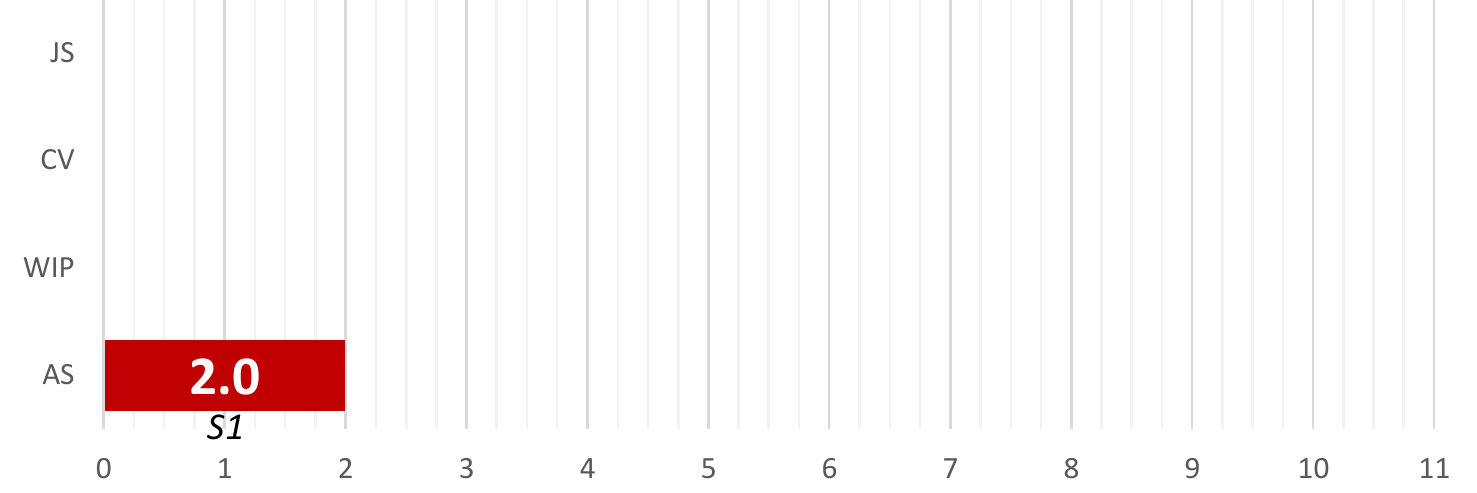}%
\label{fig:plot_12}}
\hfil
\subfloat[Input responsiveness]{\includegraphics*[width=0.6\columnwidth]{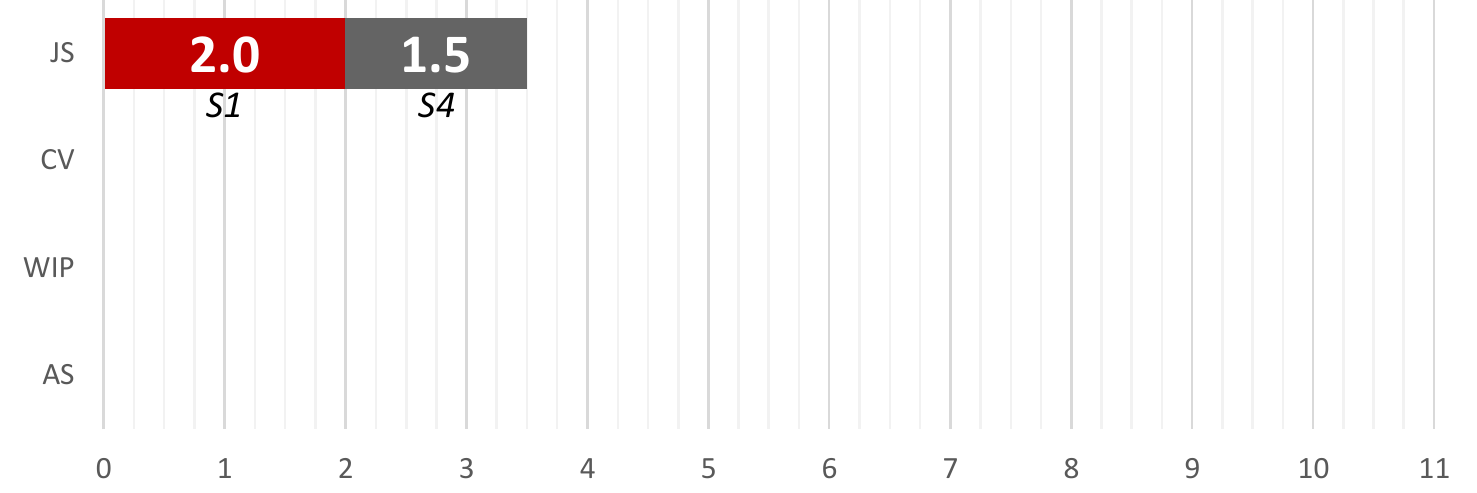}%
\label{fig:plot_13}}
\hfil
\subfloat[Ease of use]{\includegraphics*[width=0.6\columnwidth]{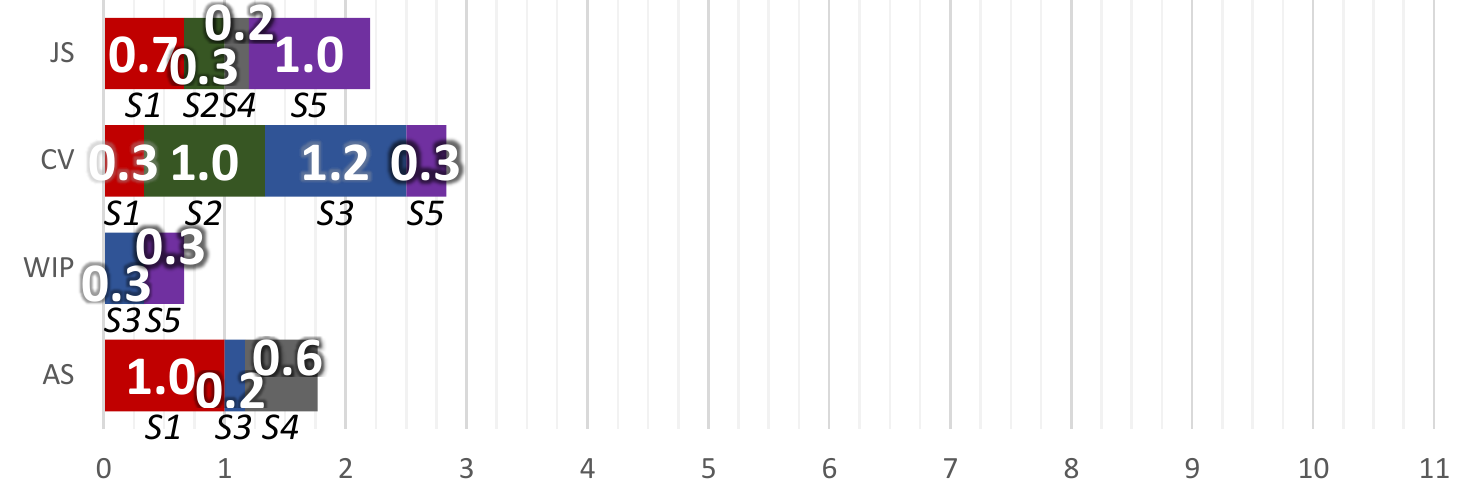}%
\label{fig:plot_4}}
\hfil
\subfloat[Perceived errors]{\includegraphics*[width=0.6\columnwidth]{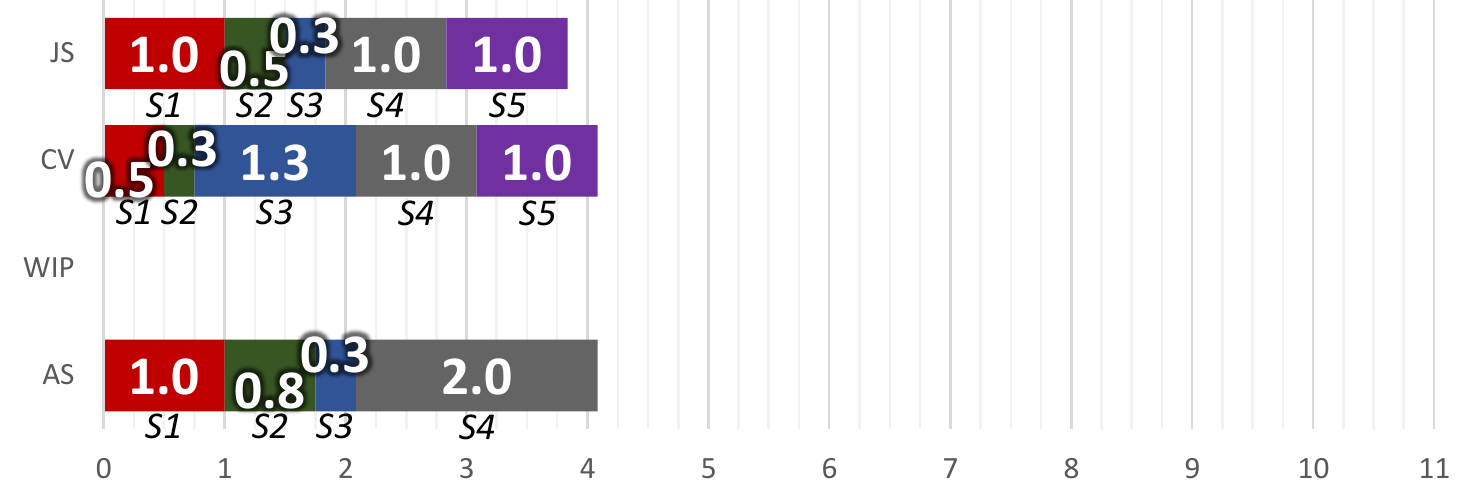}%
\label{fig:plot_7}}
\hfil
\subfloat[Appropriateness]{\includegraphics*[width=0.6\columnwidth]{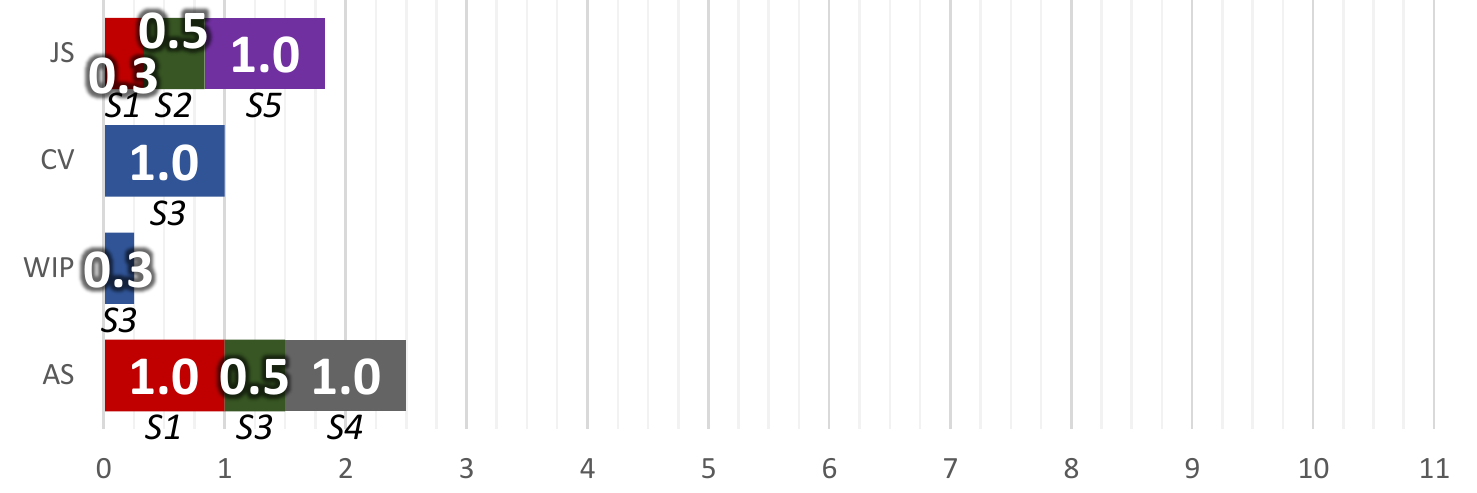}%
\label{fig:plot_10}}
\hfil
\subfloat[Satisfaction]{\includegraphics*[width=0.6\columnwidth]{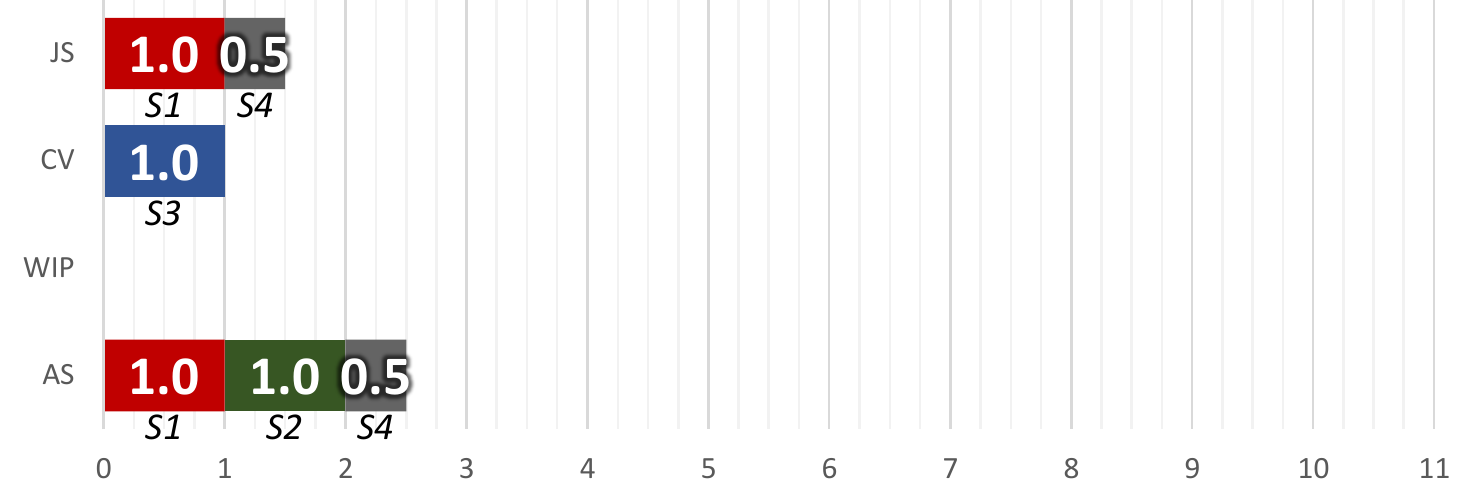}%
\label{fig:plot_11}}
\hfil
\subfloat[Mental effort]{\includegraphics*[width=0.6\columnwidth]{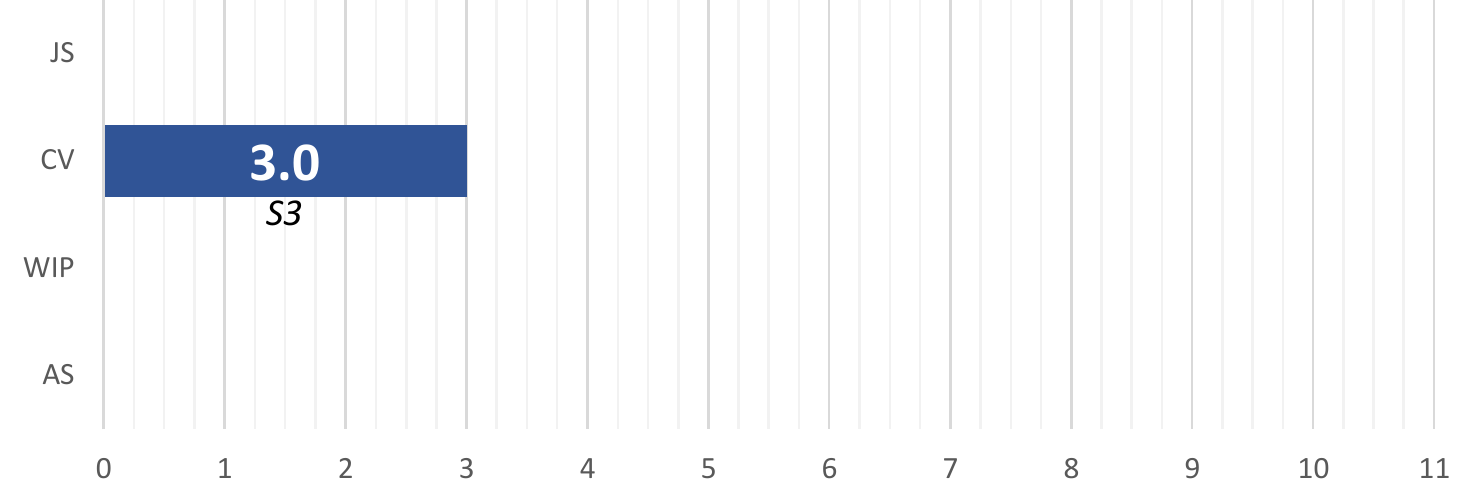}%
\label{fig:plot_6}}
\hfil
\subfloat[Perceived physical effort]{\includegraphics*[width=0.6\columnwidth]{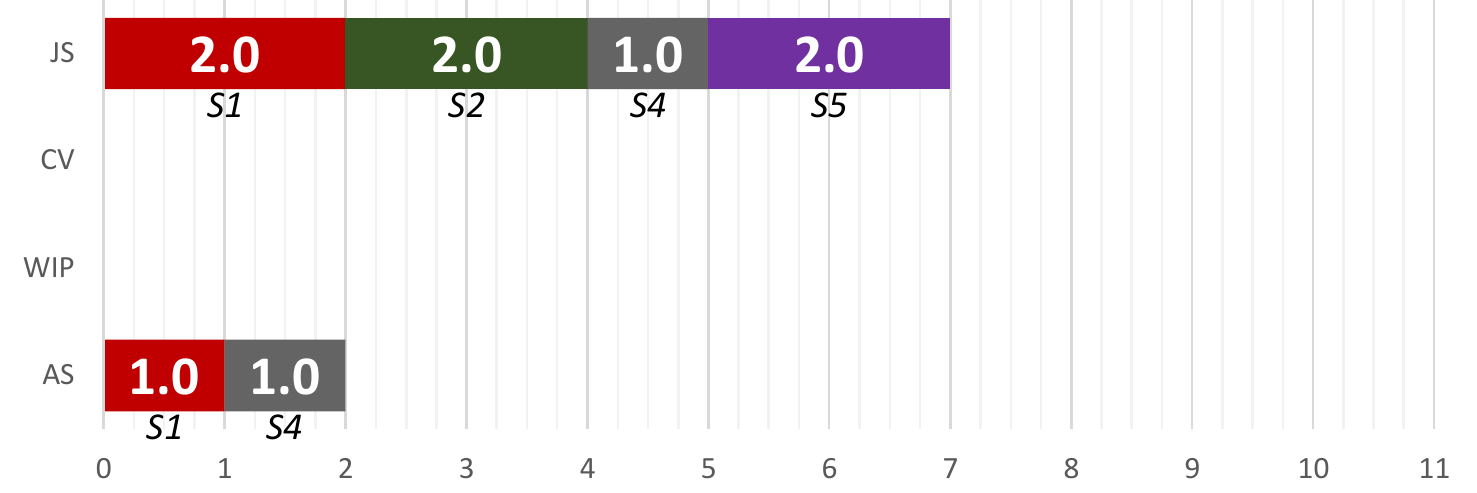}%
\label{fig:plot_9}}
\hfil
\subfloat[Naturalness]{\includegraphics*[width=0.6\columnwidth]{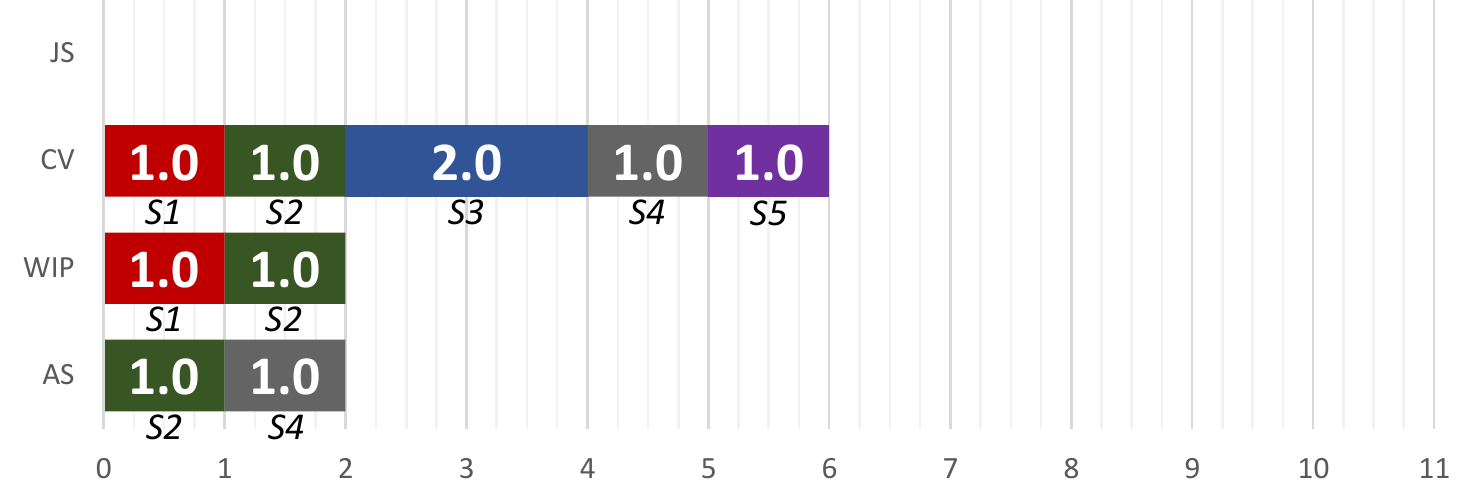}%
\label{fig:plot_5}}
\hfil
\subfloat[V/R Phys. str. similarity]{\includegraphics*[width=0.6\columnwidth]{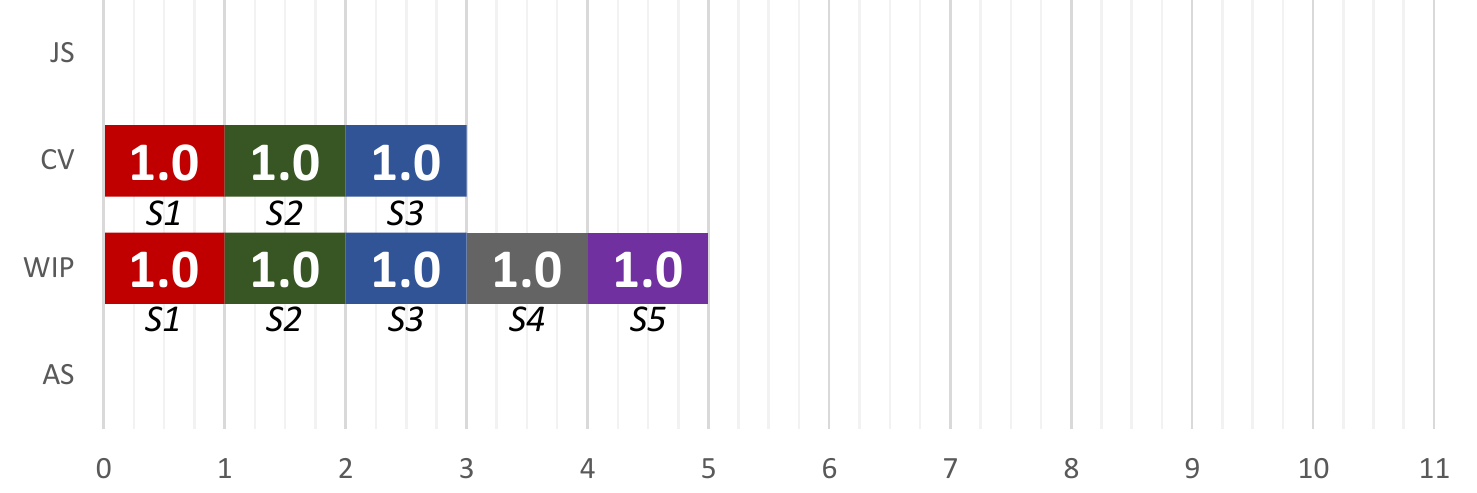}%
\label{fig:plot_8}}
\hfil
\subfloat[Self-motion compellingness]{\includegraphics*[width=0.6\columnwidth]{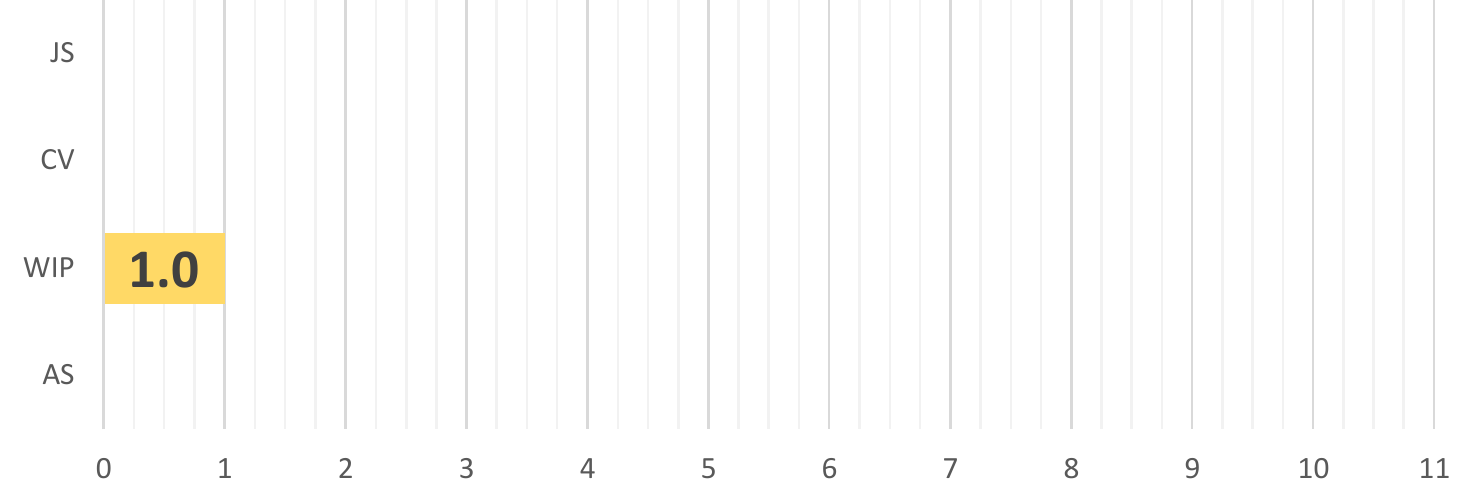}%
\label{fig:plot_16}}
\hfil
\subfloat[Acclimatisation]{\includegraphics*[width=0.6\columnwidth]{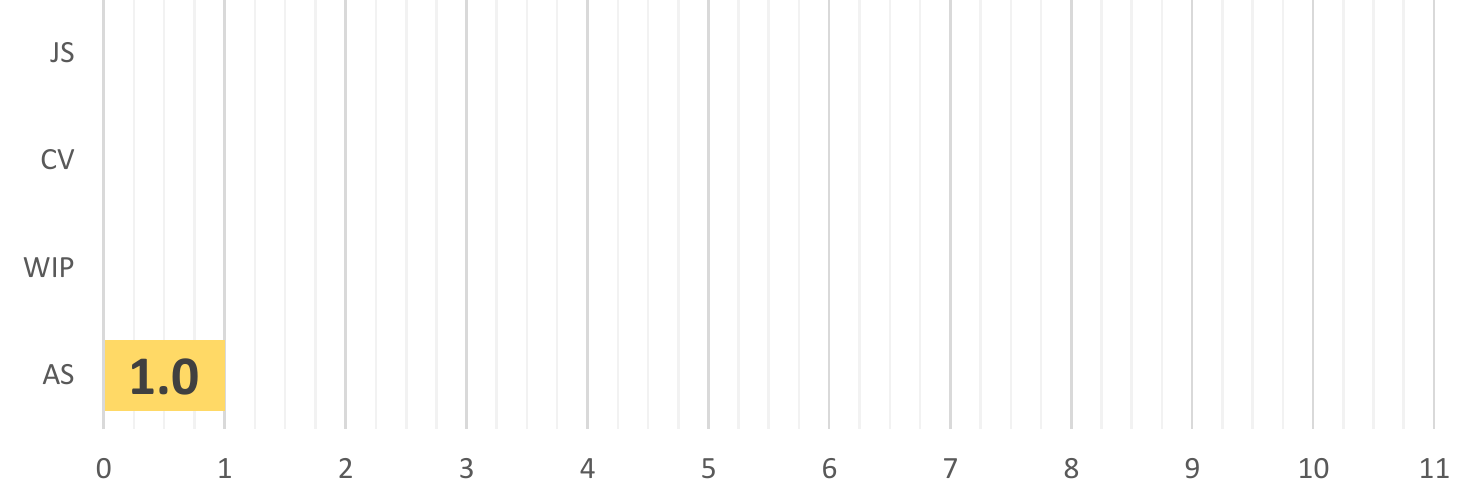}%
\label{fig:plot_17}}
\hfil
\subfloat[Control]{\includegraphics*[width=0.6\columnwidth]{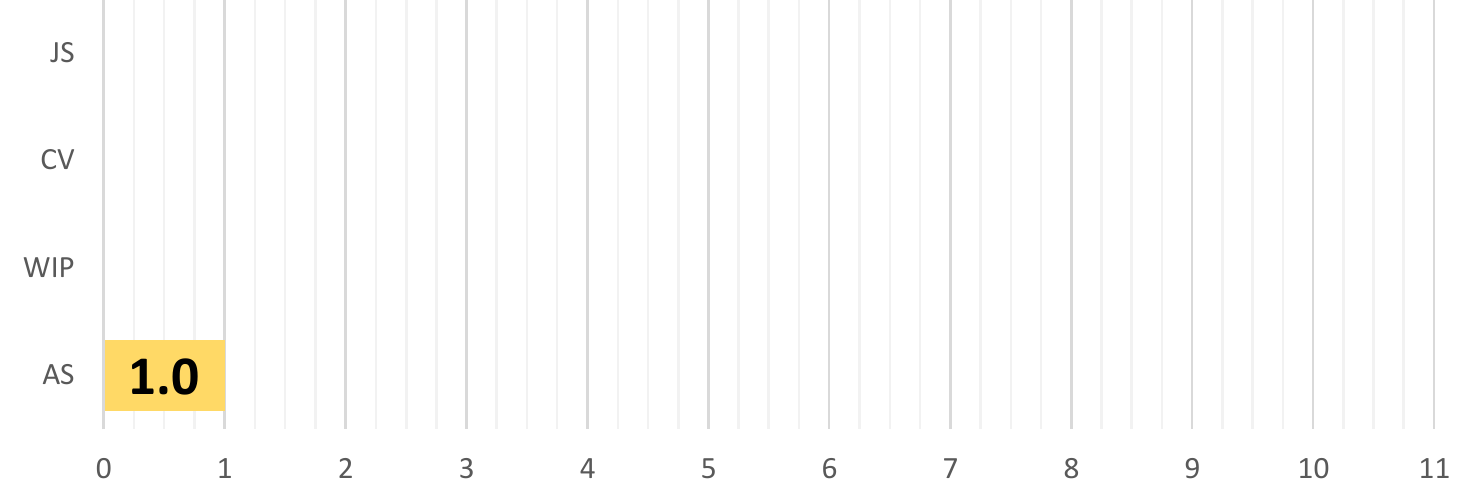}%
\label{fig:plot_15}}
\hfil
\subfloat[Presence]{\includegraphics*[width=0.6\columnwidth]{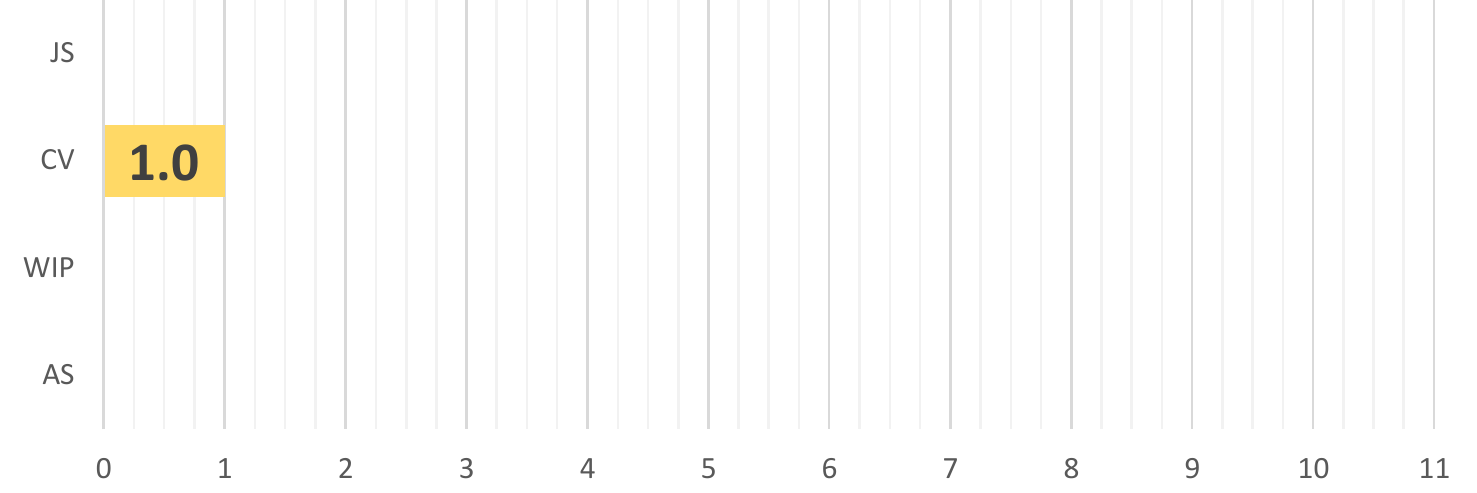}%
\label{fig:plot_18}}
\hfil
\subfloat[Learnability]{\includegraphics*[width=0.6\columnwidth]{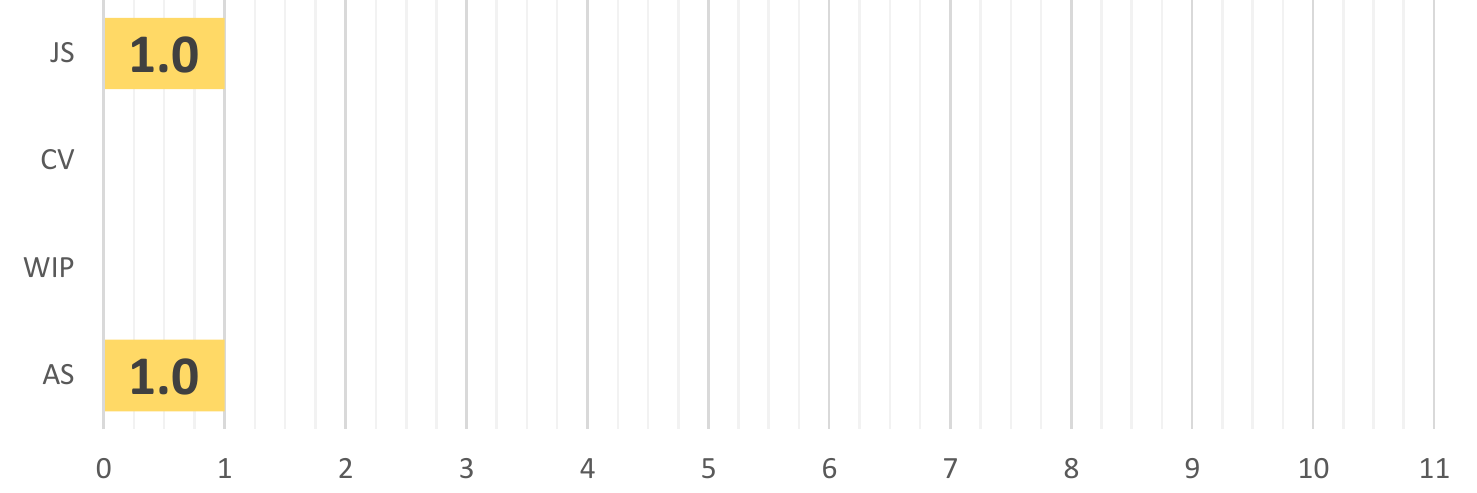}%
\label{fig:plot_19}}
\hfil
\subfloat[Intuitiveness]{\includegraphics*[width=0.6\columnwidth]{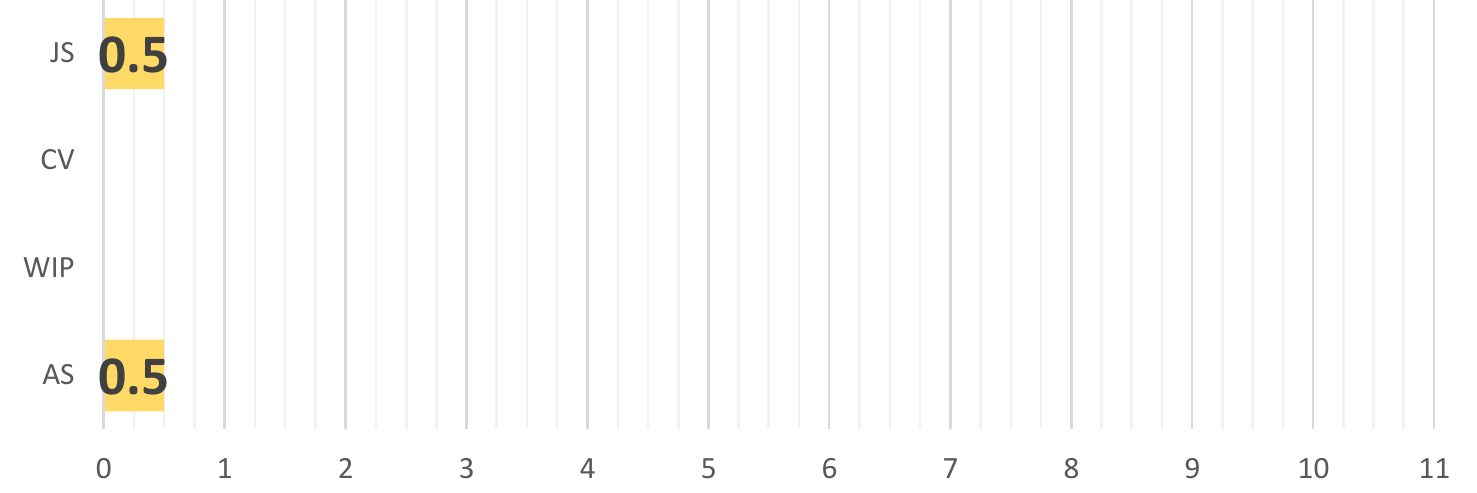}%
\label{fig:plot_20}}
\hfil
\subfloat[Overall system usability]{\includegraphics*[width=0.6\columnwidth]{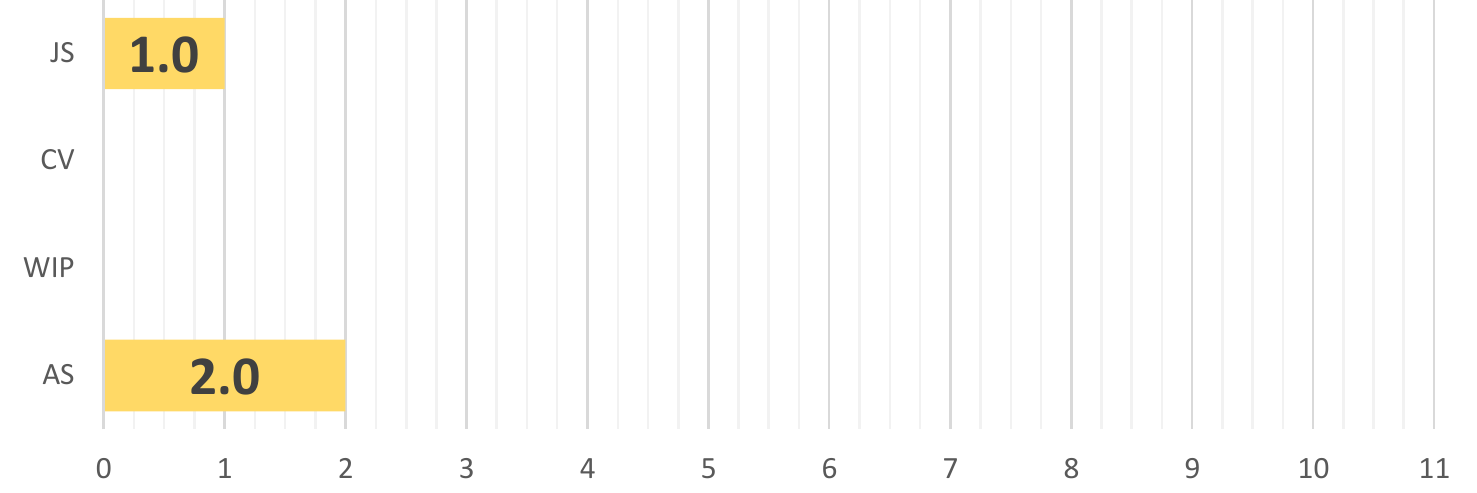}%
\label{fig:plot_23}}
\hfil
\subfloat[Nausea (SSQ)]{\includegraphics*[width=0.6\columnwidth]{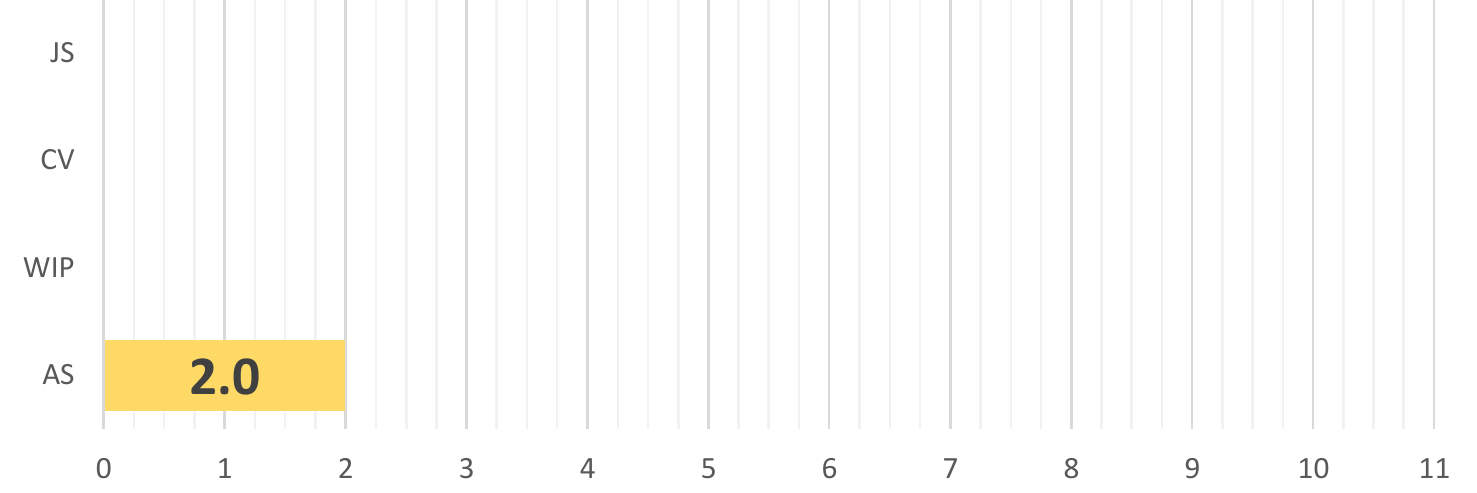}%
\label{fig:plot_24}}
\hfil
\subfloat[Oculomotor (SSQ)]{\includegraphics*[width=0.6\columnwidth]{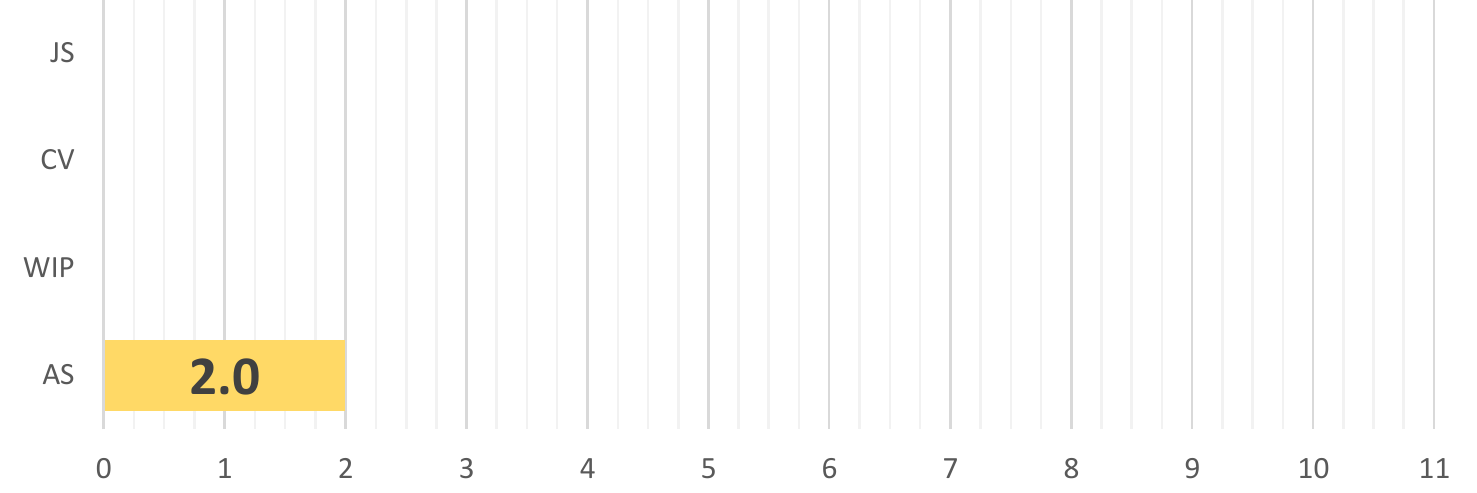}%
\label{fig:plot_25}}
\hfil
\subfloat[Disorientation (SSQ)]{\includegraphics*[width=0.6\columnwidth]{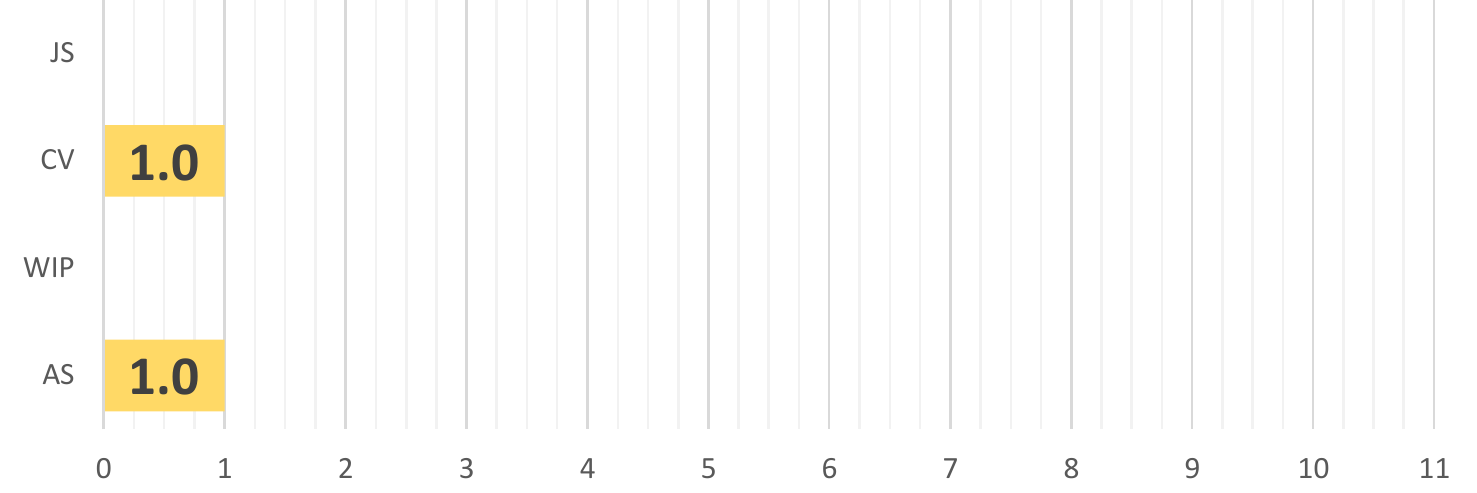}%
\label{fig:plot_26}}
\hfil
\subfloat[Total (SSQ)]{\includegraphics*[width=0.6\columnwidth]{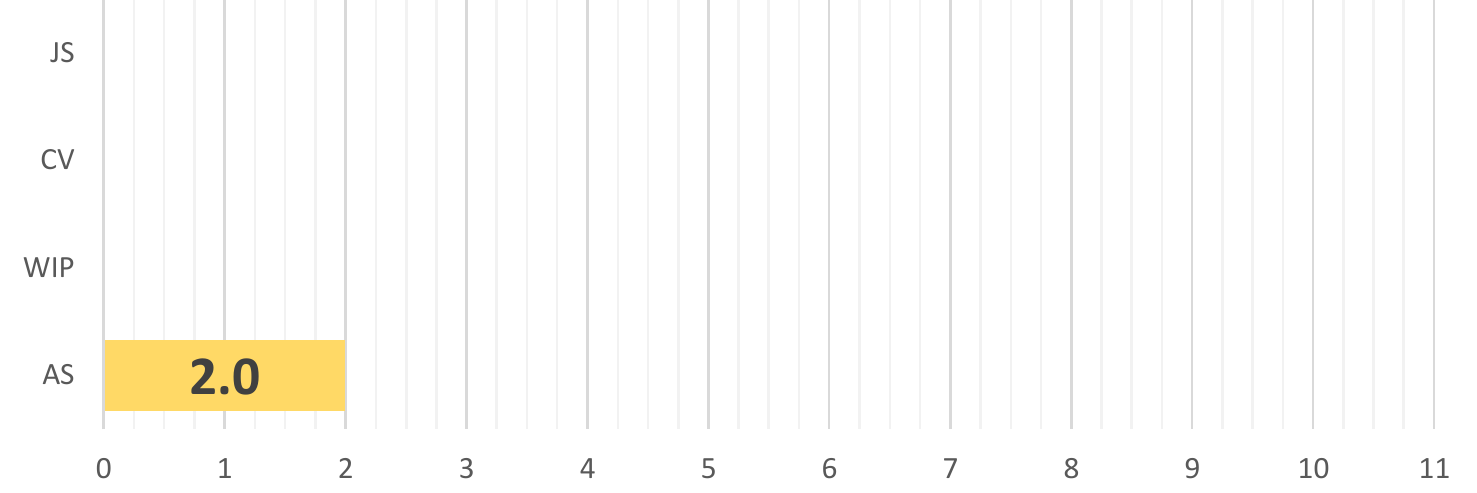}%
\label{fig:plot_27}}
\hfil
\subfloat[]{\includegraphics*[width=1.2\columnwidth]{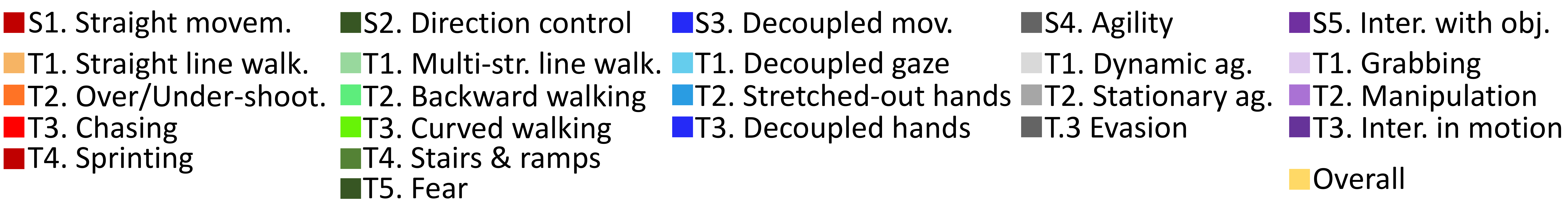}%
\label{fig:plot_0}}

\caption{Scores obtained by the four techniques for all the metrics (weights set to 1).
The same scale (0--11) was used in order to make the scores visually comparable. Values correspond to points obtained in pairwise comparisons. Missing values correspond to situations in which the given technique was never statistically better than any of the other techniques.The color coding is the same adopted in Fig.~\ref{tab:TabTasks} and Table~\ref{tab:Tab1}.}
\label{fig:plots}
\end{figure*}

Fig.~\ref{fig:plots} plots the contribution of individual metrics for each technique with all weights set to 1 (default). Different colors are used to indicate the task/scenario in which points were obtained, i.e., where differences were statistically significant and, on average, metric values were better than those of the other technique in pairwise comparisons. 
Due to the different granularity of the metrics, plots are organized as follow: Fig.~\ref{fig:plot_1}--\ref{fig:plot_3} report per-task scores, Fig.~\ref{fig:plot_14}--\ref{fig:plot_8} per-scenario scores, and Fig.~\ref{fig:plot_16}--\ref{fig:plot_27}  overall scores. With these weights, i.e., assuming that requirements have all the same importance for the application the techniques would be used into, the following ranking is obtained: JS (54.5pts), AS (53.0pts),  CV (23.6pts), and WIP (17.9pts).

It is worth remarking that both per-metric and overall scores shall not be interpreted in absolute terms: in fact, they indicate to what extent a given technique matches the weighted set of requirements compared to the other techniques included in the current evaluation. As said, by varying the subset of techniques being compared or the weights assigned to requirements, scores and ranking could change.  Moreover, it is also worth observing that, for the sake of clarity, in Fig~\ref{fig:plots} the results of the pairwise comparisons had to be omitted: thus, it is not always possible to determine which was the technique that scored better in a specific comparison. For instance, considering Fig.~\ref{fig:plot_2} it is possible to observe that, overall, JS was statistically faster than other techniques in a larger number of tasks compared to the counterparts; however, from the scores it is not possible to determine, e.g., whether JS was better than another specific technique in a given task. When the number of points for a metric in a given scenario or task (depending on the metric type) is equal to the number of considered techniques minus one, then the technique that scored significantly better than any other one can be identified: this is the case of CV for what it concerns mental effort in scenario \textit{S3} (Fig.~\ref{fig:Fig1_k}). 
To avoid ambiguities, average values and statistical significance are reported in the Appendix at \href{http://tiny.cc/quxlsz}{\textit{http://tiny.cc/quxlsz}} as well as in the provided RDB and WDB (spreadsheet above).

\subsection{Assigning Weights and Making the Selection}
\label{sec:pdb_application2}
In order to show how to use the generated WDB for selecting the technique maximizing the weighted requirements for a given scenario, a VR game named \textit{VIVECraft} \cite{vivecraft} was considered. This is a modified version of the popular game \textit{Minecraft} (a survival game with ``pixelated'' graphics) with support for VR. Reasons for choosing it are manifold: it is conveniently available, it is quite complex from the point of view of stressed FRs and NFRS and, more importantly, it natively supports more than one locomotion technique; hence, it would be reasonable to study which of the currently supported or of the existing techniques score better.

In order to set the FR weights, a testbed user should first consider that, in the game, it is crucial to move along lines and stop precisely at given points, e.g., to safely cross bridges, etc. (hence, it could set both $w_{S1.T1}$ and $w_{S1.T2}$ to 1). Control on direction change is also important, though not as much as previous requirements ($w_{S2.T1} = 0.5$). Properly managing direction on curved paths is a marginal plus ($w_{S2.T3} = 0.1$). On the contrary, backward walking is a great feature to have, especially during combat phases ($w_{S2.T2} = 1$), as much as being able to run ($w_{S1.T4} = 0.75$). The ability to react to dangerous situations like enemies appearance and to effectively engage in melee fighting ($w_{S2.T5} = 0.5$, $w_{S4.T1} = 0.5$), as well as to move nimbly and dodge attacks ($w_{S4.T2} = 1$, $w_{S4.T3} = 1$) are particularly relevant. Due to the specific implementation, decoupling gaze from movement is not mandatory, but decoupling hands can have a positive impact ($w_{S3.T1} = 0$, $w_{S3.T3} = 0.25$). Interaction is required for crafting and throwing objects, as well as for managing menus ($w_{S5.T1} = 0.5$, $w_{S5.T2} = 0.5$). Interaction in motion is a clear advantage, though not strictly mandatory ($w_{S5.T3} = 0.75$). Due the specific implementation of up/down movements which are based on stairs, corresponding weights could be set accordingly ($w_{S2.T4} = 1$, $w_{ST} = 1$, $w_{RA} = 0$). 
Concerning NFRs, based on the above discussion Input Responsiveness is very important ($w=1$), whereas Sensitivity is somehow less relevant ($w=0.25$). Operation speed ($w=0.5$) is less crucial than Accuracy and Error-proneness ($w=1$ and $w=0.75$, respectively). Since \textit{VIVECraft} is a playful application, it is preferable to have low fatigue so that users can engage in longer sessions ($w = 1$ and direction set to negative for Physical effort). Moreover, Enjoyability, Naturaless, Comfort, Self-motion compellingness, Presence are all considered equally important ($w=1$). Similar considerations apply to motion sickness and discomfort ($w=1$ for both Motion sickness total score and SUD). The weights of the Ease of use, Mental effort, Appropriateness, Acclimatisation, Learnability, Intuitiveness and Overall system usability were set to $0.5$, being all of them rather important for an application not targeted to expert VR users. Remaining requirements were set to 0.

With this arbitrary configuration, the generated WDB would produce the following ranking: JS (28.4pts), AS (25.6pts), CV (5.68pts), and WIP (4.4pts). For sake of completeness, it is worth noticing that actual implementations of the JS and AS techniques in the game slightly differ from those considered in the WDB. Hence, a new WDB shall be created with those implementations, or matching implementations must be  introduced in the game. Integrations to the WDB would be requested also to include in the comparison, e.g., AS-seated in case AS-standing is already available. As anticipated, changing the number of techniques in the comparison, scores (and ranking) could vary. For example, if CV is left out from the comparison, ranking changes as follows: AS (18.8pts), JS (18.1pts), WIP (3.7pts). This variation indicates that JS was ranked first due to points earned in pairwise comparisons with CV. 

\section{Conclusions and Future Work}
In this paper, a testbed supporting the comparison of techniques for locomotion in large-scale virtual environments from many different perspectives is proposed.  

Compared to works like, e.g., \cite{pai2017armswing, wilson2016vr, calandra2018eg, nilsson2013tapping}, which present the results of ad-hoc evaluations, the devised testbed integrates and complements a heterogeneous collection of analysis methods and tools reported in the literature by proposing a comprehensive set of both objective and subjective locomotion-related metrics to be evaluated in the execution of a variety of representative tasks. Moreover, differently than in previous works where other evaluation approaches were defined (e.g., \cite{bowman1997travel, whitton2005comparing, ferracani2016locomotion, albert2018user}), in this paper the methodology underlying the proposed testbed is accompanied by a scoring system that allows potential users to identify the locomotion techniques that best fit a set of user-weighted requirements. A use case showing how to leverage the testbed for choosing, among four known alternatives, the best techniques given a well-defined application scenario is also illustrated. The procedure followed for collecting and processing experimental data is discussed in detail in order to foster the reproducibility of results. 

The testbed is released as open source, and comes with all the results obtained applying the devised methodology. This way, comparisons could be performed among the four techniques considered so far by setting weights that match the requirements of a particular application; new experiments could be performed with the same techniques to improve statistical significance of collected measurements; alternative ways for testing significance could be also exploited; finally, new techniques could be integrated, and new tasks (with corresponding metrics) could be added as well in order to investigate other FRs (NFRs).

A present limitation of the scoring system lays in the method used to account for the contribution of the various metrics and solve the multi-criteria decision-making problem. In fact, with the selected method, adding and removing a technique to/from the analysis could lead to a different ranking. Moreover, the approach adopted to normalize the above contributions does not account for the actual magnitude of the differences between the (mean) values of a given metric for the various techniques, but it just considers statistical significance: that is, it values the fact that a technique is better than another one, but it does not consider how much it is better (for a given metric). Finally, overall results depend on the number of statistical differences found; thus, even though one may decide to simply discard metrics for which a significance is found only for a few techniques, it is indeed necessary to work towards extending the availability of experimental data in order to improve the robustness of obtained scores.       

In the future, it is planned to address the above issues by defining alternative approaches to normalization (using different significance thresholds, extending the interval of points that can be assigned, etc.) and to scoring, in general; to this aim, the possibility to collect also data about real walking and to use them in the normalization step will be investigated. Further experiments with these and other techniques will be performed, since more data are needed to extend the representativeness and significance of the publicly provided dataset. Moreover, although the testbed already supports a fair number of FRs, there are still some others missing (e.g., regarding movement in the third-dimension) which could be added. Action-specific tasks could be included as well, to better focus on interactions that are typical of specific applications (e.g., first-person shooter games). Other scenarios supporting the study of jumping, swimming, climbing and even flying configurations could be considered too. Similar considerations apply to NFRs, e.g., pertaining psychological and/or cognitive aspects, such as spatial awareness, sense of direction, navigation abilities, etc. To this purpose, a study aimed to identify possibly missing requirements and to characterize the testbed discrimination capabilities could be performed.


\ifCLASSOPTIONcompsoc
  \section*{Acknowledgments}
\else
  \section*{Acknowledgment}
\fi
This work has been supported by VR@POLITO initiative.

\ifCLASSOPTIONcaptionsoff
  \newpage
\fi

\normalem
\bibliographystyle{IEEEtran}

\bibliography{paper}

\begin{thebibliography}{10}
\providecommand{\url}[1]{#1}
\csname url@samestyle\endcsname
\providecommand{\newblock}{\relax}
\providecommand{\bibinfo}[2]{#2}
\providecommand{\BIBentrySTDinterwordspacing}{\spaceskip=0pt\relax}
\providecommand{\BIBentryALTinterwordstretchfactor}{4}
\providecommand{\BIBentryALTinterwordspacing}{\spaceskip=\fontdimen2\font plus
\BIBentryALTinterwordstretchfactor\fontdimen3\font minus
  \fontdimen4\font\relax}
\providecommand{\BIBforeignlanguage}[2]{{%
\expandafter\ifx\csname l@#1\endcsname\relax
\typeout{** WARNING: IEEEtran.bst: No hyphenation pattern has been}%
\typeout{** loaded for the language `#1'. Using the pattern for}%
\typeout{** the default language instead.}%
\else
\language=\csname l@#1\endcsname
\fi
#2}}
\providecommand{\BIBdecl}{\relax}
\BIBdecl

\bibitem{bowman20043d}
D.~Bowman, E.~Kruijff, J.~J. LaViola~Jr, and I.~P. Poupyrev, \emph{{3D} User
  interfaces: {T}heory and practice}, 2004.

\bibitem{Nilsson:2018:NWV:3181320.3180658}
N.~C. Nilsson, S.~Serafin, F.~Steinicke, and R.~Nordahl, ``Natural walking in
  virtual reality: A review,'' \emph{Computers in Entert.}, vol.~16, pp.
  8:1--8:22, 2018.

\bibitem{ruddle2006efficient}
R.~A. Ruddle and S.~Lessels, ``For efficient navigational search, humans
  require full physical movement, but not a rich visual scene,''
  \emph{Psychological Science}, vol.~17, pp. 460--465, 2006.

\bibitem{waller2013sensory}
D.~Waller and E.~Hodgson, ``Sensory contributions to spatial knowledge of real
  and virtual environments,'' in \emph{Human Walking in Virtual Environments},
  2013, pp. 3--26.

\bibitem{suma2007comparison}
E.~A. Suma, S.~Babu, and L.~F. Hodges, ``Comparison of travel techniques in a
  complex, multi-level {3D} environment,'' in \emph{Proc. IEEE Symp. on 3D UI},
  2007, pp. 147--153.

\bibitem{garg2017ares}
A.~Garg, J.~A. Fisher, W.~Wang, and K.~P. Singh, ``{ARES}: An application of
  impossible spaces for natural locomotion in {VR},'' in \emph{Proc. CHI
  Conf.}, 2017, pp. 218--221.

\bibitem{nilsson2013perceived}
N.~C. Nilsson, S.~Serafin, and R.~Nordahl, ``The perceived naturalness of
  virtual locomotion methods devoid of explicit leg movements,'' in \emph{Proc.
  Motion in Games}, 2013, pp. 155--164.

\bibitem{bowman1997travel}
D.~A. Bowman, D.~Koller, and L.~F. Hodges, ``Travel in immersive virtual
  environments: An evaluation of viewpoint motion control techniques,'' in
  \emph{Proc. IEEE Virtual Reality}, 1997, pp. 45--52.

\bibitem{whitton2005comparing}
M.~C. Whitton, J.~V. Cohn, J.~Feasel, P.~Zimmons, S.~Razzaque, S.~J. Poulton,
  B.~McLeod, and F.~P. Brooks, ``Comparing {VE} locomotion interfaces,'' in
  \emph{Proc. IEEE Virtual Reality}, 2005, pp. 123--130.

\bibitem{schuemie2005effect}
M.~Schuemie, B.~Abel, C.~van~der Mast, M.~Krijn, and P.~Emmelkamp, ``The effect
  of locomotion technique on presence, fear and usability in a virtual
  environment,'' in \emph{Proc. Europ. Conf. on Media, Communication \& Film},
  2005, pp. 129--135.

\bibitem{testbed}
D.~A. Bowman, D.~B. Johnson, and L.~F. Hodges, ``Testbed evaluation of virtual
  environment interaction techniques,'' \emph{Presence: Teleoperators \&
  Virtual Environments}, vol.~10, pp. 75--95, 2001.

\bibitem{boletsis2017new}
C.~Boletsis, ``The new era of virtual reality locomotion: a systematic
  literature review of techniques and a proposed typology,'' \emph{Multimodal
  Technologies and Interaction}, vol.~1, p.~24, 2017.

\bibitem{AlZayer2018Virtual}
M.~Al~Zayer, P.~MacNeilage, and e.~folmer, ``Virtual locomotion: {A} survey,''
  \emph{IEEE Trans. on Visualization \& Computer Graphics}, 2018.

\bibitem{anthes2016state}
C.~Anthes, R.~J. Garc{\'\i}a-Hern{\'a}ndez, M.~Wiedemann, and
  D.~Kranzlm{\"u}ller, ``State of the art of virtual reality technology,'' in
  \emph{2016 IEEE Aerospace Conf.}, 2016, pp. 1--19.

\bibitem{cardoso2019survey}
J.~C. Cardoso and A.~Perrotta, ``A survey of real locomotion techniques for
  immersive virtual reality applications on head-mounted displays,''
  \emph{Computers \& Graphics}, vol.~85, pp. 55--73, 2019.

\bibitem{lathrop2002perceived}
W.~B. Lathrop and M.~K. Kaiser, ``Perceived orientation in physical and virtual
  environments: Changes in perceived orientation as a function of idiothetic
  information available,'' \emph{Presence: Teleoperators \& Virtual
  Environments}, vol.~11, pp. 19--32, 2002.

\bibitem{templeman1999virtual}
J.~N. Templeman, P.~S. Denbrook, and L.~E. Sibert, ``Virtual locomotion:
  Walking in place through virtual environments,'' \emph{Presence}, vol.~8, pp.
  598--617, 1999.

\bibitem{bozgeyikli2016point}
E.~Bozgeyikli, A.~Raij, S.~Katkoori, and R.~Dubey, ``Point \& teleport
  locomotion technique for virtual reality,'' in \emph{Proc. Symp. on
  Computer-Human Interaction in Play}, 2016, pp. 205--216.

\bibitem{stoakley1995virtual}
R.~Stoakley, M.~J. Conway, and R.~Pausch, ``Virtual reality on a {WIM}:
  Interactive worlds in miniature,'' in \emph{Proc. SIGCHI Conf. on Human
  factors in Computing Systems}, 1995, pp. 265--272.

\bibitem{stoev2001two}
S.~L. Stoev, D.~Schmalstieg, and W.~Stra{\ss}er, ``Two-handed
  through-the-lens-techniques for navigation in virtual environments,'' in
  \emph{Immersive Projection Tech. and Virtual Env.}, 2001, pp. 51--60.

\bibitem{fiore2013towards}
L.~P. Fiore, E.~Coben, S.~Merritt, P.~Liu, and V.~Interrante, ``Towards
  enabling more effective locomotion in {VR} using a wheelchair-based motion
  platform,'' in \emph{Proc. 5th Joint VR Conf.}, 2013, pp. 83--90.

\bibitem{wang2012comparing}
J.~Wang and R.~W. Lindeman, ``Comparing isometric and elastic surfboard
  interfaces for leaning--based travel in {3D} virtual environments,'' in
  \emph{IEEE Symp. on 3D {UI}}, 2012, pp. 31--38.

\bibitem{beckhaus2007chairio}
S.~Beckhaus, K.~J. Blom, and M.~Haringer, ``Chair{IO}--the chair-based
  interface,'' \emph{Concepts and Tech. for Pervasive Games: A Reader for
  Pervasive Gaming Research}, vol.~1, pp. 231--264, 2007.

\bibitem{fels2005swimming}
S.~Fels, Y.~Kinoshita, T.-P.~G. Chen, Y.~Takama, S.~Yohanan, A.~Gadd,
  S.~Takahashi, and K.~Funahashi, ``Swimming across the pacific: {A} {VR}
  swimming interface,'' \emph{IEEE Computer Graphics \& Applications}, vol.~25,
  pp. 24--31, 2005.

\bibitem{souman2011cyberwalk}
J.~L. Souman, P.~R. Giordano, M.~Schwaiger, I.~Frissen, T.~Th{\"u}mmel,
  H.~Ulbrich, A.~D. Luca, H.~H. B{\"u}lthoff, and M.~O. Ernst, ``Cyberwalk:
  Enabling unconstrained omnidirectional walking through virtual
  environments,'' \emph{ACM TAP}, vol.~8, pp. 25:1--25:22, 2011.

\bibitem{iwata2005circulafloor}
H.~Iwata, H.~Yano, H.~Fukushima, and H.~Noma, ``Circulafloor [locomotion
  interface],'' \emph{IEEE Computer Graphics \& Applications}, vol.~25, pp.
  64--67, 2005.

\bibitem{medina2008virtusphere}
E.~Medina, R.~Fruland, and S.~Weghorst, ``Virtusphere: Walking in a human size
  {VR} `hamster ball','' in \emph{Proc. Human Factors and Ergonomics Society
  Annual Meeting}, 2008, pp. 2102--2106.

\bibitem{avila2014virtual}
L.~Avila and M.~Bailey, ``Virtual reality for the masses,'' \emph{IEEE Computer
  Graphics \& Applications}, vol.~34, pp. 103--104, 2014.

\bibitem{swapp2010implementation}
D.~Swapp, J.~Williams, and A.~Steed, ``The implementation of a novel walking
  interface within an immersive display,'' in \emph{Proc. IEEE Symp. on 3D
  {UI}}, 2010, pp. 71--74.

\bibitem{iwata1996virtual}
H.~Iwata and T.~Fujii, ``Virtual perambulator: A novel interface device for
  locomotion in virtual environment,'' in \emph{Proc. IEEE Virtual Reality Ann.
  Int. Symposium}, 1996, pp. 60--65.

\bibitem{cakmak2014cyberith}
T.~Cakmak and H.~Hager, ``Cyberith virtualizer: A locomotion device for virtual
  reality,'' in \emph{Proc. ACM SIGGRAPH}, 2014, p.~6.

\bibitem{guy2015lazynav}
E.~Guy, P.~Punpongsanon, D.~Iwai, K.~Sato, and T.~Boubekeur, ``Lazy{N}av: 3{D}
  ground navigation with non--critical body parts,'' in \emph{IEEE Symp. on 3D
  {UI}}, 2015, pp. 43--50.

\bibitem{kitson2017comparing}
A.~Kitson, A.~M. Hashemian, E.~R. Stepanova, E.~Kruijff, and B.~E. Riecke,
  ``Comparing leaning--based motion cueing interfaces for virtual reality
  locomotion,'' in \emph{IEEE Symp. {3D UI}}, 2017, pp. 73--82.

\bibitem{kruijff2016your}
E.~Kruijff, A.~Marquardt, C.~Trepkowski, R.~W. Lindeman, A.~Hinkenjann,
  J.~Maiero, and B.~E. Riecke, ``On your feet!: Enhancing vection in
  leaning-based interfaces through multisensory stimuli,'' in \emph{Proc. Symp.
  on Spatial User Interact.}, 2016, pp. 149--158.

\bibitem{de2008using}
G.~de~Haan, E.~J. Griffith, and F.~H. Post, ``Using the wii balance board™ as
  a low-cost vr interaction device,'' in \emph{Proc. ACM Symp. on Virtual
  Reality Software and Technology}, 2008, pp. 289--290.

\bibitem{harris2014human}
A.~Harris, K.~Nguyen, P.~T. Wilson, M.~Jackoski, and B.~Williams, ``Human
  joystick: Wii-leaning to translate in large virtual environments,'' in
  \emph{Proc. ACM SIGGRAPH Int. Conf. on Virtual Reality Continuum and its
  Applications in Industry}, 2014, pp. 231--234.

\bibitem{feasel2008llcm}
J.~Feasel, M.~C. Whitton, and J.~D. Wendt, ``{LLCM-WIP}: Low-latency,
  continuous-motion walking-in-place,'' in \emph{Proc. IEEE Symp. on 3D UI},
  2008, pp. 97--104.

\bibitem{pai2017armswing}
Y.~S. Pai and K.~Kunze, ``Armswing: Using arm swings for accessible and
  immersive navigation in {AR/VR} spaces,'' in \emph{Proc. Int. Conf. on Mobile
  and Ubiquitous Multimedia}, 2017, pp. 189--198.

\bibitem{suma2012impossible}
E.~A. Suma, Z.~Lipps, S.~Finkelstein, D.~M. Krum, and M.~Bolas, ``Impossible
  spaces: Maximizing natural walking in virtual environments with
  self-overlapping architecture,'' \emph{IEEE Trans. on Visualization \&
  Computer Graphics}, vol.~18, no.~4, pp. 555--564, 2012.

\bibitem{wilson2016vr}
P.~T. Wilson, W.~Kalescky, A.~MacLaughlin, and B.~Williams, ``{VR} locomotion:
  walking$>$ walking in place$>$ arm swinging,'' in \emph{Proc. 15th ACM
  SIGGRAPH Conf. on VR Continuum and Its Applications in Industry}, 2016, pp.
  243--249.

\bibitem{calandra2018eg}
D.~Calandra, M.~Billi, F.~Lamberti, A.~Sanna, and R.~Borchiellini, ``Arm
  swinging vs treadmill: A comparison between two techniques for locomotion in
  virtual reality,'' in \emph{Proc. Eurographics}, 2018, pp. 53--56.

\bibitem{calandra2019icce}
D.~Calandra, F.~Lamberti, and M.~Migliorini, ``On the usability of consumer
  locomotion techniques in serious games: Comparing arm swinging, treadmills
  and walk-in-place,'' in \emph{IEEE Int. Conf. on Consumer Electronics -
  Berlin}, 2019, pp. 1--5.

\bibitem{nilsson2013tapping}
N.~C. Nilsson, S.~Serafin, M.~H. Laursen, K.~S. Pedersen, E.~Sikstrom, and
  R.~Nordahl, ``Tapping-in-place: Increasing the naturalness of immersive
  walking-in-place locomotion through novel gestural input,'' in \emph{Proc.
  IEEE Symp. on 3D UI}, 2013, pp. 31--38.

\bibitem{sarupuri2018lute}
B.~Sarupuri, S.~Hoermann, M.~C. Whitton, and R.~W. Lindeman, ``Lute: A
  locomotion usability test environment for virtual reality,'' in \emph{Int.
  Conf. on Virtual Worlds and Games for Serious Applications}, 2018, pp. 1--4.

\bibitem{albert2018user}
J.~Albert and K.~Sung, ``User-centric classification of virtual reality
  locomotion,'' in \emph{Proc. ACM Symp. VR Soft. and Tech.}, 2018, p. 127.

\bibitem{nasatlx}
S.~G. Hart and L.~E. Staveland, ``Development of nasa-tlx (task load index):
  Results of empirical and theoretical research,'' in \emph{Human Mental
  Workload}, 1988, vol.~52, pp. 139 -- 183.

\bibitem{ferracani2016locomotion}
A.~Ferracani, D.~Pezzatini, J.~Bianchini, G.~Biscini, and A.~Del~Bimbo,
  ``Locomotion by natural gestures for immersive virtual environments,'' in
  \emph{Proc. 1st Int. Workshop on Multimedia Alt. Realities}, 2016, pp.
  21--24.

\bibitem{wilson2018object}
G.~Wilson, M.~McGill, M.~Jamieson, J.~R. Williamson, and S.~A. Brewster,
  ``Object manipulation in virtual reality under increasing levels of
  translational gain,'' in \emph{Proc. CHI Conf. on Human Factors in Computing
  Systems}, 2018, p.~99.

\bibitem{mayor2019comparative}
J.~Mayor, L.~Raya, and A.~Sanchez, ``A comparative study of virtual reality
  methods of interaction and locomotion based on presence, cybersickness and
  usability,'' \emph{IEEE Trans. on Emerging Topics in Computing}, 2019.

\bibitem{lapointe2009comparative}
J.-F. Lapointe and P.~Savard, ``A comparative study of three bimanual travel
  techniques for desktop virtual walkthroughs,'' in \emph{Proc. IEEE Work. on
  Haptic Audio Visual Env.\& Games}, 2009, pp. 182--185.

\bibitem{loup2018effects}
G.~Loup and E.~Loup-Escande, ``Effects of travel modes on performances and user
  comfort: A comparison between {ArmSwinger} and {Teleporting},'' \emph{Int.
  Jr. of Human-Computer Inter.}, pp. 1--9, 2018.

\bibitem{suma2010evaluation}
E.~Suma, S.~Finkelstein, M.~Reid, S.~Babu, A.~Ulinski, and L.~F. Hodges,
  ``Evaluation of the cognitive effects of travel technique in complex real and
  virtual environments,'' \emph{IEEE Trans. on Visualization \& Computer
  Graphics}, vol.~16, pp. 690--702, 2010.

\bibitem{nabiyouni2015comparing}
M.~Nabiyouni, A.~Saktheeswaran, D.~A. Bowman, and A.~Karanth, ``Comparing the
  performance of natural, semi-natural, and non-natural locomotion techniques
  in virtual reality,'' in \emph{Proc. IEEE Symp. on {3D UI}}, 2015, pp. 3--10.

\bibitem{UnbreakableVRRunner}
\BIBentryALTinterwordspacing
M.~Asnabrygg, ``Unbreakable {VR} runner,'' 2016. [Online]. Available:
  \url{https://store.steampowered.com/app/494310/Unbreakable_Vr_Runner}
\BIBentrySTDinterwordspacing

\bibitem{paris2017acquisition}
R.~Paris, M.~Joshi, Q.~He, G.~Narasimham, T.~P. McNamara, and B.~Bodenheimer,
  ``Acquisition of survey knowledge using walking in place and resetting
  methods in immersive virtual environments,'' in \emph{Proc. ACM Symposium on
  Applied Perception}, 2017, p.~7.

\bibitem{Fallout4VR}
\BIBentryALTinterwordspacing
{Bethesda Game Studios}, ``Fallout 4 {VR},'' 2017. [Online]. Available:
  \url{https://store.steampowered.com/app/611660/Fallout_4_VR}
\BIBentrySTDinterwordspacing

\bibitem{EchoCombat}
\BIBentryALTinterwordspacing
{Echo Combat}. [Online]. Available:
  \url{https://www.oculus.com/experiences/event/1713610428731487/}
\BIBentrySTDinterwordspacing

\bibitem{CatchandRelease}
\BIBentryALTinterwordspacing
{Metricminds {G}mbH and {C}o {KG}}, ``Catch and {R}elease,'' 2018. [Online].
  Available: \url{https://store.steampowered.com/app/679750/Catch__Release/}
\BIBentrySTDinterwordspacing

\bibitem{BeatSaber}
\BIBentryALTinterwordspacing
Beat-Games, ``Beat {S}aber,'' 2018. [Online]. Available:
  \url{https://store.steampowered.com/app/620980/Beat_Saber/}
\BIBentrySTDinterwordspacing

\bibitem{SuperHotVR}
\BIBentryALTinterwordspacing
SUPERHOT-Team, ``Super {H}ot {VR},'' 2017. [Online]. Available:
  \url{https://www.playstation.com/en-us/games/superhot-vr-ps4/}
\BIBentrySTDinterwordspacing

\bibitem{JobSimulator}
\BIBentryALTinterwordspacing
{Owlchemy Labs}, ``Job simulator,'' 2016. [Online]. Available:
  \url{https://store.steampowered.com/app/448280/Job_Simulator}
\BIBentrySTDinterwordspacing

\bibitem{fischer2004energy}
S.~L. Fischer, P.~B. Watts, R.~L. Jensen, and J.~Nelson, ``Energy expenditure,
  heart rate response, and metabolic equivalents (mets) of adults taking part
  in children's games,'' \emph{The Jr. of Sports Medicine and Physical
  Fitness}, vol.~44, p. 398, 2004.

\bibitem{westerterp2009assessment}
K.~R. Westerterp, ``Assessment of physical activity: A critical appraisal,''
  \emph{Europ. Jr. of Applied Physiology}, vol. 105, pp. 823--828, 2009.

\bibitem{hills2014assessment}
A.~P. Hills, N.~Mokhtar, and N.~M. Byrne, ``Assessment of physical activity and
  energy expenditure: An overview of objective measures,'' \emph{Frontiers in
  Nutrition}, vol.~1, p.~5, 2014.

\bibitem{kennedy1993simulator}
R.~S. Kennedy, N.~E. Lane, K.~S. Berbaum, and M.~G. Lilienthal, ``Simulator
  sickness questionnaire: An enhanced method for quantifying simulator
  sickness,'' \emph{The Int. Jr. of Aviation Psychology}, vol.~3, pp. 203--220,
  1993.

\bibitem{kalawsky1999vruse}
R.~S. Kalawsky, ``{VRUSE--A} computerised diagnostic tool: For usability
  evaluation of virtual/synthetic environment systems,'' \emph{Applied
  ergonomics}, vol.~30, pp. 11--25, 1999.

\bibitem{nonmangiarenonbere}
S.~Bruck and P.~A. Watters, ``Estimating cybersickness of simulated motion
  using the simulator sickness questionnaire ({SSQ}): A controlled study,'' in
  \emph{Proc. 6th Int. Conf. on Computer Graphics, Imaging and Visualization},
  2009, pp. 486--488.

\bibitem{armswinger}
\BIBentryALTinterwordspacing
ElectricNightOwl, ``Arm {S}winger.'' [Online]. Available:
  \url{https://github.com/ElectricNightOwl/ArmSwinger}
\BIBentrySTDinterwordspacing

\bibitem{boletsis2019vr}
C.~Boletsis and J.~E. Cedergren, ``{VR} locomotion in the new era of virtual
  reality: {A}n empirical comparison of prevalent techniques,'' \emph{Advances
  in Human-Computer Interaction}, vol. 2019, 2019.

\bibitem{speed_max_alt}
J.~B. Cronin and K.~T. Hansen, ``Strength and power predictors of sports
  speed,'' \emph{Jr. Strength \& Cond. Res.}, vol.~19, pp. 349--357, 2005.

\bibitem{speed_walk}
R.~W. Bohannon, ``Comfortable and maximum walking speed of adults aged 20—79
  years: reference values and determinants,'' \emph{Age and Ageing}, vol.~26,
  pp. 15--19, 1997.

\bibitem{pre_ssq}
B.~K. Jaeger and R.~R. Mourant, ``Comparison of simulator sickness using static
  and dynamic walking simulators,'' in \emph{Proc. of Human Factors \&
  Ergonomics Society A.M.}, vol.~45, 2001, pp. 1896--1900.

\bibitem{vivecraft}
\BIBentryALTinterwordspacing
``{VIVEC}raft,'' 2019. [Online]. Available:
  \url{http://www.vivecraft.org/how-to-play/}
\BIBentrySTDinterwordspacing

\end{thebibliography}

\vskip -2\baselineskip  plus -1fil
\begin{IEEEbiography}[{\includegraphics[width=1in,height=1.25in,clip,keepaspectratio]{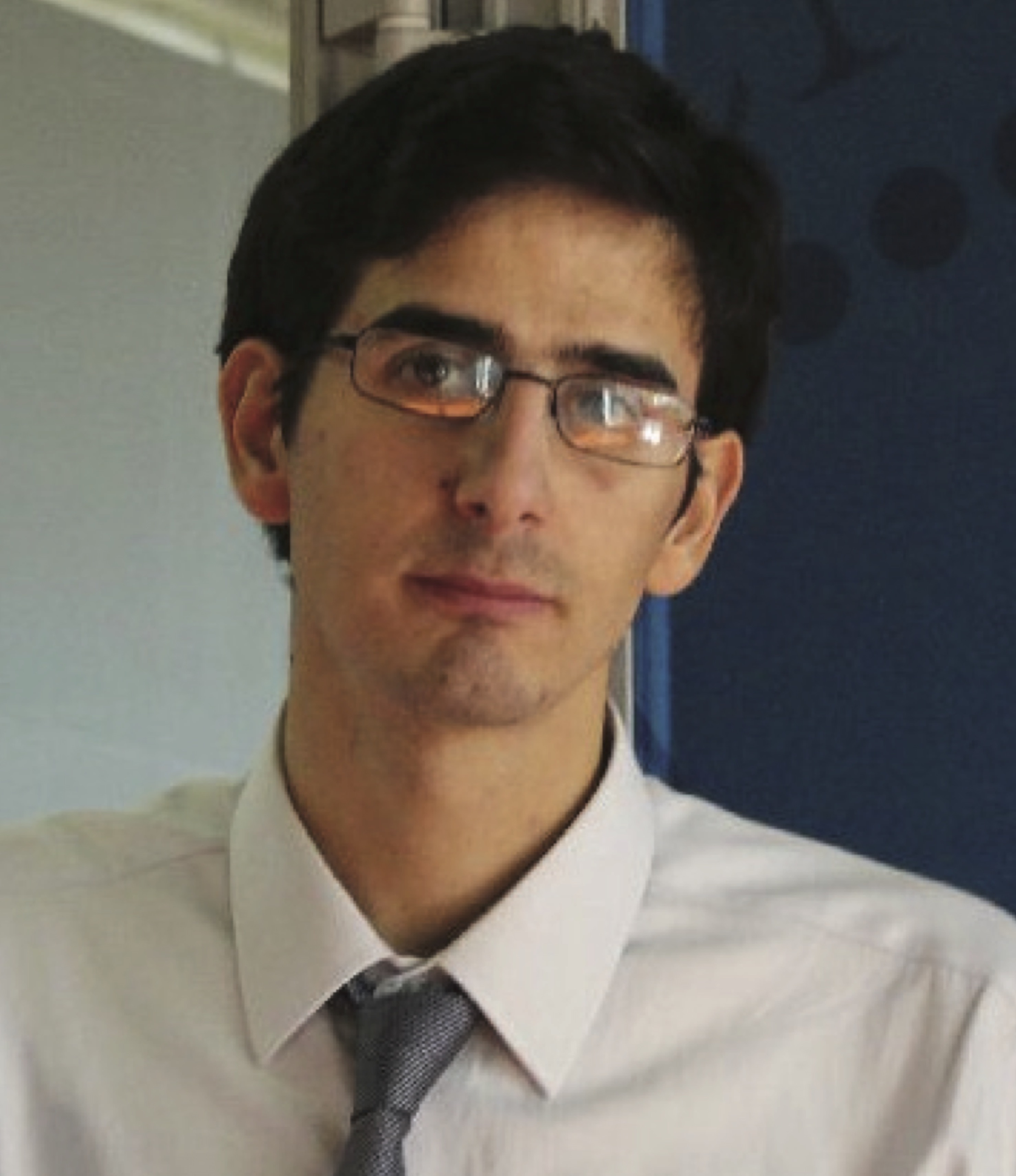}}]{Alberto Cannav\`o}
received his B.Sc. degree from University of Messina, Italy, in 2013. He received his M.Sc. and PhD. degrees in computer engineering from Politecnico di Torino, Italy, in 2015 and 2020. He now holds an assistant researcher position at the Dipartimento di Automatica e Informatica of Politecnico di Torino. His fields of interest include computer graphics and human-machine interaction.
\end{IEEEbiography}
\vskip -2\baselineskip  plus -1fil

\begin{IEEEbiography}[{\includegraphics[width=1in,height=1.25in,clip,keepaspectratio]{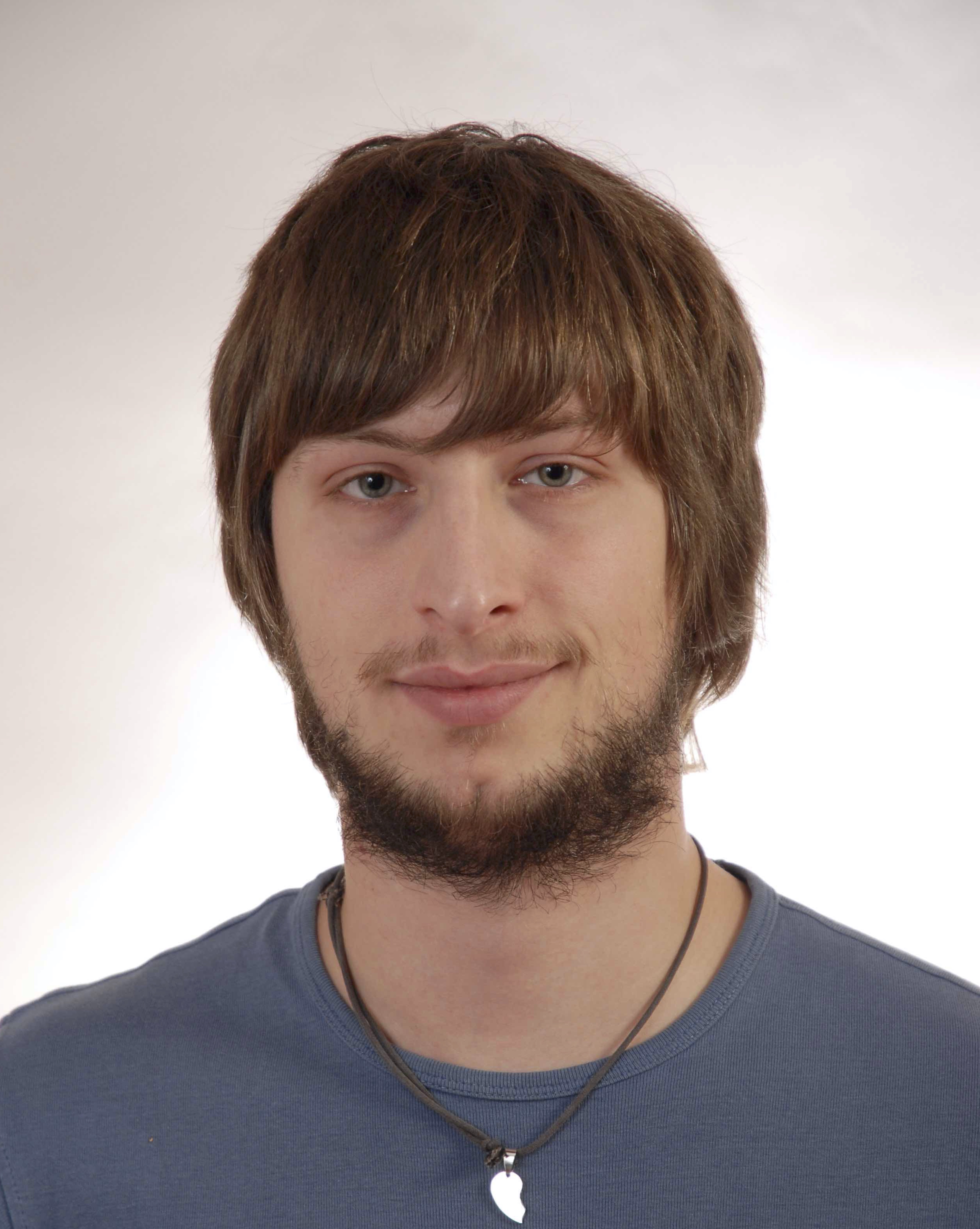}}]{Davide Calandra}
received his B.Sc. and M.Sc. degrees in computer engineering from Politecnico di Torino, Italy, in 2014 and 2017. Currently, he is a Ph.D. student at Politecnico di Torino, with interests in interactive graphics applications, virtual and augmented reality, and human-machine interaction.
\end{IEEEbiography}
\vskip -2\baselineskip  plus -1fil

\begin{IEEEbiography}[{\includegraphics[width=1in,height=1.25in,clip,keepaspectratio]{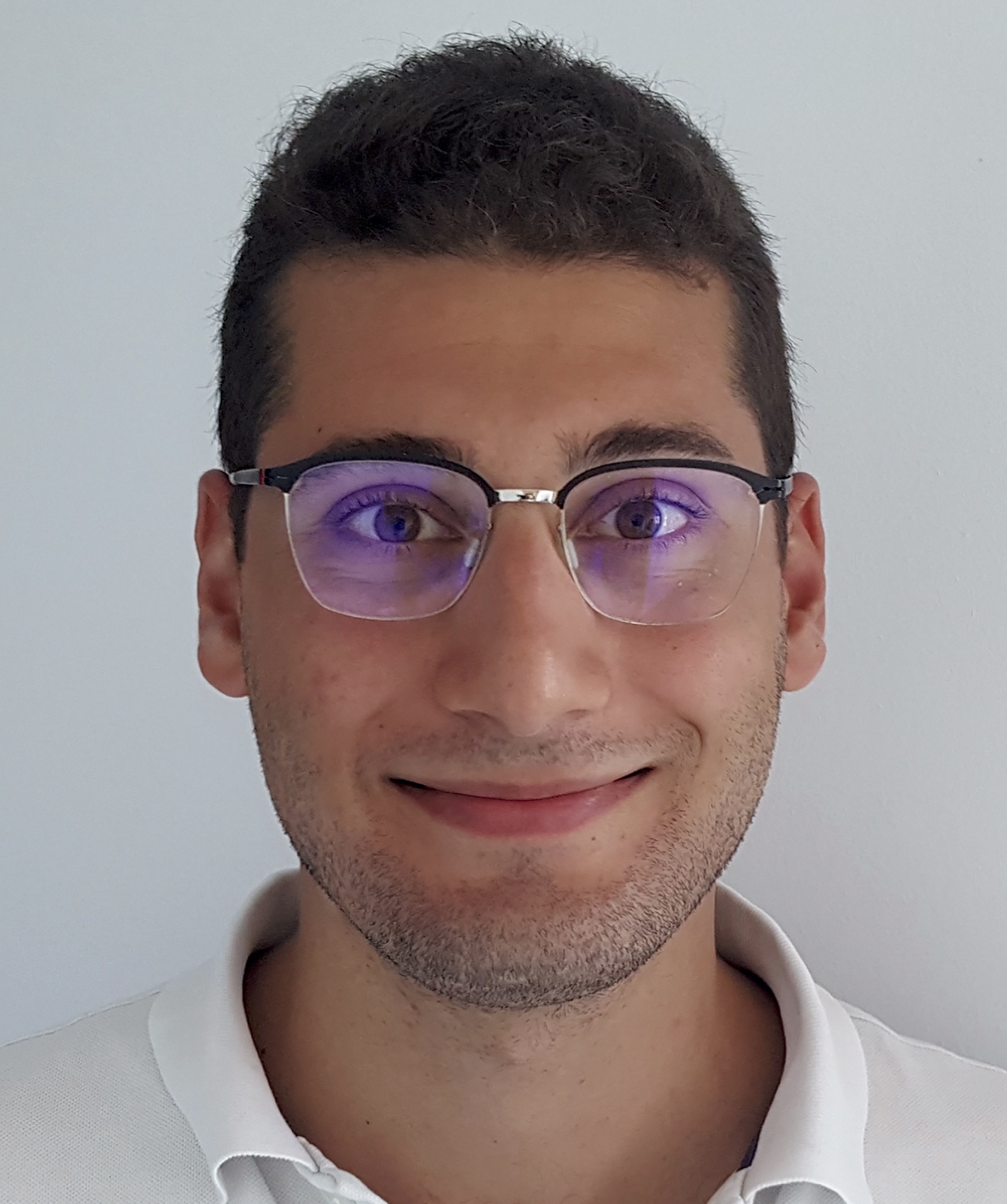}}]{Filippo Gabriele Prattic\`o}
received his B.Sc. and M.Sc. degrees in computer engineering from Politecnico di Torino, Italy, in 2014 and 2017. Currently, he is a Ph.D. student at Politecnico di Torino, where he carries out research in the areas of virtual, augmented and mixed reality, human-computer and human-robot interaction, serious games, and user experience design.
\end{IEEEbiography}
\vskip -2\baselineskip  plus -1fil

\begin{IEEEbiography}[{\includegraphics[width=1in,height=1.25in,clip,keepaspectratio]{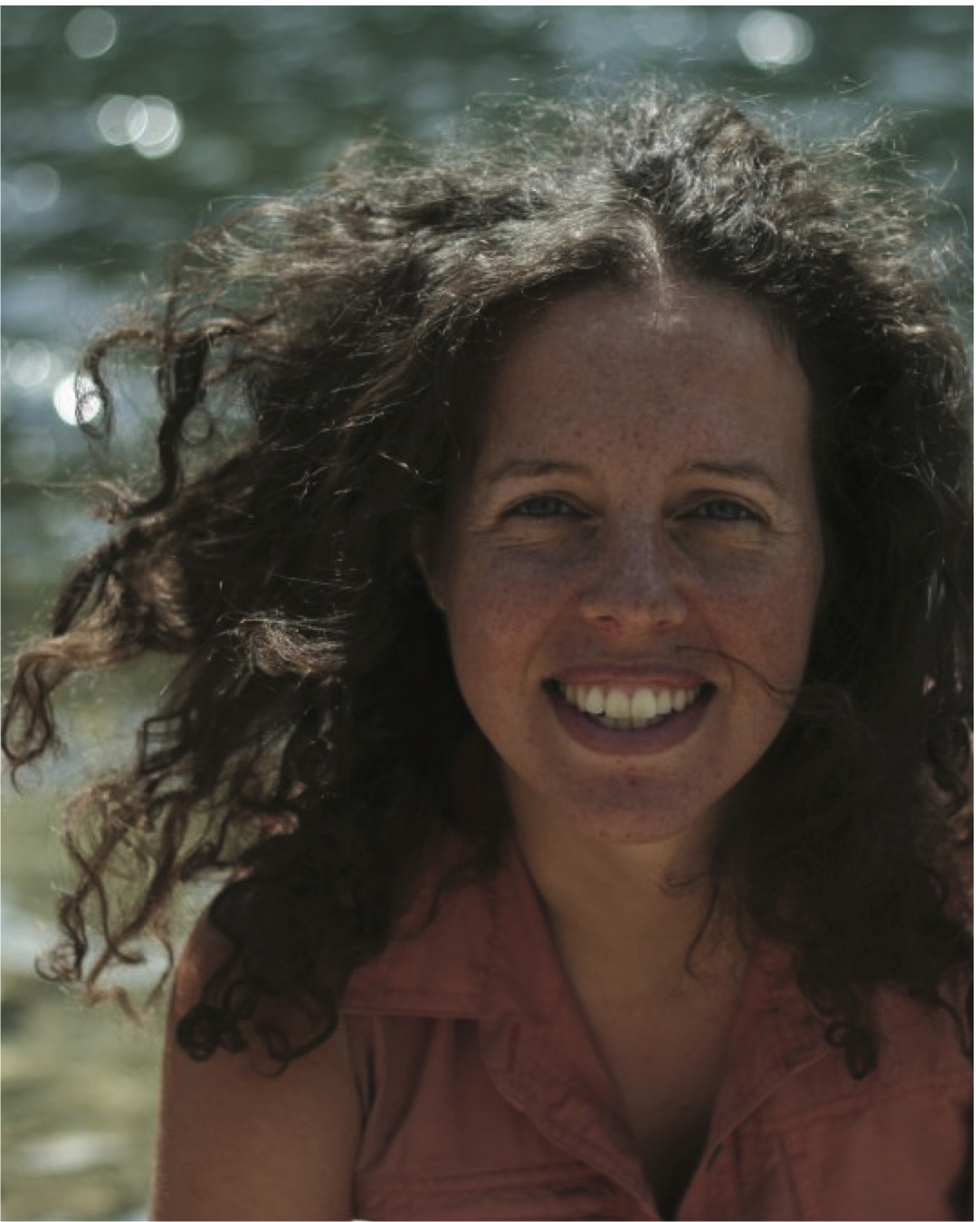}}]{Valentina Gatteschi}
received her B.Sc. and M.Sc. degrees in management engineering and her Ph.D. degree in computer engineering from Politecnico di Torino, Italy, in 2005, 2008 and 2013. She is now an assistant professor at the Dipartimento di Automatica e Informatica of Politecnico di Torino. Her interests include intelligent systems and their application to computer graphics.
\end{IEEEbiography}
\vskip -2\baselineskip  plus -1fil

\begin{IEEEbiography}[{\includegraphics[width=1in,height=1.25in,clip,keepaspectratio]{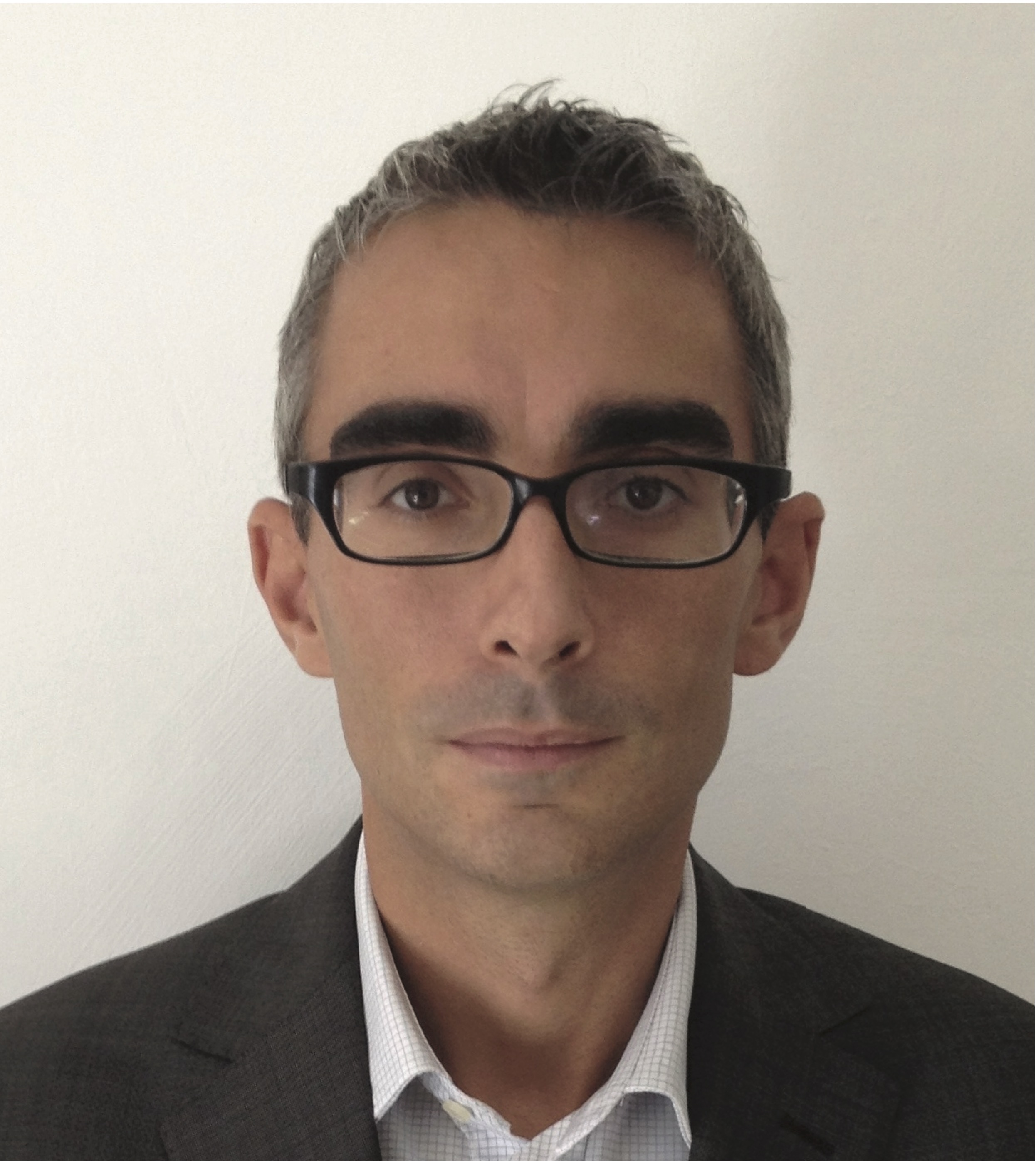}}]{Fabrizio Lamberti} received his M.Sc. and  Ph.D. degrees in computer engineering from Politecnico di Torino, Italy, in 2000 and 2005. He is now a full professor at at the Dipartimento di Automatica e Informatica of Politecnico di Torino. His interests pertain  computer graphics, human-machine interaction and computational intelligence. He serves as an Associate Editor for several journals, including the IEEE Transactions on Computers and the IEEE Transactions on Learning Technologies.
\end{IEEEbiography}





\end{document}


%
\title{An Evaluation Testbed for Locomotion in Virtual~Reality}
%
%
%
%

\author{Alberto Cannav\`o, \textit{Student Member,~IEEE}, Davide Calandra, F. Gabriele Prattic\`o, \textit{Student Member,~IEEE}, Valentina Gatteschi and Fabrizio Lamberti, \textit{Senior Member,~IEEE}
\IEEEcompsocitemizethanks{\IEEEcompsocthanksitem The authors are with the GRAINS -- GRAphics And INtelligent Systems group at the Dipartimento di Automatica e Informatica of Politecnico di Torino, 10129 Torino, Italy. e-mail: (see http://grains.polito.it/people.php).}
\thanks{Manuscript received XXX XX, XXXX; revised XXX XX, XXXX.}}

%
%

\markboth{IEEE TRANSACTIONS ON VISUALIZATION AND COMPUTER GRAPHICS,~Vol.~XX, No.~X, XXXX~XXXX}%
{Shell \MakeLowercase{\textit{et al.}}: Bare Demo of IEEEtran.cls for Computer Society Journals}
%


\onecolumn
\maketitle
\section*{Appendix A}
This Appendix shows results obtained in the user study carried out by considering four locomotion techniques: arm swinging (AS), walk-in-place (WIP), Cyberith's Virtualizer (CV) and joystick (JS).

Data collected in the pre-test section of the questionnaire including by the demographic questions and Simulator Sickness Questionnaire (SSQ) are reported in Table~\ref{tab:Demograph} and Table~\ref{tab:Before}, respectively. Values in Table~\ref{tab:Demograph} represent the percentage of users who selected the given option, whereas values in Table~\ref{tab:Before} indicate the average score assigned by the users (top of the cell) and the standard deviation (bottom of the cell). Columns $p$-value reports the results of the statistical analysis performed with ANOVA or Kruskal-Wallis. The outcome of the pairwise comparisons performed with the Tuckey's or Dunn's tests is reported using the * and ** symbols (to indicates a $p$-value less than or equal to $0.05$ and $0.01$, respectively).

Objectives and subjective data collected after each scenario are reported in Tables~\ref{tab:S1obj} --~\ref{tab:S5subj}. For objective metrics, the units of measurement are those given in the paper. 
Table~\ref{tab:heart-rate} reports average differences in users’ heart rate measured before and after the experimentation of the various scenarios.
Overall metrics, based on questions in the post-test section of the questionnaire, are reported in Table~\ref{tab:Overall}. For data reported in Tables~\ref{tab:S1obj} --~\ref{tab:Overall}, column $p$-value has the same meaning discussed above.

\section*{Appendix B}
This Appendix includes additional information required to prepare the raw data in the Raw Database (RDB) and compute the weighted scores in the Weighted Database (WDB) using the provided spreadsheet. Computation relies on the logs of the experimental activity (which are automatically generated by the framework application). Although for most of the tasks the generated output can be directly inserted in the RDB, in some cases a pre-processing is required. 

In particular, 
for task \textit{S2.T1}, the application generates six different lines in the log file, one for each target (artwork) on which the participant has to perform the assigned task. The six targets do not differ in terms of difficulty, so the value of a specific metric $m$ defined for the task (\textit{ComplTime}, \textit{InitAngErr}, \textit{EstPathLen} \textit{RecallTime}) for the $i$-user is calculated as:
\textbf{\begin{equation}
m_{i} = \frac{1}{6} \sum_{j=1}^{6} m_{i_{j} } 
\label{eq:fn1}
\end{equation}}

For tasks \textit{S2.T2}, \textit{S3.T1}, \textit{S3.T2}, and \textit{S3.T3}, metrics \textit{AccuracyBck}, \textit{AccuracyGazeUnc}, \textit{AccuracyStrc}, and \textit{AccuracyHandsUnc} are defined as compound. The spreadsheet requires the framework user to compute them using the equations provided in the paper. More precisely, since a normalization of the \textit{STPathDev} metric is necessary, the maximum distance available to the participant on the left/right side of the path (\textit{MaxDist}) are required. Values are as follows: 7m for \textit{S2.T2}, 5m for \textit{S3.T1}, \textit{S3.T2}, and \textit{S3.T3}. For task \textit{S3.T3}, the total number of coins (50) is used to calculate the \textit{ScoreRate} (\%) from the \textit{Score} (necessary for \textit{AccuracyHandsUnc}).

Similarly to \textit{S2.T1}, for tasks \textit{S1.T2} and \textit{S5.T1} the application log contains multiple lines. However, differently than with data collected for \textit{S2.T1}, data cannot be simply averaged, since they refer to measures collected under different conditions (three target sizes for \textit{S1.T2}, three different configurations for \textit{S5.T1}). Hence they have to be maintained as separate in the computation, and scores could be possibly combined afterwards.

For task \textit{S2.T4}, the participant's choice is logged as ``ST'' for the stairs, and ``SL'' for the ramp (slope).

\begin{center}
\begin{longtable}{|p{5.0cm}|p{1.5cm}|p{1.5cm}|p{1.5cm}|p{1.5cm}|p{1.5cm}|}
    \caption{Results based on demographic section of the questionnaire.} \label{tab:Demograph}\\
    \hline
          & Never & Little & Sometimes & Often & {Very often} \\
    \hline
    1) How often do you use Virtual Reality system such as HTC-VIVE, Oculus Rift, Cardboard? & 47.917 & 27.083 & 12.500 & 4.167 & 8.333 \\
    \hline
    2) How familiar are you with the locomotion technique you will use in this test?  & 70.833 & 20.833 & 8.333 & 0.000 & 0.000 \\
    \hline
    3) How often do you play 3D video-games using interaction technologies different than VR? & 47.917 & 14.583 & 6.250 & 12.500 & 18.750 \\
    \hline
    4) How often do you use 3D software such as Blender, Unity, 3DSmax, Maya? & 50.000 & 8.333 & 4.167 & 22.917 & 14.583 \\
    \hline
    5) How often do you suffer from motion sickness? & 77.083 & 18.750 & 4.167 & 0.000 & 0.000 \\
    \hline
    \end{longtable}%
\end{center}%

\newpage

\begin{center}
    \begin{longtable}{|c|c|c|c|c|c|c|c|}
    \caption{Subjective results based on pre-test section of the questionnaire.} \label{tab:Before}\\
    \hline
    \textbf{Task} & \textbf{Metric} & \textbf{\textit{p}-value} & \multicolumn{5}{c|}{\textbf{Techniques}} \\
    \hline
    Mot. sickness & Nausea  & 0.9364 &       & \textbf{AS} & \textbf{WIP} & \textbf{CV} & \textbf{JS} \\
\cline{4-8}          &       &       & \textbf{AS} & 3.975 &       &       &  \\
          &       &       &       & \textit{8.589} &       &       &  \\
\cline{4-8}          &       &       & \textbf{WIP} &       & 3.975 &       &  \\
          &       &       &       &       & \textit{8.589} &       &  \\
\cline{4-8}          &       &       & \textbf{CV} &       &       & 4.770 &  \\
\hline
    \end{longtable}%
\end{center}%

\begin{center}
    \begin{longtable}{|c|c|c|c|c|c|c|c|}
    \caption{Objective results for Scenario 1. Straight movements.} \label{tab:S1obj}\\
    \hline
    \textbf{Task} & \textbf{Metric} & \textbf{\textit{p}-value} & \multicolumn{5}{c|}{\textbf{Techniques}} \\
    \hline
    T1. Straight line walking & Accuracy (STPathDev) & 0.0019 &       & \textbf{AS} & \textbf{WIP} & \textbf{CV} & \textbf{JS} \\
\cline{4-8}          &       &       & \textbf{AS} & 0.345 &       &       &  \\
          &       &       &       & \textit{0.245} &       &       &  \\
\cline{4-8}          &       &       & \textbf{WIP} & *     & 1.532 &       &  \\
          &       &       &       &       & \textit{1.628} &       &  \\
\cline{4-8}          &       &       & \textbf{CV} & *     &       & 1.213 &  \\
\hline
    \end{longtable}%
\end{center}

\begin{center}
    \begin{longtable}{|c|p{4.5cm}|c|c|c|c|c|c|}
    \caption{Subjective results for Scenario 1. Straight movements.} \label{tab:S1subj}\\
    \hline
    \textbf{Metric} & \textbf{Statement} & \textbf{\textit{p}-value} & \multicolumn{5}{c|}{\textbf{Techniques}} \\
    \hline
    Ease of use & \multirow{2}{*}{\parbox{4.5cm}{22) I found the interface easy to use}} & 0.0154 &       & \textbf{AS} & \textbf{WIP} & \textbf{CV} & \textbf{JS} \\
\cline{4-8}          &       &       & \textbf{AS} & 4.500 &       &       &  \\
          &       &       &       & \textit{0.522} &       &       &  \\
\cline{4-8}          &       &       & \textbf{WIP} & *     & 3.583 &       &  \\
          &       &       &       &       & \textit{0.793} &       &  \\
\cline{4-8}          &       &       & \textbf{CV} &       &       & 3.917 &  \\
\hline
    \end{longtable}%
\end{center}%

\begin{center}
    \begin{longtable}{|c|c|c|c|c|c|c|c|}
    \caption{Objective results for Scenario 2. Direction control.} \label{tab:S2obj}\\
    \hline
    \textbf{Task} & \textbf{Metric} & \textbf{\textit{p}-value} & \multicolumn{5}{c|}{\textbf{Techniques}} \\
    \hline
    T1. Multi-straight line walking & Accuracy (InitAngErr) & 0.0001 &       & \textbf{AS} & \textbf{WIP} & \textbf{CV} & \textbf{JS} \\
\cline{4-8}          &       &       & \textbf{AS} & 5.680 &       &       &  \\
          &       &       &       & \textit{3.463} &       &       &  \\
\cline{4-8}          &       &       & \textbf{WIP} &       & 8.365 &       &  \\
          &       &       &       &       & \textit{6.116} &       &  \\
\cline{4-8}          &       &       & \textbf{CV} & **    &       & 30.428 &  \\
\hline
    \end{longtable}%
\end{center}%

\begin{center}
    \begin{longtable}{|c|p{4.5cm}|c|c|c|c|c|c|}
    \caption{Subjective results for Scenario 2. Direction control.} \label{tab:S2subj}\\
    \hline
    \textbf{Metric} & \textbf{Statement} & \textbf{\textit{p}-value} & \multicolumn{5}{c|}{\textbf{Techniques}} \\
    \hline
    Ease of use & 22) I found the interface easy to use & 0.7498 &       & \textbf{AS} & \textbf{WIP} & \textbf{CV} & \textbf{JS} \\
\cline{4-8}          &       &       & \textbf{AS} & 4.333 &       &       &  \\
          &       &       &       & \textit{0.492} &       &       &  \\
\cline{4-8}          &       &       & \textbf{WIP} &       & 4.083 &       &  \\
          &       &       &       &       & \textit{0.793} &       &  \\
\cline{4-8}          &       &       & \textbf{CV} &       &       & 4.167 &  \\
\hline
    \end{longtable}%
\end{center}%

\begin{center}
    \begin{longtable}{|c|c|c|c|c|c|c|c|}
        \caption{Objective results for Scenario 3. Decoupled movements.} \label{tab:S3obj}\\
        \hline
        \textbf{Task} & \textbf{Metric} & \textbf{\textit{p}-value} & \multicolumn{5}{c|}{\textbf{Techniques}} \\
        \hline
        T1. Decoupled gaze  & Accuracy (AccuracyGazeUnc) & 0.0007 &       & \textbf{AS} & \textbf{WIP} & \textbf{CV} & \textbf{JS} \\
    \cline{4-8}          &       &       & \textbf{AS} & 0.971 &       &       &  \\
              &       &       &       & \textit{0.016} &       &       &  \\
    \cline{4-8}          &       &       & \textbf{WIP} &       & 0.953 &       &  \\
              &       &       &       &       & \textit{0.032} &       &  \\
    \cline{4-8}          &       &       & \textbf{CV} & **    &       & 0.884 &  \\
\hline
    \end{longtable}%
\end{center}%

\begin{center}
    \begin{longtable}{|c|p{4.5cm}|c|c|c|c|c|c|}
    \caption{Subjective results for Scenario 3. Decoupled movements.} \label{tab:S3subj}\\
    \hline
    \textbf{Metric} & \textbf{Statement} & \textbf{\textit{p}-value} & \multicolumn{5}{c|}{\textbf{Techniques}} \\
    \hline
    Ease of use & 22) I found the interface easy to use & 0.0062 &       & \textbf{AS} & \textbf{WIP} & \textbf{CV} & \textbf{JS} \\
\cline{4-8}          &       &       & \textbf{AS} & 3.833 &       &       &  \\
          &       &       &       & \textit{0.718} &       &       &  \\
\cline{4-8}          &       &       & \textbf{WIP} &       & 4.250 &       &  \\
          &       &       &       &       & \textit{0.866} &       &  \\
\cline{4-8}          &       &       & \textbf{CV} &       &       & 4.417 &  \\
\hline
    \end{longtable}%
\end{center}%

\begin{center}
    \begin{longtable}{|c|c|c|c|c|c|c|c|}
    \caption{Objective results for Scenario 4. Agility.} \label{tab:S4_obj}\\
    \hline
    \multicolumn{1}{|c|}{\textbf{Task}} & \textbf{Metric} & \textbf{\textit{p}-value} & \multicolumn{5}{c|}{\textbf{Techniques}} \\
    \hline
    \multicolumn{1}{|l|}{T1. Dynamic agility} & Operation speed (ComplTime) & 0.0001 &       & \textbf{AS} & \textbf{WIP} & \textbf{CV} & \textbf{JS} \\
\cline{4-8}          &       &       & \textbf{AS} & 17.046 &       &       &  \\
          &       &       &       & \textit{6.585} &       &       &  \\
\cline{4-8}          &       &       & \textbf{WIP} &       & 35.007 &       &  \\
          &       &       &       &       & \textit{16.081} &       &  \\
\cline{4-8}          &       &       & \textbf{CV} & **    &       & 89.481 &  \\
\hline
    \end{longtable}%
\end{center}%

\begin{center}
    \begin{longtable}{|c|p{4.5cm}|c|c|c|c|c|c|}
        \caption{Subjective results for Scenario 4. Agility.} \label{tab:S4subj}\\
        \hline
        \textbf{Metric} & \textbf{Statement} & \textbf{\textit{p}-value} & \multicolumn{5}{c|}{\textbf{Techniques}} \\
        \hline
        Ease of use & \multirow{2}{*}{\parbox{4.5cm}{22) I found the interface easy to use}} & 0.0115 &       & \textbf{AS} & \textbf{WIP} & \textbf{CV} & \textbf{JS} \\
    \cline{4-8}          &       &       & \textbf{AS} & 4.250 &       &       &  \\
              &       &       &       & \textit{0.754} &       &       &  \\
    \cline{4-8}          &       &       & \textbf{WIP} &       & 3.750 &       &  \\
              &       &       &       &       & \textit{0.866} &       &  \\
    \cline{4-8}          &       &       & \textbf{CV} &       &       & 3.417 &  \\
\hline
    \end{longtable}%
\end{center}%

\begin{center}
    \begin{longtable}{|c|c|c|c|c|c|c|c|}
    \caption{Objective results for Scenario 5. Interaction with objects.} \label{tab:S5obj}\\
    \hline
    \textbf{Task} & \textbf{Metric} & \textbf{\textit{p}-value} & \multicolumn{5}{c|}{\textbf{Techniques}} \\
    \hline
    T1. Grabbing & Operation speed (ComplTime\_1) & 0.0162 &       & \textbf{AS} & \textbf{WIP} & \textbf{CV} & \textbf{JS} \\
\cline{4-8}          &       &       & \textbf{AS} & 19.074 &       &       &  \\
          &       &       &       & \textit{11.405} &       &       &  \\
\cline{4-8}          &       &       & \textbf{WIP} &       & 23.778 &       &  \\
          &       &       &       &       & \textit{14.071} &       &  \\
\cline{4-8}          &       &       & \textbf{CV} &       &       & 19.552 &  \\
\hline
    \end{longtable}%
\end{center}%

\begin{center}
    \begin{longtable}{|c|p{4.5cm}|c|c|c|c|c|c|}
    \caption{Subjective results for Scenario 5. Interaction with objects.} \label{tab:S5subj}\\
    \hline
    \textbf{Metric} & \textbf{Statement} & \textbf{\textit{p}-value} & \multicolumn{5}{c|}{\textbf{Techniques}} \\
    \hline
    Ease of use & \multirow{2}{*}{\parbox{4.5cm}{22) I found the interface easy to use}} & 0.0115 &       & \textbf{AS} & \textbf{WIP} & \textbf{CV} & \textbf{JS} \\
\cline{4-8}          &       &       & \textbf{AS} & 4.250 &       &       &  \\
          &       &       &       & \textit{0.754} &       &       &  \\
\cline{4-8}          &       &       & \textbf{WIP} &       & 3.750 &       &  \\
          &       &       &       &       & \textit{0.866} &       &  \\
\cline{4-8}          &       &       & \textbf{CV} &       &       & 3.417 &  \\
\hline
    \end{longtable}%
\end{center}%

\begin{center}
    \begin{longtable}{|c|c|c|c|c|c|c|}
    \caption{Average differences in users' heart rate measured before and after the experimentation of the various scenarios.} \label{tab:heart-rate}\\
    \hline
    \textbf{Scenario} & \textbf{\textit{p}-value} & \multicolumn{5}{c|}{\textbf{Techniques}} \\
    \hline
    Straight movements & 0.0001 &       & \textbf{AS} & \textbf{WIP} & \textbf{CV} & \textbf{JS} \\
\cline{3-7}          &       & \textbf{AS} & 11.667 &       &       &  \\
          &       &       & \textit{15.078} &       &       &  \\
\cline{3-7}          &       & \textbf{WIP} & *     & 23.667 &       &  \\
          &       &       &       & \textit{7.536} &       &  \\
\cline{3-7}          &       & \textbf{CV} & *     &       & 25.667 &  \\
\hline
    \end{longtable}%
\end{center}%

\begin{center}
    \begin{longtable}{|c|p{4.5cm}|c|c|c|c|c|c|}
    \caption{Subjective results based on post-test section of the questionnaire.} \label{tab:Overall}\\
    \hline
    \textbf{Metric} & \textbf{Statement} & \textit{p-value} & \multicolumn{5}{c|}{\textbf{Techniques}} \\
    \hline
    Self-motion compellingness & \multirow{2}{*}{\parbox{4.5cm}{54) Rate whether you indeed felt as if you were moving while walking and interacting with the virtual environment, overall}}& 0.0060 &       & \textbf{AS} & \textbf{WIP} & \textbf{CV} & \textbf{JS} \\
\cline{4-8}          &       &       & \textbf{AS} & 2.667 &       &       &  \\
          &       &       &       & \textit{0.985} &       &       &  \\
\cline{4-8}          &       &       & \textbf{WIP} &       & 3.583 &       &  \\
          &       &       &       &       & \textit{0.669} &       &  \\
\cline{4-8}          &       &       & \textbf{CV} &       &       & 3.500 &  \\
\hline
    \end{longtable}%
\end{center}%

\twocolumn
